\documentclass[a4paper,11pt]{article}
\usepackage[compatibility=false]{caption}
\usepackage{jheppub}
\usepackage{graphicx}
\usepackage{subcaption}
\usepackage{braket}
\usepackage{multirow}
\usepackage{comment}

\newcommand{\dmsq}[1]{\Delta m^{2}_{#1}}
\newcommand{\dmsqeff}[1]{\Delta m^{2\,\mathrm{eff}}_{#1}}  
\newcommand{\eff}{^{\mathrm{eff}}}

\newcommand{\dCP}{\delta_{\mathrm{CP}}}
\newcommand{\PMNS}{U_{\text{PMNS}}}
\newcommand{\diag}{\text{diag}}

\newcommand{\ACC}{A_{CC}}
\newcommand{\Amm}{A_{\mu \mu}}
\newcommand{\Att}{A_{\tau \tau}}
\newcommand{\Aaa}{A_{\alpha \alpha}}

\newcommand{\Dee}{\Delta_{ee}}
\newcommand{\Dmm}{\Delta_{\mu\mu}}
\newcommand{\Dtt}{\Delta_{\tau \tau}}
\newcommand{\Daa}{\Delta_{\alpha \alpha}}

\newcommand{\DeeS}{\Delta^{S}_{ee}}
\newcommand{\AeeS}{A^{S}_{ee}}
\newcommand{\DmmS}{\Delta^{S}_{\mu\mu}}
\newcommand{\AmmS}{A^{S}_{\mu\mu}}
\newcommand{\DttS}{\Delta^{S}_{\tau\tau}}
\newcommand{\AttS}{A^{S}_{\tau\tau}}
\newcommand{\DaaS}{\Delta^{S}_{\alpha\alpha}}
\newcommand{\AaaS}{A^{S}_{\alpha\alpha}}

\newcommand{\dmee}{\delta m_{ee}}
\newcommand{\dmmm}{\delta m_{\mu \mu}}
\newcommand{\dmtt}{\delta m_{\tau \tau}}
\newcommand{\dmaa}{\delta m_{\alpha \alpha}}

\title{The Scalar MSW Effect: Compact analytical formulas for neutrino oscillations with large matter effects, including $\mu -\mu$ and $\tau -\tau$ dominant matter potentials }

\author{Sandhya Choubey, Andreas Lund}
\affiliation{Department of Physics, School of Engineering Sciences, KTH Royal Institute of Technology, AlbaNova University Center, Roslagstullsbacken 21, SE--106~91 Stockholm, Sweden}
\affiliation{The Oskar Klein Centre for Cosmoparticle Physics, AlbaNova University Center, Roslagstullsbacken 21, SE--106 91 Stockholm, Sweden}

\emailAdd{choubey@kth.se}
\emailAdd{alund7@kth.se}

\abstract{Deriving simple and accurate three-flavor neutrino oscillation probability formulas in matter is critical for understanding the phenomenology of upcoming experiments. In this article, we extend the Jacobi diagonalization method for deriving effective parameter mappings for neutrino oscillations in matter. The key insight behind the extension is to purposefully parameterize the mixing matrix, $\PMNS$, using particular arrangements of Euler rotations. This novel extension allows for deriving simple and accurate effective mixing parameter mappings for exotic types of matter potentials. 
First, we revisit the Jacobi diagonalization method for neutrino oscillations with standard matter effects and use this to demonstrate our generalization. 
Subsequently, we apply the Jacobi diagonalization method to scalar non-standard interactions (SNSI) and derive effective parameter mappings for all three single diagonal coupling cases. The resulting formulas reveal that all diagonal SNSI couplings could give rise to a scalar MSW (SMSW) effect, that is an energy independent resonant enhancement analogous to the MSW effect. Finally, we study the accuracy of the derived SNSI formulas for various contemporary and upcoming experiments. For $\dmee$, the accuracy is excellent for all considered experiments. For $\delta m_{\mu\mu}$ and $\dmtt$ the accuracy is worse since $\ACC$ must be neglected, and first order perturbative corrections are applied to amend this. }

\begin{document}
\maketitle
\flushbottom

\section{Introduction}

Neutrino flavor oscillations have been a subject of great interest in the past decades. Neutrino oscillation experiments have provided a good measurement on most of the neutrino oscillation parameters \cite{NuFit6.0}. However, a few loose ends still remain. The most important among those are the CP phase in the leptonic sector and the neutrino mass ordering. Bigger and better experiments are planned to measure these remaining parameters as well as to improve the measured sensitivity on the other previously measured oscillation parameters. These highly precise experiments will also have the capability to look for signs of new physics beyond the standard model (BSM) three-flavor neutrino oscillations. It is therefore important to have analytical formulas for the neutrino oscillation probabilities in the presence of BSM physics, which can help to better understand their signature in future experiments.
A significant part of BSM physics scenarios give rise to additional contributions to neutrino matter effects.
Therefore, one needs neutrino oscillation formulas that can take into account these new matter effects within the full three-generation framework. 

Finding correct analytical solutions to the three-generation neutrino oscillation probabilities in the presence of matter effects constitutes an important area of research in its own right. 
Exact three-flavor oscillation probability formulas in presence of matter effects are exceptionally impenetrable to calculate analytically, even in the simplest case with only standard model interactions, and so one needs to rely on approximate methods to obtain insightful formulas. Perhaps the most prominent method for calculating neutrino oscillation probabilities has been standard quantum mechanical time-independent perturbation theory \cite{Schrodinger:1926gei,Sakurai:2011zz,Freund:2001pn,Akhmedov:2004ny,Minakata:2015gra,Li:2016pzm,Chaves:2018sih}. In this work, we will focus on another technique, known as the Jacobi diagonalization method, and in particular its use for finding the neutrino oscillation probabilities in the presence of matter effects coming from BSM physics. 

The Jacobi diagonalization method \cite{Jacobi:1845} has garnered newfound interest for understanding the three-flavor neutrino propagation Hamiltonian in matter \cite{Honda:2006hp, Honda:2006gv,Agarwalla:2013tza,Blennow:2013rca, Denton:2016wmg, Ge:2016dlx, Ioannisian:2018qwl,Denton:2018fex}. The basic procedure is to rotate the 3x3 Hamiltonian matrix to an approximately diagonal basis. The Jacobi method constructs a set of rotations such that after every rotation one gets either zero or increasingly negligible elements in the off-diagonal positions. Neglecting these small elements returns a completely diagonal Hamiltonian containing the neutrino mass eigenvalues in matter. This ultimately returns an orthogonal transformation that depends on three independent rotation angles. These rotation angles constitute the so-called ``effective mixing angles" in presence of matter, which can be expressed in terms of the intermediate Jacobi rotation angles. Finally, the effective mass squared differences in matter can be obtained using the diagonal elements of the Hamiltonian matrix. Note that these effective parameter mappings are generally applicable to all flavor oscillation probabilities, $P_{\alpha \beta}$.


The strength of using the Jacobi diagonalization method for understanding neutrino oscillations, as opposed to standard perturbation theory, is that the approximation holds excellently for high matter densities, and for a broad range of baselines and energies \cite{Ge:2016dlx}. Consequently, the method yields resonant enhancement relations for both the atmospheric and solar sectors in a unified three-generation framework.
On the other hand, a significant limitation of this version of the Jacobi diagonalization method has been that it requires a particular matrix structure for the matter potential, {\it viz.,} the matrix must have a non-zero contribution only for the $e$--$e$ element. This works perfectly for matter effects with standard $W-$mediated weak interactions since normal matter contains only electrons and the $W-$boson only allows $\nu_e$--$e$ scattering. However, when new physics is taken into account, it is possible to have contributions coming from scatterings of any of the neutrino flavors with matter. This could give rise to matter effect contributions in every element of the matter potential matrix, both diagonal as well as off-diagonal, which is completely at odds with the Jacobi diagonalization procedure as employed in the literature. 

In this work, we generalize the Jacobi diagonalization method to include matter potential matrices that have a contribution in {\it  any one} of the three diagonal elements. That is, we allow for matter effects in  
the $e$--$e$ element or the $\mu$--$\mu$ element or the $\tau$--$\tau$ element, taken one at a time, and use the Jacobi method to diagonalize the full Hamiltonian to find effective mixing angles and mass squared difference. 
This is achieved by carefully considering the parameterization of the neutrino mixing matrix, $\PMNS$, based on the structure of the specific matter potential matrix under consideration. We stress that the standard Particle Data Groups (PDG) \cite{ParticleDataGroup:2024cfk} choice of  parametrizating the PMNS matrix, that is $U^e = R_{23} U_\delta R_{13} {U_\delta}^\dagger R_{12}$, is but only one possible way to parametrize $\PMNS$. One may in principle choose other valid parameterizations whenever appropriate. 
Different options for parameterization of the mixing matrix have been classified and discussed by Fritzsch and Xing \cite{Fritzsch:1997st, Fritzsch:2001ty}. The benefit of using the PDG parameterization in particular has been motivated in previous literature \cite{Rodejohann:2011vc, Denton:2020igp}.
Indeed, the PDG parameterization has de facto become the lingua franca of neutrino oscillation phenomenology. 
In this context, our work highlights the utility of freely choosing other parameterizations based on the phenomenology at hand.
We will show which parametrization is needed for which matter potential structure, and why. 


In the present work, we particularly focus on scalar non-standard interactions (SNSI), which is a BSM physics scenario that would affect neutrino propagation via a specific type of matter potential. 
In previous articles \cite{Gupta:2023wct, Singha:2023set}, the Jacobi method was successfully employed to diagonalize the neutrino matter Hamiltonian in the presence of a single SNSI parameter, $\dmee$. The default Jacobi diagonalization works out in this case because the matter potential comes in the $e$--$e$ element, just like standard matter effects, for which the PDG parametrization is appropriate. 
We generalize these results for SNSI by doing the calculation for $\delta m_{\mu\mu}$ and $\delta m_{\tau\tau}$ as well. We stress that, in this work, we only consider one dominant SNSI parameter at a time. Still, this is the first time in the literature that analytical expressions for neutrino oscillation probabilities in presence of the SNSI parameters $\delta m_{\mu\mu}$ and $\delta m_{\tau\tau}$ have been derived using the Jacobi diagonalization method. Our results for these parameters are distinct from the ones previously derived with other methods \cite{Denton:2022pxt,Bezboruah:2024yhk,Denton_2024} in that we have accounted for the full three-flavor picture and that the formulas remain valid even for very large matter potentials, including resonant effects. Since the SNSI-driven resonant enhancement of the mixing angle is similar, but in various ways distinct, from the one we see for standard matter effects, we refer to such resonances as the scalar MSW (SMSW) effect.

We will use the Jacobi method only when there is contribution at one of the diagonal elements of the neutrino matter potential matrix. Note that when new physics contributes to the matter potential matrix, we also have a contribution from the standard matter effect. Therefore, in realistic situations in general, the considered SNSI scenarios result in at least two terms in the matter potential matrix simultaneously. 
Since the standard matter effects come in the $e$--$e$ element, for the case of SNSI parameter $\dmee$, we have both contributions in the $e$--$e$ element. Therefore, in this case we can take the standard and SNSI matter effects simultaneously and employ the Jacobi diagonalization method to get the effective neutrino oscillation parameters. These parameters can then be inserted into the standard three-generation neutrino oscillation probability expressions for constant matter density. However, for the $\dmmm$ and $\dmtt$ cases, the standard matter potential comes in the $e$--$e$ element while the SNSI matter potential comes in the $\mu$--$\mu$ and $\tau$--$\tau$ elements, respectively. Therefore, for these cases we will do the following. We will include SNSI matter effects via the Jacobi diagonalization method, while standard matter effects will be taken as first order perturbative corrections. We will show that amongst the forthcoming experiments, the perturbativity condition holds only well for the JUNO experiment \cite{JUNO:2022mxj}. For the long-baseline experiments we can still use our method, however, for them we insert our effective oscillation parameters in presence of SNSI into expressions for probabilities obtained in \cite{Martinez-Soler:2019nhb}. These probability formulas already include the standard matter effects and yield good approximations for low energy long-baseline experiments such as Hyper-Kamiokande \cite{Hyper-Kamiokande:2018ofw} and ESSnuSB \cite{ESSnuSB:2023ogw}. 



This article is structured as follows. In Sec.~\ref{sec:standardderivation}, we review in detail the Jacobi diagonalization procedure for deriving effective oscillation parameter mappings for the standard $e$--$e$ dominant matter potential. Subsequently, in Sec.~\ref{sec:generalizing} we generalize the Jacobi diagonalization procedure to vector-like $\mu$--$\mu$ and $\tau$--$\tau$ dominant matter potentials. In Sec.~\ref{sec:SNSImain}, we consider scalar non-standard interactions and employ our generalized Jacobi diagonalization method for each of the three single diagonal element cases. In Sec.~\ref{sec:approximationaccuracyandcorrections}, we numerically showcase the accuracy of the formulas derived in Sec.~\ref{sec:SNSImain}. We further work employ first order perturbative corrections to these expressions, which is primarily used to improve the accuracy of the $\mu$--$\mu$ and $\tau$--$\tau$ formulas in certain regimes. Finally, in Sec.~\ref{sec:discussionandconclusion} we discuss our results and present our conclusions. The supplementary material provides more detail on mixing matrix parameterization and on perturbation theory.

\section{The Jacobi diagonalization Method for Standard Matter Effects} \label{sec:standardderivation}
We consider the neutrino Hamiltonian in matter written as \cite{10.1093/acprof:oso/9780198508717.001.0001}
\begin{equation} \label{eq:HamiltonianStandard}
    H = \frac{\PMNS M^2 \PMNS^\dagger + A}{2 E},
\end{equation}
where we define $M^2=\diag(m_{1}^2,m_2^2,m_3^2)$ and $A= \diag(\ACC,0,0)$. Here, $m_i$ are the neutrino masses and $\ACC = \pm 2 E V_{CC}$ is the charged current matter potential, where we define $V_{CC}=\sqrt{2}G_FN_AY_e\rho $, with $G_F$ the Fermi constant, $N_A$ the Avogadro's constant, $Y_e$ the electron fraction in matter and $\rho$ the density of matter. The positive sign of the matter term is relevant for neutrinos and the negative sign for antineutrinos. 
Note that we use $\PMNS$ to mean the neutrino mixing matrix, for which we will take different parameterizations in different parts of this work. 
In this section it is parameterized with the conventional PDG choice
\begin{equation} \label{eq:Uee}
    U^e(\theta_{12},\theta_{13}, \theta_{23}, \dCP) = R_{23} U_\delta R_{13} U_\delta^\dagger R_{12},
\end{equation}
where $R_{ij}$ are the Euler rotation matrices
\begin{equation} \label{eq:Eulerrotations}
    R_{12}= \begin{pmatrix}
        c_{12} & s_{12} & 0 \\
         -s_{12} & c_{12} & 0 \\
          0 & 0 & 1 \\
    \end{pmatrix}, \quad R_{13}= \begin{pmatrix}
        c_{13} & 0 & s_{13} \\
         0 & 1 & 0 \\
          -s_{13} & 0 & c_{13} \\
    \end{pmatrix}  \quad R_{23}= \begin{pmatrix}
        1 & 0 & 0 \\
        0 & c_{23} & s_{23} \\
        0 & -s_{23} & c_{23} \\
    \end{pmatrix},
\end{equation}
with $c_{ij} \equiv \cos \theta_{ij}$ and  $s_{ij} \equiv \sin \theta_{ij}$, and $U_\delta = \diag(1,1, e^{i \dCP})$.
As we shall see, the ability of the Jacobi method to elegantly diagonalize Eq.~\eqref{eq:HamiltonianStandard} relies intricately on the parameterization, where $U^e$ turns out to be the appropriate choice only for this specific matter potential matrix structure. 

In what follows, we shall rotate the Hamiltonian three times to reach the diagonal basis. For the first rotation, 
we choose an \textit{auxiliary basis},
\begin{equation} \label{eq:Uae}
    U_{a_e}=R_{23} U_{\delta}  R_{13},
\end{equation}
such that 
\begin{align}
    H' &= U_{a_e}^\dagger H U_{a_e} = \frac{U_\delta R_{12}^\dagger M^2 R_{12} U_\delta^\dagger + U_{a_e}^\dagger \diag(\ACC,0,0) U_{a_e}}{2 E} \nonumber \\
\label{eq:Hpstandard}
    &= \frac{1}{2 E}\begin{pmatrix}
        s_{12}^2 \Delta m_{21}^2 + c_{13}^2 \ACC & s_{12} c_{12} \Delta m_{21}^2 & c_{13} s_{13} \ACC \\
        s_{12} c_{12} \Delta m_{21}^2 & c_{12}^2 \Delta m_{21}^2 & 0 \\
        c_{13} s_{13} \ACC & 0 & \Delta m_{31}^2 + s_{13}^2 \ACC \\
    \end{pmatrix}.
\end{align}
The choice of this auxiliary basis will be thoroughly discussed in Sec.~\ref{sec:generalizing}. Here we spell out the method itself. 
Subsequently, we rotate the Hamiltonian as $H''= R_{13}'^\dagger H' R_{13}'$ such that 
\begin{align}
    &H''= \nonumber \\&\begin{pmatrix}
        {c_{13}'}^2 H_{11}' + {s_{13}'}^2 H_{33}'-2 c_{13}' s_{13}' H_{13}'  & 
        c_{13}' H_{12}' - s_{13}' H_{23}' &  
        H_{13}'({c_{13}'}^2-{s_{13}'}^2)+c_{13}' s_{13}' (H_{11}'-H_{33}')\\
        c_{13}' H_{12}' - s_{13}' H_{23}' & 
        H_{22}' & 
        s_{13}' H_{12}' + c_{13}' H_{23}' \\
        H_{13}'(c_{13}'^2-s_{13}'^2)+c_{13}' s_{13}' (H_{11}'-H_{33}') & 
        s_{13}' H_{12}' + c_{13}' H_{23}' & 
        {c_{13}'}^2 H_{33}' + {s_{13}'}^2 H_{11}'+2 c_{13}' s_{13}' H_{13}'
    \end{pmatrix} .\label{eq:H13rotated}
\end{align}
We specify this rotation angle, $\theta_{13}'$, such that 
\begin{equation} \label{eq:t13prequirement}
    H''_{13}=H''_{31}= H_{13}'({c_{13}'}^2-{s_{13}'}^2)+c_{13}' s_{13}' (H_{11}'-H_{33}')=0,
\end{equation}
which has the solution
\begin{equation} \label{eq:t13eff}
    \tan2 \theta_{13}'=\frac{2H_{13}'}{H_{33}'-H_{11}'}.
\end{equation}
Evaluating Eq.~\eqref{eq:t13eff} here explicitly yields
\begin{equation} \label{eq:SIdt13p}
    \tan2 \theta_{13}'= \frac{ \ACC \sin2 \theta_{13}}{\Dee-\ACC \cos{2 \theta_{13}}} ,
\end{equation}
where we use the definition \cite{Nunokawa:2005nx,Parke:2016joa,Denton:2016wmg}
\begin{equation} \label{eq:Dee}
\Dee = \Delta m^2_{31} - \Delta m^2_{21} s_{12}^2
  = \Delta m^2_{31}c_{12}^2 + \Delta m^2_{32} s_{12}^2.
\end{equation}
With this pair of off-diagonal elements set to zero, we proceed with the {\it only} necessary approximation of this procedure, that is neglecting the elements
\begin{equation} \label{eq:Neglectstandard}
    H''_{23}=H_{32}''=s_{13}' H_{12}' + c_{13}' H_{23}' = \frac{ \Delta m_{21}^2s_{13}' s_{12} c_{12} }{2 E} \equiv \epsilon_{e}.
\end{equation}
This approximation holds by virtue of the smallness of $\Delta m_{21}^2 \ll \Delta m_{3i}^2$, which gets further suppressed by a factor of $s_{13}' c_{12} s_{12}$. 
Furthermore, a crucial detail here is the degree to which the small elements, $H''_{23}=H_{32}''=\epsilon_e$, affects the final results. 
In the Jacobi method, $\epsilon_e$ enters both in the final eigenvalues and eigenvector only to second order \cite{Schoenhage1961,Wilkinson1962,Denton:2018fex}, meaning that the final expressions are accurate up the order of $O(\epsilon_e^2)\approx 10^{-8}\text{--}10^{-4}$ for relevant neutrino energies \cite{Agarwalla:2013tza,Denton:2018fex}.
Since this is the only approximation taken, the final results are remarkably accurate and in principle valid for any magnitude of $A_{CC}$\footnote{In practice, the resulting formulas are valid for high matter potentials up to the point where other physical phenomena start to dominate over coherent elastic forward scattering.}.






Now, to diagonalize away the final pair of off-diagonal elements we apply the rotation $H^d=R_{12}'^\dagger H''_{\rm{SNSI}}R_{12}'$ yielding 
\begin{align}
    &H^d = \nonumber \\&\begin{pmatrix}
        {c_{12}'}^2 H_{11}'' + {s_{12}'}^2 H_{22}''-2 c_{12}' s_{12}' H_{12}''  & 
        H_{12}''({c_{12}'}^2-{s_{12}'}^2)+c_{12}' s_{12}' (H_{11}''-H_{22}'')  &  
        0\\
        H_{12}''({c_{12}'}^2-{s_{12}'}^2)+c_{12}' s_{12}' (H_{11}''-H_{22}'') & 
        {c_{12}'}^2 H_{22}'' + {s_{12}'}^2 H_{11}''+2 c_{12}' s_{12}' H_{12}''   & 
        0 \\
        0 & 
        0 & H_{33}''
    \end{pmatrix}, \label{eq:H12rotated}
\end{align}
where we can now choose $\theta_{12}'$ so that
\begin{equation} \label{eq:t12prequirement}
    H_{21}''=H_{12}''=H_{12}''({c_{12}'}^2-{s_{12}'}^2)+c_{12}' s_{12}' (H_{11}''-H_{22}'')=0,
\end{equation} 
which has the solution
\begin{equation} \label{eq:t12eff}
    \tan2\theta_{12}'=\frac{2H_{12}''}{H_{22}''- H_{11}''}.
\end{equation}
Evaluating Eq.~\eqref{eq:t12eff} explicitly yields
\begin{equation}
    \tan2 \theta_{12}'= \frac{ \Delta m^2_{21} \sin2\theta_{12} c_{13}'}{\Delta m^2_{21} \cos2\theta_{12} -\ACC(c_{13}\eff)^2-\Dee s_{13}'^2 }.
\end{equation}
At this point the Hamiltonian, $H^d$, is fully diagonal, and it is possible to extract the effective mass squared differences from the eigenvalues as
\begin{equation} \label{eq:Hd_eigenvalues}
    \frac{\Delta {m_{ij}^{2,\mathrm{eff}}}}{2E} = H_{ii}^d - H_{jj}^d.
\end{equation}
Collecting the rotations, we see that
\begin{equation} \label{eq:Ufinalee}
    \tilde U^{\text{eff}} =U_{a_e} R_{13}' R_{12}' = R_{23} U_{\delta}  R_{13} R_{13}' R_{12}'.
\end{equation}
In order to compare Eq.~\eqref{eq:Ufinalee} with the original vacuum matrix decomposition in Eq.~\eqref{eq:Uee} we add a final $U_{\delta}^\dagger$ rotation, which does not affect the diagonal $H^d$ but yields the effective mixing matrix
\begin{equation} \label{eq:Ueffee}
    U^{\text{eff}} = \tilde U^{\text{eff}}U_{\delta}^\dagger  = R_{23} U_{\delta}  R_{13} R_{13}' U_{\delta}^\dagger R_{12}',
\end{equation}
where the last two matrices commute since $U_{\delta}^\dagger$ enters only in the $3$--$3$ position decoupled from the $R_{12}'$ rotation.
Comparing Eq.~\eqref{eq:Ueffee} with the original vacuum matrix decomposition in Eq.~\eqref{eq:Uee} yields the simple relations
\begin{equation} \label{eq:primetoeff_ee}
    \theta_{13}^{\text{eff}}=\theta_{13}+\theta_{13}', \quad \theta_{12}^{\text{eff}}=\theta_{12}', \quad  \theta_{23}^{\text{eff}}=\theta_{23} \quad \text{and} \quad \dCP^{\text{eff}}=\dCP.
\end{equation}
Before we present the final parameter mappings, it is worth noting some relations from the rotation in Eq.~\eqref{eq:SIdt13p}. Since $\theta_{13}'$ describes the triangle $h_{13}^2=(\Dee-\ACC \cos2\theta_{13})^2+\ACC^2$ we have the relations
\begin{equation}
    \sin 2 \theta_{13}' = {\ACC \sin 2 \theta_{13} \over h_{13}}, \quad \cos 2 \theta_{13}' = {\Dee-\ACC \cos 2 \theta_{13} \over h_{13}}, \label{eq:t13ptrigs}
\end{equation}
and using Eq.~\eqref{eq:primetoeff_ee} for $\theta_{13}\eff$ this implies the triangle $a_{13}^2+b_{13}^2=h_{13}^2$ with $a_{13}=\Dee \cos 2 \theta_{13} - \ACC$ and $b_{13}=\Dee \sin2\theta_{13}$, and thus
\begin{equation}
    \sin 2 \theta_{13}\eff = {\Dee \sin 2 \theta_{13} \over h_{13}}, \quad \cos 2 \theta_{13}\eff = {\Dee \cos 2 \theta_{13}-\ACC  \over h_{13}}\label{eq:t13efftrigs}.
\end{equation}
These trigonometric relations can be used to express the diagonal elements of $H''$ as
\begin{align*}
    H_{11}'' =& {c_{13}'}^2 H_{11}' + {s_{13}'}^2 H_{33}'-2 c_{13}' s_{13}' H_{13}',  \\=& 
    \frac{H_{11}'+H_{33}' -h_{13}}{2},
\end{align*}
and similarly one finds $H_{33}'' = (H_{11}'+H_{33}' +h_{13})/2$. This implies that 
\begin{equation} \label{eq:Hypotenuse}
    H_{33}''-H_{11}'' = h_{13}.
\end{equation}
Analogous relations to Eq.~\eqref{eq:Hypotenuse}, \eqref{eq:t13ptrigs} and \eqref{eq:t13efftrigs} may be found for $\theta_{12}'=\theta_{12}\eff$, which describes the triangle $a_{12}^2+b_{12}^2=h_{12}^2$ with $a_{12}=\Delta m^2_{21} \cos2\theta_{12} -\ACC(c_{13}\eff)^2-\Dee s_{13}'^2 $ and $b_{12}=\Delta m^2_{21} \sin2\theta_{12} c_{13}'$. Taken together, these relations can be used to simplify the expressions for the effective mass squared differences. In particular, for the 1--2 sector we can write
\begin{equation}
    \Delta {m_{21}^{2,\mathrm{eff}}} = h_{12}= \sqrt{a_{12}^2+b_{12}^2}= \frac{b_{12}}{\cos 2 \theta_{12}\eff}=\frac{a_{12}}{\sin 2 \theta_{12}\eff},
\end{equation}
and for the 1--3 sector
\begin{equation}
    \Delta {m_{31}^{2,\mathrm{eff}}} = H_{33}^d - H_{11}^d = (H_{33}''-H_{11}'')+(H_{11}'' -  H_{11}^d)= \frac{a_{13}}{\sin 2 \theta_{13}\eff}+\Delta {m_{21}^{2,\mathrm{eff}}}(s_{12}\eff)^2,
\end{equation}
where we used that $\Delta {m_{21}^{2,\mathrm{eff}}}=H_{22}^d-H_{11}^d$.

Finally, we can write down all the effective parameter mappings
\begin{align}
    \tan2 \theta_{13}\eff&= {\sin2\theta_{13} \over  \cos2\theta_{13} - \ACC/\Dee},\label{eq:t13effStandard} \\
    \tan2 \theta_{12}\eff&= \frac{ \Delta m^2_{21} \sin2\theta_{12} c_{13}'}{\Delta m^2_{21} \cos2\theta_{12} -\ACC(c_{13}\eff)^2-\Dee s_{13}'^2 } ,\label{eq:t12effStandard} \\
    \Delta {m_{21}^{2,\mathrm{eff}}}  &= \Delta m_{21}^2 \frac{\sin 2\theta_{12} c_{13}'}{\sin 2\theta_{12}\eff}, \label{eq:dm21effStandard} \\
    \Delta {m_{31}^{2,\mathrm{eff}}}  &= \Dee \frac{\sin 2\theta_{13}}{\sin 2\theta_{13}\eff}+ \Delta {m_{21}^{2,\mathrm{eff}}}(s_{12}\eff)^2. \label{eq:dm31effStandard} 
\end{align}
where Eq.~\eqref{eq:dm31effStandard} can also be interpreted as
\begin{equation}
     \Dee\eff  = \Dee \frac{\sin 2\theta_{13}}{\sin 2\theta_{13}\eff}. 
\end{equation}
Note that $\dCP$ and $\theta_{23}$ are exactly unchanged from their vacuum values here, regardless of the size of the matter potential, albeit in an exact framework they may be subtly altered \cite{Krastev:1988yu,Toshev:1991ku}.
Eq.~\eqref{eq:t13effStandard} exhibits the resonant enhancement of $\theta_{13}\eff$ due to the standard matter potential $\ACC$, with the resonance condition
\begin{equation} \label{eq:t13resonanceStandard}
    \ACC^{13,\text{res}}=\Dee \cos2\theta_{13},
\end{equation}
which occurs at $10\,$GeV for $\rho_\oplus=2.6\,\text{g}/\text{cm}^3$.
Similarly, Eq.~\eqref{eq:t12effStandard} shows the 3-flavor resonant enhancement of $\theta_{12}\eff$ with the resonance condition
\begin{equation} \label{eq:t12resonanceStandard}
     \ACC^{12,\text{res}}=\frac{\Delta m^2_{21} \cos2\theta_{12}-\Dee s_{13}'^2}{(c_{13}\eff)^2} \approx\frac{\Delta m^2_{21} \cos2\theta_{12}}{c_{13}^2},
\end{equation}
which occurs around $0.13\,$GeV for $\rho_\oplus=2.6\,\text{g}/\text{cm}^3$ and $2.0\,$MeV for $\rho_\odot =150\,\text{g}/\text{cm}^3$. Note that we neglected $\theta_{13}'$ in Eq.~\eqref{eq:t12resonanceStandard}, which is a good assumption since the $1$--$2$ and $1$--$3$ resonances are separated by a factor $\sim \dmsq{31}/\dmsq{21}\approx10^2$, resulting in $\theta_{13}'\sim0.1^\circ$ near the $1$--$2$ resonance. Further note that, when we take $\theta_{13}= 0$, Eq.~\eqref{eq:t12effStandard} reduces to the recognizable 2-flavor MSW formula
\begin{equation}
    \tan2 \theta_{12}\eff= { \Delta m^2_{21} \sin2\theta_{12}  \over  \Delta m^2_{21} \cos2\theta_{12}-\ACC}.
\end{equation}
Indeed, the Jacobi method yields both effective mixing angle resonance relations and effective mass squared differences to high accuracy in the full 3-flavor framework. This has been discussed extensively in Refs.~\cite{Agarwalla:2013tza, Denton:2016wmg}. 


Note that we thus far have implicitly assumed $\ACC>0$, which is the case for neutrinos. The provided formulas are equally applicable for antineutrinos where $\ACC<0$. A more general resonance condition including both neutrinos and antineutrinos, e.g. for the 1--2 sector, reads as ${\rm sign}(\ACC) = {\rm sign}(\Delta m^2_{21}\cos2\theta_{12}/(c_{13}\eff)^2)$. This is satisfied for neutrinos and not antineutrinos only when $\theta_{12}<\pi/4$ as is imposed by global oscillation data \cite{NuFit6.0}. 



Overall, the compactness of the above expressions, in addition to the high accuracy due to only neglecting quadratic contributions of Eq.~\eqref{eq:Neglectstandard}, marks out this method as remarkably well suited for analyzing neutrino oscillations with large matter effects. 
Furthermore, this method allows for a unified derivation of the resonant enhancement of the mixing angles in both the 1-2 and 1-3 sectors with excellent analytical accuracy. In the following section, we will consider how this method can be generalized to other structures of the matter potentials while still yielding the same level of accuracy, clarity and insight. 

\section{Generalizing the Jacobi Diagonalization Method} \label{sec:generalizing}
The accuracy and simplicity of the results derived in Sec.~\ref{sec:standardderivation} is quite remarkable, yet it seems almost by accident that it worked out so well. The fact that we could isolate only the smallest terms in one pair of elements, $H_{23}''=H_{32}''=\epsilon_{e}$, relies both on the structure of the matter potential matrix, $A$, and the {\it seemingly ad hoc} choice of the auxiliary basis, $U_{a_e}$. In what follows, we aim to untangle these details and, in so doing, outline how the method can be generalized to other conceivable matrix structures of the matter potential.

It is well known that the PMNS matrix satisfies $\PMNS \in U(3)$
\cite{10.1093/acprof:oso/9780198508717.001.0001}. Its real rotational part can
be parametrised using the subgroup $SO(3) \subset U(3)$, where
$SO(3) = SO(2)_i \, SO(2)_j \, SO(2)_k$ with $i,j,k \in \{x,y,z\}$, is
understood as a product of rotations about the $x$-, $y$- and $z$-axes. The
order of the factors depends on the chosen parameterisation. In the standard PDG 
convention, $(i,j,k) = (x,y,z)$, corresponding to rotations in the $(23)$,
$(13)$ and $(12)$ planes, respectively, and the CP-violating phases are
introduced on top of them.  
Consider the conventional charged-current matter potential matrix,
$A = A_{\rm CC}\,\mathrm{diag}(1,0,0)$, which exhibits a continuous
$SO(2)_x$ symmetry within $SO(3)$. One of the primary purposes of the
auxiliary basis, $U_{a_e}$, is to apply the $SO(2)_x$ rotation, which is
both one of the rotations in the parametrisation of $\PMNS$ and a symmetry of
the matter potential $A$, as is well recognised in the literature
(see e.g.\ \cite{Akhmedov:1999uz,Denton:2021vtf}).
Without this simplification, the procedure in Sec.~\ref{sec:standardderivation} would be considerably more involved. Note that $\PMNS$ also contains the $SO(2)_y$ and $SO(2)_z$ rotations, which can
be utilized in the same way whenever these are continuous symmetries of $A$,
provided that the parametrization of $\PMNS$ is ordered such that the
corresponding rotation is the leftmost factor.  Taking $SO(2)_z$ as an example, the matrices that exhibit this continuous symmetry are
\begin{equation}
    \mathcal A_z=\begin{pmatrix}
        a & ib & 0 \\
        -ib & a & 0 \\
        0 & 0 & c \\
    \end{pmatrix}, \quad a,b,c\in \mathbb{R}.
\end{equation}
Considering the diagonal case $a=b=0, \, c\neq0$, i.e. $\mathcal A_z\propto \rm{diag}(0,0,1)$, we see that any matter potential with a dominant contribution in the $\tau$--$\tau$ element is unaffected by the $SO(2)_z$ component of $\PMNS$. Analogously, one could consider $\mathcal A_y\propto \rm{diag}(0,1,0)$, which is the single-element potential invariant under $SO(2)_y$. Since the choice of parameterizing $\PMNS$ is arbitrary, we are free to choose the parameterization such that the rotation under which the specific
matter potential is invariant is the leftmost factor.

Now, a second point warrants attention regarding the approximation in Eq.~\eqref{eq:Neglectstandard} where $\epsilon_e = s_{13}' s_{12} c_{12} \Delta m_{21}^2$ was neglected. We were able to neglect this element by virtue of the hierarchy $\Delta m_{21}^2 \ll \Delta m_{3i}^2$. It is thus essential that precisely the 1-2 sector rotates the $M^2$ matrix in the auxiliary basis, since this avoids mixing the non-negligible $\propto \Delta m_{31}^2$ terms into off-diagonal elements. This is the case whenever the rightmost rotation in our parameterization of $\PMNS$ is $R_{12}$. 
With these points in mind, the auxiliary basis will always be chosen to include the first two rotations in the PMNS matrix and not the final $R_{12}$ rotation. With this choice, the auxiliary rotation consists of one rotation that is an invariant of the matter potential, and a second rotation that constitutes the first step of the Jacobi diagonalization procedure.

We conclude from these insights that the specific PDG parameterization of the PMNS matrix was essential for the method of Sec.~\ref{sec:standardderivation} to work out so well for neutrinos in the presence of the standard matter potential with the structure $A \propto \diag(1,0,0)$. The following conditions underpin the requirements on the parameterization choice of $\PMNS$ for our generalization of the Jacobi diagonalization method:
\begin{enumerate} \label{conditions}
    \item The PMNS matrix is parameterized using a decomposition with three Euler rotations: $\PMNS \sim R_{ij} R_{kl} R_{mn}$, as well as an appropriately placed $U_\delta$.
    \item The leftmost rotation matrix in this decomposition ($R_{ij}$) needs to be a continuous symmetry of the matter potential matrix, $A$. For example, in the standard case $R_{23}$ is a continuous symmetry of $A \propto \mathrm{diag}(1,0,0)$.
    \item The rightmost rotation ($R_{mn}$) needs to be $R_{12}$. This rotation applies to the intrinsic neutrino masses in the auxiliary basis, and thus when chosen to be $R_{12}$ it yields off-diagonal elements $\propto s_{12}c_{12}\Delta m^2_{21} \ll \dmsq{31}$ that can be safely neglected.
\end{enumerate}
Applying the above conditions to the standard potential, $A\propto \diag(1,0,0)$, we are restricted to choose $\PMNS \sim R_{23} R_{kl} R_{12}$, which yields a valid representation of $SO(3) \subset U(3)$ only for the PDG parameterization $\PMNS \sim R_{23} R_{13} R_{12}$. Interestingly however, we may now employ these conditions to extend the method to other matter potentials, specifically $A\propto \diag(0,1,0)$ and $A\propto \diag(0,0,1)$. These two cases will be demonstrated in sections \ref{sec:Vmmtoy} and \ref{sec:Vtttoy}, and will turn out to be particularly useful for understanding the phenomenology of the SNSI model to be discussed in Sec.~\ref{sec:SNSImain}.
We emphasize that the above conditions are in principle independent of the numerical values of the mixing matrix angles and the matter potential, and rely solely on the smallness of $\Delta m_{21}^2/\Delta m_{31}^2$ and the structure of the matter potential matrix.

Before we proceed, let us comment on how our non-standard choices of mixing matrix parameters are related to the PDG parameters in practice. We show in Tab.~\ref{tab:nufitvalues} the Nufit 6.0 \cite{NuFit6.0} central values expressed in the three parameterization schemes that will be used in this work. These parameterizations are defined in Eqs.~\eqref{eq:Uee},\eqref{eq:Umm} and \eqref{eq:Utt} respectively. The $U^e$ values come directly from Ref.~\cite{NuFit6.0}, and we use conversion formulas between the parameterizations (see Appendix \ref{app:parameterization}) to give the same central values for the other two parameterization choices, that is $U^\mu$ and $U^\tau$. We generally find that the numerical variation between the parameters in each section is quite small, except for quadrant shifts like $\delta_{CP}^\mu \approx 2\pi-\delta_{CP}$. The 1-2 sector is most numerically sensitive to a change of basis. The smallness of the variation between the parameters inside each of the other two sectors can be largely understood from the smallness of $\theta_{13}$, which translates to a comparable smallness of $\theta_{13}^\mu$ and $\tilde \theta_{12}^\tau$. If one performs a series expansion for the conversion relations in terms of the smallest mixing angle, one obtains simple and accurate relations between the parameterization schemes as shown in Appendix  \ref{app:approximateconversionrelations}. This serves as a quite useful way of interpreting the derivations that follow in the context of all previous literature that almost exclusively uses the PDG parameterization.

\begin{table}[]
    \centering
    \caption{Nufit 6.0 central values for all parameters in the three parameterization choices of $\PMNS$ used in this work. Normal mass ordering is assumed.}
    \begin{tabular}{|c|c|c|c|c|}
        \hline
        Parameterization & 1-2 sector & 1-3 sector & 2-3 sector & CP-violation  \\
        \hline
        $U^e$ & $\theta_{12}=33.68^{\circ}$ & $\theta_{13}=8.56^{\circ}$ & $\theta_{23}=43.3^{\circ}$ & $\delta_{CP}=212^{\circ}$\\
        $U^\mu$ & $\theta_{12}^\mu=27.08^{\circ}$ & $\theta_{13}^\mu=11.68^{\circ}$ & $\theta_{23}^\mu=42.7^{\circ}$ &  $\delta_{CP}^\mu=143^{\circ}$ \\
        $U^\tau$ & $ \theta_{12}^\tau=41.41^{\circ}$ & $\tilde \theta_{12}^\tau=12.38^{\circ}$ & $\theta_{23}^\tau=43.97^{\circ}$ & $\delta_{CP}^\tau=210^{\circ}$  \\
        \hline
    \end{tabular}
    \label{tab:nufitvalues}
\end{table}

\subsection{\texorpdfstring{$\Amm$}{Amm}} \label{sec:Vmmtoy}
We now consider the Hamiltonian in Eq.~\eqref{eq:HamiltonianStandard} with the modified potential term $A = \rm{diag}(0,\Amm,0)$. We primarily work with this Hamiltonian to showcase the generalization outlined above. Phenomenologically, this scenario corresponds to a neutrino propagating through muon-dominated matter, which is not realized in any current terrestrial experiment (although muons can be present in dense astrophysical environments). The more pertinent application is as a representative example for when non-standard effects give rise to a $\mu$--$\mu$ dominated matter potential. Note that this Hamiltonian differs from that of vector-like NSI with $V_{\text{NSI}}=\text{diag}(0,\varepsilon_{\mu \mu},0)$ only in that the standard $\ACC$ here has been disregarded.

Considering the conditions outlined 
above, the matter potential structure $A \propto(0,1,0)$ suggests parameterizing the PMNS matrix as $\PMNS\sim R_{13} R_{kl} R_{12}$. The only valid Euler rotation decomposition that satisfies this is $R_{13} R_{23} R_{12}$ and so we choose the parameterization
\begin{equation} \label{eq:Umm}
    U^{\mu}(\theta_{13}^\mu, \theta_{23}^\mu,\theta_{12}^\mu, \delta_{\text{CP}}^\mu)=R_{13} U_\delta R_{23} {U_\delta}^\dagger R_{12}.
\end{equation}
where $U_\delta=\text{diag}(1,1,e^{i \delta_{CP}^\mu})$ and $R_{ij}$ are the rotation matrices in Eq.~\eqref{eq:Eulerrotations}.
The corresponding choice of auxiliary basis is 
\begin{equation}
    U_{a_\mu}=R_{13} U_\delta R_{23}.
\end{equation}
The only difference between the PDG parameterization decomposition and the one in Eq.~\eqref{eq:Umm} is that $R_{13}$ and $R_{23}$ have switched places. This changes the definition of all mixing parameters. Naturally, one can use the invariance of $|U_{\alpha i}|$ to convert between the angles in two different parameterizations, which we show explicitly in App.~\ref{app:parameterization}.

We now proceed as in Sec.~\ref{sec:standardderivation}. Moving to the auxiliary basis $H'=U_{a_\mu}^\dagger H U_{a_\mu}$ yields
\begin{equation}
    H' = {1 \over 2E}\begin{pmatrix}
        (s_{12}^\mu)^2 \Delta m_{21}^2 & s_{12}^\mu c_{12}^\mu \Delta m_{21}^2 & 0 \\
        s_{12}^\mu c_{12}^\mu \Delta m_{21}^2 & (c_{12}^\mu)^2 \Delta m_{21}^2 + (c_{23}^\mu)^2  \Amm & c_{23}^\mu s_{23}^\mu \Amm \\
        0 & c_{23}^\mu s_{23}^\mu \Amm & \Delta m_{31}^2 + (s_{23}^\mu)^2 \Amm \\
    \end{pmatrix}.
\end{equation}
Note that we superscript each trigonometric function as $s_{ij}^\mu\equiv \sin\theta_{ij}^\mu$ and $ c_{ij}^\mu\equiv\cos\theta_{ij}^\mu$ to clearly distinguish these from the conventional PDG parameterization. We proceed by applying a $\theta_{23}'^\mu$ rotation as $H''=(R_{23}'^{\mu})^\dagger H' R_{23}'^\mu$, where we set an off-diagonal pair of elements to zero
\begin{equation} \label{eq:H23pVmm}
    H_{32}''=H''_{23}=H_{23}'((c_{23}'^\mu)^2-(s_{23}'^\mu)^2)+c_{23}'^\mu s_{23}'^\mu (H_{22}'-H_{33}')=0.
\end{equation}
Eq.~(\ref{eq:H23pVmm}) has the solution
\begin{equation} \label{eq:t23pVmm}
    \tan 2\theta_{23}'^\mu = \frac{A_{\mu\mu}\,\sin 2\theta_{23}^\mu}
  {\Dmm - A_{\mu\mu}\cos 2\theta_{23}^\mu},
\end{equation}
where we defined
\begin{equation}
    \Dmm = \Delta m^2_{31} - \Delta m^2_{21} (c_{12}^\mu)^2
  = \Delta m^2_{31} (s_{12}^\mu)^2 + \Delta m^2_{32} (c_{12}^\mu)^2.
\end{equation}
Another pair of off-diagonal elements is neglected, specifically
\begin{equation}
    H_{31}''=H_{13}'' = s_{23}'^\mu s_{12}^\mu c_{12}^\mu \Delta m_{21}^2 \equiv \epsilon_\mu,
\end{equation}
which is motivated similarly to the discussion around $\epsilon_e$ in Sec.~\ref{sec:standardderivation}.
The last pair of off-diagonal elements, $H_{12}''=H_{21}''$, can now be rotated away with $H^d = (R_{12}'^{\mu})^\dagger H'' R_{12}'^\mu$, which diagonalizes $H^d$ for $\theta_{12}'^\mu$ given by Eq.~\eqref{eq:t12eff}.

With $H^d$ fully diagonal, the  eigenvalues could be extracted directly. However, the effective mass squared differences can be expressed more compactly using the trigonometric relations associated with $R_{23}'^\mu$ and $R_{12}'^\mu$, analogously to what was shown explicitly in Sec.~\ref{sec:standardderivation}.
Collecting the rotations, we see that the effective mixing matrix is given by 
\begin{equation} 
    \tilde U^{\mu,\text{eff}} =U_{a_\mu} R_{23}'^\mu R_{12}'^\mu=R_{13} U_\delta R_{23}R_{23}'^\mu R_{12}'^\mu.
\end{equation}
As before, we add a final ${U_\delta}^\dagger$ rotation in order to facilitate the comparison with the vacuum parameters
\begin{equation} \label{eq:Ueffmm}
    U^{\mu,\text{eff}}=\tilde U^{\mu,\text{eff}} {U_\delta}^\dagger =R_{13} U_\delta R_{23}R_{23}'^\mu {U_\delta}^\dagger R_{12}'^\mu.
\end{equation}
Eq.~\eqref{eq:Ueffmm} can be compared with Eq.~\eqref{eq:Umm} to yield the simple relations
\begin{equation} \label{eq:primetoeff_mm}
    \theta_{23}^{\mu,\text{eff}}=\theta_{23}^\mu+\theta_{23}'^\mu, \quad \theta_{12}^{\mu,\text{eff}}=\theta_{12}'^\mu, \quad \theta_{13}^{\mu,\text{eff}}=\theta_{13}^\mu \quad \text{and} \quad \dCP^{\mu,\text{eff}}=\dCP^\mu.
\end{equation}
Finally, the effective parameter mappings are
\begin{align}
\tan 2{\theta_{23}^{\mu,\text{eff}}} &= \frac{\sin 2\theta_{23}^\mu}
  {\cos 2\theta_{23}^\mu - A_{\mu\mu}/\Dmm}, \label{eq:t23effVmm} \\
\tan 2{\theta_{12}^{\mu,\text{eff}}} &= \frac{\Delta m^2_{21}\,\sin 2\theta_{12}^\mu c_{23}'^\mu}
  {\Delta m^2_{21}\cos 2\theta_{12}^\mu
   + A_{\mu\mu} (c_{23}^{\mu,\text{eff}})^2
   + \Dmm (s_{23}'^\mu)^2} \label{eq:t12effVmm}, \\
\Delta {m_{21}^{2,\mathrm{eff}}} &= \Delta m^2_{21}\,
  \frac{\sin 2\theta_{12}^\mu c_{23}'^\mu }
       {\sin 2{\theta_{12}^{\mu,\text{eff}}}} \label{eq:dm21effVmm}, \\
 \Delta {m_{31}^{2,\mathrm{eff}}} &= \Dmm\,
  \frac{\sin 2\theta_{23}^\mu}{\sin 2{\theta_{23}^{\mu,\text{eff}}}}
  + \Delta {m_{21}^{2,\mathrm{eff}}} (c_{12}^{\mu,\text{eff}})^2. \label{eq:dm31effVmm}
\end{align}
where Eq.~\eqref{eq:dm31effVmm} can also be interpreted as
\begin{equation}
     \Dmm\eff = \Dmm\,
  \frac{\sin 2\theta_{23}^\mu}{\sin 2{\theta_{23}^{\mu,\text{eff}}}}.
\end{equation}
Note that, in this approximate framework, $\theta_{13}^\mu$ and $\dCP^\mu$ are exactly unchanged from their vacuum values, although they are expected to be slightly altered for very high $\Amm$ as in the standard $\ACC$ case. Let us emphasize that if these relations are converted to PDG parameters they will exhibit a strong dependence on all six oscillation parameters including $\theta_{13}$ and $\delta_{CP}$.
Now, note the striking similarity between Eqs.~\eqref{eq:t23effVmm}--\eqref{eq:dm31effVmm} and the previous
results for $A=\mathrm{diag}(A_{\rm CC},0,0)$. Indeed, replacing
$(\theta^\mu_{12},\theta^\mu_{23},A_{\mu\mu},\Delta_{\mu\mu})$ with
$(\theta_{12},\theta_{13},A_{\rm CC},\Delta_{ee})$ yields \textit{almost} the relations
in Eqs.~\eqref{eq:t13effStandard}--\eqref{eq:dm31effStandard}. The differences are that (i) the signs of the two
terms $\propto A$ and $\propto\Delta$ in the denominator of Eq.~\eqref{eq:t12effVmm} are
flipped, and (ii) $\Dmm$ and the last term of Eq.~\eqref{eq:dm31effVmm} contains $(c_{12}^{\mu,\rm eff})^2$
whereas $\Dee$ and Eq.~\eqref{eq:dm31effStandard} has $(s_{12}^{\rm eff})^2$.

The $\theta_{23}^\mu$ resonance condition can be extracted from the denominator in Eq.~\eqref{eq:t23effVmm} yielding
\begin{equation} \label{eq:t23resonanceVmm}
    A_{\mu\mu}^{23,\text{res}}=\Dmm\cos 2\theta_{23}^\mu.
\end{equation}
Note here that $\cos2\theta^\mu_{23} = 1-2|U_{\mu3}|^2 = 1-2s_{23}^2c_{13}^2
\approx 0.08$ with the contemporary global best-fit values. This means that the $\theta_{23}^{\mu, \text{eff}}$ resonance in Eq.~\eqref{eq:t23resonanceVmm} is satisfied 
for a value of
$A_{\mu\mu}$ about an order of magnitude smaller than the value of $A_{\rm CC}$
that satisfies the $\theta^{\rm eff}_{13}$ resonance.
Further, since the vacuum value of $\theta^\mu_{23}$ is measured to be
almost maximal, this resonance amplification would have very little impact. More interestingly, for $|A_{\mu\mu}|\gg\Delta_{\mu\mu}$ the $\theta^{\mu,\rm eff}_{23}$ tends away from maximal towards 0 ($\pi/2$) for antineutrinos (neutrinos).

Similarly, the $\theta_{12}^\mu$ resonance condition for $\Amm$ can be seen from Eq.~\eqref{eq:t12effVmm} and written as
\begin{equation} \label{eq:t12resonanceVmm}
    A_{\mu\mu}^{12,\text{res}}=-\frac{\Dmm (s_{23}'^\mu)^2+\Delta m^2_{21}\cos 2\theta_{12}^\mu}{(c_{23}^{\mu,\text{eff}})^2} \approx -\frac{\Delta m^2_{21}\cos 2\theta_{12}^\mu}{(c_{23}^{\mu})^2}.
\end{equation}
We assumed in Eq.~\eqref{eq:t12resonanceVmm} that $\theta_{23}'^\mu$ is small near the $\theta_{12}^\mu$ resonance, which holds at the $1\%$ level even though we earlier noted that the two resonances are significantly closer here than in the $\ACC$ case.  
We further observe that only antineutrinos could satisfy the $\theta_{12}^\mu$ resonance for $\theta_{12}^\mu<\pi/4$, a complete reversal of the standard $\ACC$ case. 
Conversely, for the neutrino case where $A_{\mu\mu}>0$ the effective angle, $\theta_{12}^{\mu,\text{eff}}$, would decrease asymptotically towards zero.




\subsection{\texorpdfstring{$\Att$}{Att}} \label{sec:Vtttoy}
We now consider the Hamiltonian in Eq.~\eqref{eq:HamiltonianStandard} with the potential term $A = \rm{diag}(0,0,\Att)$. 
Similarly to Sec.~\eqref{sec:Vmmtoy}, this scenario corresponds to a neutrino propagating
through $\tau$-dominated matter, which is not realized in any current terrestrial
experiment, but is a representative example of the plausible non-standard
phenomenology of a $\tau$--$\tau$ dominated matter potential.
For this structure of the matter potential, the conditions of our method requires the parameterization choice $U\sim \tilde R_{12} R_{kl} R_{12}$. Note that, since two 1--2 rotations are used here, we use the notation $\tilde \theta_{12}^\tau$ and $\theta_{12}^\tau$ to distinguish the first 1--2 rotation from the latter. Now, there are two valid Euler decompositions that satisfy the conditions from Sec.~\ref{sec:generalizing}. These are $(k,l)=(1,3)$ and $(k,l)=(2,3)$. 
The two seemingly distinct parameterizations are related via a symmetry \cite{Fritzsch:1997st}, meaning that the choice between the two is inconsequential for the final results. We choose the latter of these parameterizing the PMNS matrix as
\begin{equation}\label{eq:Utt}
    U^{\tau}(\tilde \theta_{12}^\tau,\theta_{23}^\tau, \theta_{12}^\tau, \dCP^\tau) = \tilde R_{12} R_{23} U_\delta^\tau R_{12} 
\end{equation}
where $U_\delta^\tau=\diag(e^{i\delta_{\text{CP}}^\tau},1,1)$ and $R_{ij}$ are the Euler rotation matrices defined in Eq.~\eqref{eq:Eulerrotations}. Note that the phase here is placed in the slot left
untouched by $R_{23}$, so that it commutes with $R_{23}$ and $H'$ remains real. The corresponding choice of auxiliary basis is
$U_{a_\tau} = \tilde R_{12} R_{23} U_\delta^\tau$. Evaluating the Hamiltonian in this basis,
$H'=U_{a_\tau}^\dagger H U_{a_\tau}$, yields
\begin{equation} \label{eq:HpVtt}
    H' = {1 \over 2 E}\begin{pmatrix}
        (s_{12}^\tau)^2 \Delta m_{21}^2 & s_{12}^\tau c_{12}^\tau \Delta m_{21}^2 & 0 \\
        s_{12}^\tau c_{12}^\tau \Delta m_{21}^2 & (c_{12}^\tau)^2 \Delta m_{21}^2 + (s_{23}^\tau)^2 \Att & -s_{23}^\tau c_{23}^\tau \Att \\
        0 & - s_{23}^\tau c_{23}^\tau \Att & \Delta m_{31}^2 + (c_{23}^\tau)^2 \Att \\
    \end{pmatrix}.
\end{equation}
The diagonalization now proceeds as in Sec.~\ref{sec:Vmmtoy}, that is we apply a rotation $H''=(R_{23}'^{\tau})^\dagger H' R_{23}'^\tau$ where the angle $\theta_{23}'^\tau$ is fixed by imposing
\begin{equation} \label{eq:H23prequirementVtt}
    H_{32}''=H''_{23}=H_{23}'((c_{23}'^\tau)^2-(s_{23}'^\tau)^2)+c_{23}'^\tau s_{23}'^\tau (H_{22}'-H_{33}')=0.
\end{equation}
Eq.~(\ref{eq:H23prequirementVtt}) has the solution
\begin{equation} \label{eq:t23pVtt}
    \tan 2\theta_{23}'^\tau = \frac{-\,A_{\tau\tau}\,\sin 2\theta_{23}^\tau}
  {\Dtt + A_{\tau\tau}\cos 2\theta_{23}^\tau},
\end{equation}
where we defined
\begin{equation}
\Dtt =  \Delta m^2_{31} - \Delta m^2_{21} (c_{12}^\tau)^2
  = \Delta m^2_{31} (s_{12}^\tau)^2 + \Delta m^2_{32} (c_{12}^\tau)^2.
\end{equation}
Another pair of off-diagonal elements are neglected, in particular
\begin{equation}
    H_{31}''=H_{13}'' = s_{23}'^\tau s_{12}^\tau c_{12}^\tau \Delta m_{21}^2 \equiv \epsilon_{\tau},
\end{equation}
where the small term $\epsilon_\tau$ would only enter quadratically in the final results as noted in Sec.~\ref{sec:standardderivation}.
The final step for diagonalization is to apply the rotation
$H^d=(R_{12}'^{\tau})^\dagger H'' R_{12}'^\tau$ and fix the angle $\theta_{12}'^\tau$ using
Eq.~\eqref{eq:t12eff}.

With $H^d$ fully diagonal, the mass squared differences could be extracted directly. However, the effective mass squared differences can be expressed more compactly using the trigonometric relations associated with $R_{23}'^\tau$ and $R_{12}'^\tau$, analogously to what was shown explicitly in Sec.~\ref{sec:standardderivation}.
Collecting the rotations, we see that the effective mixing matrix is given by 
\begin{equation} \label{eq:Uefftt}
    U^{\tau,\text{eff}} =U_{a_\tau} R_{23}'^\tau R_{12}'^\tau = \tilde R_{12}^\tau R_{23}^\tau U_\delta^\tau R_{23}'^\tau R_{12}'^\tau.
\end{equation}
Noting that $U_\delta$ here commutes with $R_{23}'^\tau$, we may can compare Eq.~\eqref{eq:Uefftt} with Eq.~\eqref{eq:Utt} to see that
\begin{equation} \label{eq:primetoeff_tt}
    \theta_{23}^{\tau,\text{eff}}=\theta_{23}'^\tau+\theta_{23}^\tau, \quad \theta_{12}^{\tau,\text{eff}}=\theta_{12}'^\tau, \quad \tilde \theta_{12}^{\tau,\text{eff}}=\tilde \theta_{12}^\tau \quad \text{and} \quad \dCP^{\tau,\text{eff}}=\dCP^\tau.
\end{equation}
Finally, the effective parameter mappings are
\begin{align}
\tan 2{\theta_{23}^{\tau,\text{eff}}} &= \frac{\sin 2\theta_{23}^\tau}
  {\cos 2\theta_{23}^\tau + A_{\tau\tau}/\Dtt},\label{eq:t23effVtt}\\
\tan 2{\theta_{12}^{\tau,\text{eff}}} &= \frac{\Delta m^2_{21} \sin 2\theta_{12}^\tau c_{23}'^\tau 
   }
  {\Delta m^2_{21}\cos 2\theta_{12}^\tau
   + A_{\tau\tau}(s_{23}^{\tau,\text{eff}})^2
   + \Dtt (s_{23}'^\tau)^2} \label{eq:t12effVtt}, \\
\Delta {m_{21}^{2,\mathrm{eff}}} &= \Delta m^2_{21}\,
  \frac{\sin 2\theta_{12}^\tau c_{23}'^\tau}
       {\sin 2{\theta_{12}^{\tau,\text{eff}}}},  \label{eq:dm21effVtt} \\
\Delta {m_{31}^{2,\mathrm{eff}}} &= \Dtt\,
  \frac{\sin 2\theta_{23}^\tau}{\sin 2{\theta_{23}^{\tau,\text{eff}}}}
  +\Delta {m_{21}^{2,\mathrm{eff}}}  (c_{12}^{\tau,\text{eff}})^2,  \label{eq:dm31effVtt}
\end{align}
where Eq.~\eqref{eq:dm31effVtt} can also be interpreted as
\begin{equation}
     \Dtt\eff = \Dtt\,
  \frac{\sin 2\theta_{23}^\tau}{\sin 2{\theta_{23}^{\tau,\text{eff}}}}. 
\end{equation}
The parameters $\tilde\theta_{12}^\tau$ and
$\dCP^\tau$ are left exactly unchanged by the matter potential, $\Att$, in this approximate framework.
We here see a correspondence between the parameters $(\theta_{12}^\tau, \theta_{23}^\tau, \Att)$ for and the parameters $(\theta_{12}, \theta_{23}, \ACC)$. 
With this replacement,
Eqs.~\eqref{eq:t23effVtt}--\eqref{eq:dm31effVtt} become \textit{almost} the same as  Eqs.~\eqref{eq:t13effStandard}--\eqref{eq:dm31effStandard}. The
differences are that (i) all terms in the denominators of Eqs.~\eqref{eq:t12effVtt} and
\eqref{eq:t23effVtt} are positive, (ii) the $A_{\tau\tau}$ term in Eq.~\eqref{eq:t12effVtt} contains $(s^{\tau,\rm eff}_{23})^2$ whereas Eq.~\eqref{eq:t12effStandard} has $(c^{\rm eff}_{13})^2$, and
(iii) $\Delta_{\tau\tau}$ and the last term of Eq.~\eqref{eq:dm31effVtt} involve $c_{12}^\tau$
whereas $\Delta_{ee}$ and Eq.~\eqref{eq:dm31effStandard} involve $s_{12}$. 
Consequently, both the
$\theta^{\tau,\rm eff}_{23}$ resonance at
\begin{equation}\label{eq:t23resonanceVtt}
    A_{\tau\tau}^{23,\text{res}}=-\Dtt\cos 2\theta_{23}^\tau
\end{equation}
and the $\theta_{12}^{\tau,\text{eff}}$ resonance at 
\begin{equation}\label{eq:t12resonanceVtt}
    A_{\tau\tau}^{12,\text{res}}=-\frac{\Delta m^2_{21}\cos 2\theta_{12}^\tau + \Dtt (s_{23}'^\tau)^2}{(s_{23}^{\tau,\text{eff}})^2} \approx -\frac{\Delta m^2_{21}\cos 2\theta_{12}^\tau}{(s_{23}^{\tau})^2}
\end{equation}
would occur for $\Att<0$, i.e. antineutrinos, assuming the vacuum values of the mixing angles are in the first octant.
Similarly to the $A_{\mu\mu}$ case, the resonant enhancement of
$\theta^{\tau,\rm eff}_{23}$ due to $A_{\tau\tau}$ would be quite insignificant
since the vacuum value $\theta^\tau_{23}\approx\theta_{23}$ is already almost
maximal. Further, in this case $\theta^\tau_{12}$ is also near maximal for the Nufit 6.0
values (see Tab.~\ref{tab:nufitvalues}), meaning that its resonance relation would be similarly insignificant and occur at a similarly small value of $|A_{\tau\tau}|$. As such, in the antineutrino case an increase of $|A_{\tau\tau}|$ beyond each respective resonance pushes the respective mixing angle towards minimal mixing at $\pi/2$.
Converesely, in the neutrino case, an increase in $|\Att|$ decreases both $\theta_{23}\eff$ and $\theta_{12}\eff$ asymptotically towards zero.


Taking a step back, we recognize that all three single element diagonal cases of the matter potential yield very similar relations. As a matter of fact, we can use the mixing matrix elements to summarize the resonance conditions for the three cases in a parameterization-independent way. We have
\begin{equation} \label{eq:AaaMixingmatrix}
     \Aaa^{12,\text{res}}\approx\Delta m_{21}^2\frac{|U_{\alpha1}|^2-|U_{\alpha2}|^2}{(1-|U_{\alpha3}|^2)^2}, \qquad \Aaa^{i3,\text{res}}=\Daa(1-2|U_{\alpha3}|^2),
\end{equation}
where $\alpha=e,\mu,\tau$; $i=1$ for $\alpha=e$ otherwise $i=2$; and where
\begin{equation} \label{eq:Daamixingmatrix}
    \Daa=\Delta m_{31}^2 - \Delta m_{21}^2\frac{|U_{\alpha2}|^2}{1-|U_{\alpha3}|^2}.
\end{equation}
Using Eqs.~\eqref{eq:AaaMixingmatrix}--~\eqref{eq:Daamixingmatrix}, enables us to comment on the numerical dependence of the sign of $\Aaa^{\text{res}}$. Taking the Nufit6.0 best-fit values, we have $|U_{e1}|^2>|U_{e2}|^2$ meaning that the $1$--$2$ resonance occurs for a positive $\ACC$, whereas $|U_{\mu1}|^2<|U_{\mu2}|^2$ and $|U_{\tau1}|^2<|U_{\tau2}|^2$ meaning that the $1$--$2$ resonance would occur for negative $\Amm$ or $\Att$. Furthermore, since $|U_{e3}|<0.5$ and $|U_{\mu3}|<0.5$ the $1$--$3$ ($2$--$3$) resonance occurs for positive $\ACC$ ($\Amm$), whereas $|U_{\tau3}|>0.5$ meaning that the $2$--$3$ resonance would occur for negative $\Att$.
All of these signs turn out to be in total agreement with the signs in Eqs.~~\eqref{eq:t13resonanceStandard}, \eqref{eq:t12resonanceStandard},  \eqref{eq:t23resonanceVmm}, \eqref{eq:t12resonanceVmm},\eqref{eq:t23resonanceVtt} and \eqref{eq:t12resonanceVtt}.
This means that, for the contemporary global best fit values, every analytical resonance condition presented occurs with the sign as given here. In other words, all mixing angles in all these parameterizations are restricted to the first octant and so the $\cos2\theta$ expressions never flips the sign of $\Aaa^{\text{res}}$. This is also corroborated by Tab.~\ref{tab:nufitvalues}.

\section{Scalar Non-Standard Interactions} \label{sec:SNSImain}
The presence of a BSM scalar field, $\phi'$, with couplings to the SM fermions may affect neutrino propagation. This field could arise from an extended Higgs sector \cite{Dutta:2020scq,Dutta:2024hqq,Dutta:2025rxh}. When a neutrino propagates through matter, $\phi'$ would act as a mediator for interactions between the neutrino and the medium, which would give rise to an effective interaction of the form
\begin{equation}\label{eq:Leff}
    \mathcal{L}_{\rm{SNSI}}
    = \frac{y_f y_{\alpha \beta}}{m_{\phi'}^2} (\bar{\nu_\alpha}\nu_\beta)(\bar{f}f),
\end{equation}
where $y_{\alpha \beta}$ is the Yukawa coupling between $\phi'$ and the neutrinos of flavors $\alpha$ and $\beta$; and $y_f$ are the couplings between $\phi'$ and the other standard model fermions \cite{Ge:2018uhz}. 
Similarly to the standard model Yukawa terms, this coupling would yield an effective mass term for neutrinos. However, these mass terms are distinct from an inherent neutrino mass in that they depend on the matter density. The neutrino flavor evolution during propagation would be governed by the following effective Hamiltonian 
\begin{equation} \label{eq:HSNSI}
    H_{\rm{SNSI}}=\frac{(UMU^\dagger+\delta M)(UMU^\dagger+\delta M)^\dagger +A}{2 E},
\end{equation}
where $M=\diag(m_1,m_2,m_3)$ is the neutrino mass matrix and $A=\diag(\ACC,0,0)$ is the standard charged current matter potential discussed in Sec.~\ref{sec:standardderivation}. The contribution from SNSI, $\delta M$, is usually expressed with either of the following parameterizations
\begin{equation}
    \delta M =  
    \begin{pmatrix}
    \dmee & \delta m_{e \mu} & \delta m_{e \tau} \\
    \delta m_{e \mu}^* & \dmmm & \delta m_{\mu \tau} \\
    \delta m_{e \tau}^* & \delta m_{\mu \tau}^* & \dmtt
    \end{pmatrix} = \sqrt{|\Delta m^2_{31}|} 
    \begin{pmatrix}
    \eta_{ee} & \eta_{e \mu} & \eta_{e \tau} \\
    \eta_{e \mu}^* & \eta_{\mu \mu} & \eta_{\mu \tau} \\
    \eta_{e \tau}^* & \eta_{\mu \tau}^* & \eta_{\tau \tau}
    \end{pmatrix},
\end{equation}
where
\begin{equation}
    \delta m_{\alpha \beta} = \frac{y_{\alpha \beta}}{m_{\phi'}^2} \sum_f N_f y_f = \frac{y_{\alpha \beta}}{m_{\phi'}^2}  ~\rho ~ N_A\sum_f  Y_f y_f ,
\end{equation}
where $N_A$ is the Avogadro's constant, $\rho$ the matter density of the propagation medium and $Y_f$ the fraction of fermion $f$ in the medium. 
Note for the $\eta$ notation that there are different choices for the normalization with respect to the mass squared difference. While most papers have normalized it with respect to $\sqrt{|\Delta m^2_{31}|}$, some works on solar neutrinos have used $\sqrt{|\Delta m^2_{21}|}$ instead. The actual value of this mass squared difference also changes over time since one uses the current best-fit for this parameter. We avoid these ambiguities by using the $\delta m_{\alpha \beta}$ parameterization instead. A comparison with existing literature can be very easily made by dividing our $\delta m_{\alpha \beta}$ with $\sqrt{|\Delta m^2_{31}|}$ (or $\sqrt{|\Delta m^2_{21}|}$, wherever relevant). 

\subsection{\texorpdfstring{$\dmee$}{dmee}} \label{sec:SNSI_ee}
In the subsequent derivation we diagonalize Eq.~\eqref{eq:HSNSI} in the special case where $\delta M = \text{diag}(\dmee,0,0)$. This resembles previous literature \cite{Gupta:2023wct, Singha:2023set}, which proceeds much in the same way as in Sec.~\ref{sec:standardderivation}. Our results here stand out in that we
arrive at compact expressions that are analogous to those in Sec.~\ref{sec:standardderivation}. Further, we use these expressions to derive resonance relations for the mixing angles.

We here parameterize the lepton mixing matrix, $\PMNS$, using the PDG convention as given by Eq.~\eqref{eq:Uee}. As in Sec.~\ref{sec:standardderivation}, we consider the rotated frame $H'=U_{a_e}^\dagger H_{\rm{SNSI}} U_{a_e}$ with the auxiliary matrix $U_{a_e}=R_{23} U_{\delta} R_{13}$ and $H_{\rm{SNSI}}$ as defined in Eq.~(\ref{eq:HSNSI}). With the notation $\mathcal H'=2E~H'$, the Hamiltonian evaluated in this basis reads
\begin{align}
    \mathcal H_{11}' &= m_1^2 c_{12}^2 + m_2^2 s_{12}^2 + c_{13}^2(\,\dmee(2m_++\dmee)+\ACC), \nonumber \\ 
    \mathcal H_{22}' &= m_1^2 s_{12}^2 + m_2^2 c_{12}^2, \nonumber \\ 
    \mathcal H_{33}' &= m_3^2 + s_{13}^2(\dmee(2 m_3 + \dmee)+\ACC) , \nonumber \\ 
    \mathcal H_{12}' =\mathcal H_{21}' &= c_{12} s_{12} \Delta m_{21}^2 + c_{13}^2\,\dmee\, m_-, \nonumber \\ 
    \mathcal H_{13}'=\mathcal H_{31}'  &= s_{13} c_{13}(\dmee(m_++m_3+\dmee)+\ACC), \nonumber \\ 
    \mathcal H_{23}'=\mathcal H_{32}'  &= s_{13} c_{13}\,\dmee\, m_-, \label{eq:Hamiltoniandmee}
\end{align}
where we defined $m_+=m_1 c_{12}^2+m_2 s_{12}^2$ and $m_-=s_{12} c_{12} (m_2-m_1)$.
Note that the Hamiltonian expression in Eq.~\eqref{eq:Hamiltoniandmee} is more complicated than the one in Sec.~\ref{sec:standardderivation}. Here for example $H_{23}' \neq 0$. Nonetheless, we proceed with applying the rotation $H''=(R_{13}')^\dagger H' R_{13}'$. The angle $\theta_{13}'$ is fixed by imposing Eq.~\eqref{eq:t13prequirement}, which now has the solution
\begin{equation} \label{eq:t13pSNSIee}
    \tan 2\theta_{13}' = \frac{\AeeS\,\sin 2\theta_{13}}{\DeeS - \AeeS \cos 2\theta_{13}},
\end{equation}
where we defined 
\begin{equation}
    A_{ee}^S=\ACC+\dmee (m_++m_3) + \dmee^2
\end{equation}
and 
\begin{align} \label{eq:DeeS}
    \Dee^{S}&= (m_3-m_1)(m_1+m_3+\dmee)- (m_2-m_1)(m_1+m_2+\dmee) s_{12}^2\nonumber \\&= \Delta_{ee}+\delta m_{ee} (m_3-m_+),
\end{align}
with $\Dee$ as defined in Eq.~\eqref{eq:Dee}.
A second pair of off-diagonal elements can be set to zero with the assumption
\begin{equation} \label{eq:dmeeNeglect}
    H_{23}''=H_{32}'' = s_{13}'\left(c_{12}s_{12}\Delta m_{21}^2 + c_{13}^2 \dmee m_-\right)
             + c_{13}'\, s_{13} c_{13}\,\dmee\, m_-\equiv \epsilon^S_{e}.
\end{equation}
Neglecting $\epsilon^S_{e}$ is a more complicated assumption than in previous cases, as it involves additional terms $\propto \dmee m_-$ that are suppressed by either $s_{13}'$ or $s_{13}$. Note that $m_-\sim m_2-m_1$ is generally small since $m_2-m_1 =  \Delta m_{21}^2/(m_2+m_1) \ll \sqrt{|\Delta m_{31}^{2}|}$. Furthermore, the final eigenvalues and eigenvectors depend only on $O((\epsilon^S_e)^2)$, as was explained in Sec.~\ref{sec:standardderivation}. With these points in mind, we will numerically show in Sec.~\ref{sec:approximationaccuracy} that this assumption holds remarkably well in practice. 

Now, the final pair of off-diagonal elements are diagonalized via the rotation $H^d=R_{12}'^\dagger H'' R_{12}'$, where $\theta_{12}'$ is given by Eq.~\eqref{eq:t12eff}. After this final rotation, the Hamiltonian $H^d$ is diagonal and the eigenvalues can be extracted as in Sec.~\ref{sec:standardderivation}. The effective mixing matrix is here composed of the aforementioned Jacobi rotation angles, in addition to a final $U_\delta^{\dagger}$ rotation, exactly as in Eq.~\eqref{eq:Ueffee}. With same comparison between Eq.~\eqref{eq:Ueffee} and Eq.~\eqref{eq:Uee} we conclude that the relations in Eq.~\eqref{eq:primetoeff_ee} also hold here.
Finally, we arrive at the effective parameter mappings
\begin{align}
\tan 2{\theta_{13}^{\text{eff}}} &= \frac{\DeeS \sin 2 \theta_{13}}{\DeeS \cos 2 \theta_{13} - \AeeS}, \label{eq:t13effee} \\
\tan2{\theta_{12}^{\text{eff}}}&= \frac{\Delta m^2_{21}\sin 2\theta_{12}c_{13}'
   + 2\,\dmee\, m_- c_{13}c_{13}^{\text{eff}}}
  {\Delta m^2_{21}\cos 2\theta_{12}
   - \AeeS (c_{13}^{\text{eff}})^2
   - \DeeS (s_{13}')^2
   + \dmee (m_3 - m_+)c_{13}^2},\label{eq:t12effee} \\
\Delta {m_{21}^{2,\mathrm{eff}}} &=
  \frac{\Delta m^2_{21}\sin 2\theta_{12} c_{13}'
   + 2\,\dmee\, m_- c_{13}c_{13}^{\text{eff}}}{\sin 2{\theta_{12}^{\text{eff}}}},\label{eq:dm21effee} \\
\Delta {m_{31}^{2,\mathrm{eff}}} &= \DeeS\,
  \frac{\sin 2\theta_{13}}{\sin 2{\theta_{13}^{\text{eff}}}}
  + \Delta {m_{21}^{2,\mathrm{eff}}} (s_{12}^{\text{eff}})^2,\label{eq:dm31effee}
\end{align}
where Eq.~\eqref{eq:dm31effee} can also be interpreted as
\begin{equation}
     \Dee\eff  =  \DeeS\,
  \frac{\sin 2\theta_{13}}{\sin 2{\theta_{13}^{\text{eff}}}}.
\end{equation}
As in Sec.~\ref{sec:standardderivation}, $\dCP$ and $\theta_{23}$ retain their vacuum values. Overall, Eqs.~\eqref{eq:t13effee}--\eqref{eq:dm31effee} are analogous to Eqs.~\eqref{eq:t13effStandard}--\eqref{eq:dm31effStandard}, the differences being that $(A_{CC},\Delta_{ee}) \to (A_{CC}^S,\Delta_{ee}^S)$ and that additional terms appear in the numerator and denominator of Eq.~\eqref{eq:t12effee} and the numerator of Eq.~\eqref{eq:dm21effee}. While these additional terms are of the order $\sim \epsilon_e^S$ they are relevant to include since they enter linearly, unlike the neglected terms from Eq.~\eqref{eq:dmeeNeglect} that only enter quadratically.

The condition for resonant enhancement of $\theta_{13}\eff$ here reads
\begin{equation} \label{eq:t13resonancedmee}
    A_{CC}(E)+2 \delta m_{ee} M_++ \delta m_{ee}^2= \Dee \cos2\theta_{13}.
\end{equation}
where we defined 
\begin{equation}
    M_+\equiv c_{13}^2 m_++s_{13}^2 m_3=c_{13}^2 (m_1 c_{12}^2+m_2 s_{12}^2)+s_{13}^2 m_3.
\end{equation}
We emphasize that the only energy dependent term in Eq.~\eqref{eq:t13resonancedmee} is the standard matter potential $A_{CC}(E)$.
With this in mind, we present two distinct ways to view this resonance condition.
One interpretation is to write it as a shifted MSW resonance condition for $A_{CC}$, i.e.
\begin{equation} 
    A_{CC}^{13,\text{res}} = \Dee \cos2\theta_{13}-2 \delta m_{ee} M_+-\delta m_{ee}^2, 
    \label{eq:t13dmeeMSW}
\end{equation}
which can be compared to Eq.~\eqref{eq:t13resonanceStandard}.
Another interpretation is to instead write this as a resonance condition for $\dmee$, i.e.
\begin{equation}\label{eq:t13eeSMSW}
    \dmee^{13,\pm}= -M_+ \pm \sqrt{M_+^2+\Dee \cos2\theta_{13}-\ACC}. 
\end{equation}
Remarkably, even when setting $\ACC=0$, the SNSI coupling can by itself induce a resonant enhancement of the mixing angle, that is a Scalar MSW effect (SMSW). For negligible $\ACC$, this effect is energy independent. If both $\ACC$ and $\dmee$ are large however, the SMSW resonance condition in Eq.~\eqref{eq:t13eeSMSW} becomes shifted in an energy dependent way, or, put differently, the MSW condition in  Eq.~\eqref{eq:t13dmeeMSW} gets shifted in an energy independent way. 
Unlike the standard MSW effect, the SMSW effect would \textbf{not} manifest as a spectral signature. The only way to deduce or exclude the presence of the SMSW effect is by comparing the effective mixing parameters in neutrino propagation environments with different matter densities. Notice also that this resonance condition intricately depends on the values of the intrinsic neutrino masses $m_i$.
Further note that the potential arising from a scalar coupling has the same sign for a neutrino and an antineutrino, and so, unlike $\ACC$, $\dmee$ is either always positive or always negative \cite{Ge:2018uhz}. 


Another important aspect of Eq.~\eqref{eq:t13eeSMSW} is that it has two roots. The positive root, $|\delta m_{ee}|\ll M_+$, is a genuine scalar MSW resonance since the 1--3 mixing is maximal while $\Delta^S_{ee}$ remains of similar size, so $\nu_e$ oscillations have maximal amplitude and a normal oscillation length. The negative root, $\delta m_{ee}\simeq-(m_3+m_+)$, also makes the 1--3 mixing maximal for $A_{CC}\neq0$. However, it simultaneously drives $\Delta^S_{ee}$ to a very small value, $\mathcal{O}(m_-^2)$ for $A_{CC}=0$, or otherwise to $\simeq(m_3-m_+)A_{CC}/(2 M_+)$. As a result, the oscillation length of $\nu_e$ becomes very large, so $\nu_e$ oscillations effectively vanish at finite baseline, yielding a level crossing without mixing. 

The resonance condition in Eq.~\eqref{eq:t13eeSMSW} may simplify 
with the assumption that
$\dmee$ is small enough to where the two roots approximately coincide to the linear solution
\begin{equation} \label{eq:t13eeSMSWlinear}
    \dmee^{13,\text{res}}=\frac{\Dee \cos2\theta_{13}-\ACC}{2M_+}.
\end{equation}




Moving on, the resonance condition for $\theta_{12}\eff$ can be seen from Eq.~\eqref{eq:t12effee} and reads
\begin{equation}
    \AeeS (c_{13}^{\text{eff}})^2
   +\DeeS (s_{13}')^2 - \dmee (m_3 - m_+)c_{13}^2 = \Delta m_{21}^2\cos2\theta_{12}.
\end{equation}
This relation can be simplified if we assume that the $1$--$2$ resonance occurs for much smaller values of the matter potential than the $1$--$3$, as is the case for the standard MSW scenario, in which case $\theta_{13}'\approx0$ here and we may write
\begin{equation} \label{eq:t12resonanceSNSIee}
    A_{CC}(E)+2\dmee m_++\dmee^2 = \frac{\Delta m_{21}^2 \cos2\theta_{12}}{c_{13}^2}.
\end{equation}
Analogously to Eq.~\eqref{eq:t13resonancedmee}, Eq.~\eqref{eq:t12resonanceSNSIee} exhibits a shifted MSW relation for $A_{CC}$ as
\begin{equation}
    A_{CC}^{12,\text{res}}\approx \frac{\Delta m_{21}^2 \cos2\theta_{12}}{c_{13}^2}-2\dmee m_+-\dmee^2,
\end{equation}
or alternatively, one can put it as a SMSW condition for $\dmee$, yielding
\begin{equation} \label{eq:t12eeSMSW}
    \dmee^{12,\pm}=-m_+\pm\sqrt{m_+^2+\frac{\Delta m_{21}^2 \cos2\theta_{12}}{c_{13}^2}-\ACC}.
\end{equation}
Eqs.~\eqref{eq:t12resonanceSNSIee}--\eqref{eq:t12eeSMSW} were derived assuming $\theta'_{13}\approx0$, which requires $|\delta m_{ee}|\ll m_3+m_+$. This holds for the positive root but not for the negative one, where $\delta m_{ee}\simeq-2m_+$ and $A^S_{ee}\simeq-\Delta_{ee}$ drives $\theta^{\rm eff}_{13}\to0$. Retaining $\theta'_{13}$ one finds that at this point both the numerator and the denominator of Eq.~\eqref{eq:t12eff} vanish: the levels cross without mixing, so there is no 1--2 resonance.
With this in mind, the linearized form of Eq.~\eqref{eq:t12eeSMSW}, valid for small $\dmee$, reads
\begin{equation} \label{eq:t12eeSMSWlinear}
    \dmee^{12,\text{res}} =\frac{\Delta m_{21}^2 \cos 2\theta_{12}}{2 m_+c_{13}^2},
\end{equation}
which was identified previously in Ref.~\cite{Choubey:2026jiq}.

\subsection{\texorpdfstring{$\dmmm$}{dmmm}} \label{sec:SNSI_mm}
In this section, we consider the SNSI Hamiltonian given by Eq.~\eqref{eq:HSNSI} in the special case $\delta m = \diag(0,\dmmm,0)$. We proceed as in Sec.~\ref{sec:Vmmtoy} taking the $U^\mu$ parameterization of the PMNS matrix as given by Eq.~\eqref{eq:Umm}.
Since $\ACC$ and $\delta m$ are in different elements here we cannot employ the method without neglecting one or the other.
Thus, we will treat $\ACC$ as subdominant for this section and proceed as before. Considering this caveat, the resulting formulas of this section should be employed judiciously in experimental settings where $\ACC \gtrsim \Delta m_{21}^{2,\text{eff}}$. In Sec.~\ref{sec:1stordercorrections}, we detail how $\ACC$ can be accounted for to some extent using standard perturbation theory. 

Now, evaluating the rotated Hamiltonian, $H'=U_{a_\mu}^\dagger H_{SNSI} U_{a_\mu}$, using the auxiliary basis $U_{a_\mu}=R_{13}U_\delta R_{23}$ yields
\begin{align} 
    \mathcal H_{11}' &= m_1^2 (c_{12}^\mu)^2 + m_2^2 (s_{12}^\mu)^2, \nonumber \\ 
    \mathcal H_{22}' &= m_1^2 (s_{12}^\mu)^2 + m_2^2 (c_{12}^\mu)^2 + (c_{23}^\mu)^2 \dmmm(2m_+^{\mu} +\dmmm),\nonumber \\ 
    \mathcal H_{33}' &= m_3^2 + (s_{23}^\mu)^2 \dmmm(2 m_3 + \dmmm) , \nonumber\\ 
    \mathcal H_{12}'=\mathcal H_{21}' &= s_{12}^\mu c_{12}^\mu \Delta m_{21}^2 + (c_{23}^\mu)^2 \dmmm m_-^{\mu} \,\nonumber \\ 
    \mathcal H_{13}'=\mathcal H_{31}' &=  c_{23}^\mu s_{23}^\mu \dmmm m_-^{\mu}, \nonumber\\ 
    \mathcal H_{23}'=\mathcal H_{32}' &=c_{23}^\mu s_{23}^\mu \dmmm(m_+^{\mu} +m_3+\dmmm), \label{eq:Hpmm}
\end{align}
where we used the shorthands $\mathcal H'=H'/2E$, $m_+^{\mu} = m_1 (s_{12}^\mu)^2+m_2 (c_{12}^\mu)^2$ and $m_-^{\mu} = c_{12}^\mu s_{12}^\mu (m_2-m_1)$. The derivation now follows exactly as in Sec.~\ref{sec:Vmmtoy} where we first apply the rotation $H''=(R_{23}')^\dagger H' R_{23}'$. One pair of off-diagonal elements are then set to zero by assigning $\theta_{23}'$ with Eq.~\eqref{eq:H23pVmm}, which here has the solution
\begin{equation}
    \tan 2\theta_{23}' = \frac{\AmmS\,\sin 2\theta_{23}}{\DmmS - \AmmS \cos 2\theta_{23}}, 
\end{equation}
where we defined 
\begin{equation}
    \AmmS=\dmmm(m_+^{\mu}+m_3)+\dmmm^2
\end{equation}
and
\begin{align} \label{eq:DmmS}
    \DmmS &= (m_3-m_1)(m_1+m_3+\dmmm) - (m_2-m_1)(m_1+m_2+\dmmm)(c_{12}^\mu)^2\nonumber \\&= \Dmm+\dmmm (m_3-m_+^\mu).
\end{align}
Another pair of off-diagonal elements are set to zero by neglecting the element
\begin{equation} \label{eq:dmmmNeglect}
    H_{13}'' = H_{31}'' = s_{23}'\left(c_{12}s_{12}\Delta m_{21}^2 + c_{23}^2 \dmmm m_-^{\mu}\right)
             + c_{23}'\, s_{23} c_{23}\,\dmmm\, m_-^{\mu} \equiv \epsilon^S_{\mu},
\end{equation}
which can be motivated analogously to Eq.~\eqref{eq:dmeeNeglect}.
The final pair of off-diagonal elements are set to zero via the rotation $H^d = (R_{12}'^{\mu})^\dagger H'' R_{12}'^\mu$, where $\theta_{12}'^{\mu}$ is fixed by Eq.~\eqref{eq:t12eff}. After this, the Hamiltonian, $H^d$, is diagonal and the eigenvalues can be extracted.
As in Sec.~\ref{sec:Vmmtoy}, the rotations are collected and a final $U_\delta^\dagger$ phase is added to express the effective mixing matrix as given by Eq.~\eqref{eq:Ueffmm}, which can be compared with the vacuum mixing matrix in Eq.~\eqref{eq:Umm} to conclude the relations in Eq.~\eqref{eq:primetoeff_mm}.
The effective parameter mappings can finally be written as
\begin{align}
\tan 2{\theta_{23}^{\mu,\text{eff}}} &= \frac{\DmmS \sin 2 \theta_{23}^\mu}{\DmmS \cos 2 \theta_{23}^\mu - \AmmS},\label{eq:t23effmm} \\
\tan2{\theta_{12}^{\mu,\text{eff}}}&= \frac{\Delta m^2_{21}\sin 2\theta_{12}^\mu c_{23}'^\mu
   + 2\,\dmmm\, m_-^{\mu} c_{23}^\mu c_{23}^{\mu,\text{eff}}}
  {\Delta m^2_{21}\cos 2\theta_{12}^\mu
   + \AmmS (c_{23}^{\mu,\text{eff}})^2
   + \DmmS (s_{23}'^\mu)^2
   - \dmmm (m_3 - m_+^{\mu})(c_{23}^\mu)^2},\label{eq:t12effmm} \\
\Delta {m_{21}^{2,\mathrm{eff}}} &=
  \frac{\Delta m^2_{21}\sin 2\theta_{12} c_{23}'^\mu
   + 2 \dmmm m_-^{\mu} c_{23} c_{23}^{\mu,\text{eff}}}{\sin 2{\theta_{12}^{\mu,\text{eff}}}},\label{eq:dm21effmm} \\
\Delta {m_{31}^{2,\mathrm{eff}}} &= \DmmS\,
  \frac{\sin 2\theta_{23}^\mu}{\sin 2{\theta_{23}^{\mu,\text{eff}}}}
  + \Delta {m_{21}^{2,\mathrm{eff}}} (c_{12}^{\mu,\text{eff}})^2\label{eq:dm31effmm},
\end{align}
where Eq.~\eqref{eq:dm31effmm} can also be interpreted as
\begin{equation}
     \Dmm\eff = \DmmS \frac{\sin 2\theta_{23}^\mu}{\sin 2{\theta_{23}^{\mu,\text{eff}}}}.
\end{equation}
As in Sec.~\ref{sec:Vmmtoy}, $\dCP^\mu$ and $\theta_{23}^\mu$ retain their vacuum values.
Overall, the expressions for the effective parameters in Eqs.~\eqref{eq:t23effmm}--\eqref{eq:dm31effmm} are analogous to the ones in earlier sections. The difference between these $\dmmm$ results as compared to the ones for $\Amm$ are very similar to the differences we found between the $\dmee$ and $\ACC$ cases, and so we refer the reader to the detailed discussions in Secs.~\ref{sec:Vmmtoy} and \ref{sec:SNSI_ee}.
The noteworthy difference here is that we have not accounted for the standard potential, $\ACC$, which considerably limits the applicability of the conclusions we draw. The previously seen interplay between the $\ACC$ MSW relation and $\dmee$ SMSW relation here corresponds to an interplay between the $\Amm$ MSW relation and the $\dmmm$ SMSW relation. Such discussion will be omitted due to the lack of experimental support for vector-like NSI with a dominant $\Amm$ and a sub-leading $\ACC$ \cite{Ohlsson:2012kf,Coloma:2023ixt}, although one could straightforwardly work it out like in the previous section by taking $\AmmS \mapsto \Amm+\AmmS$.
Nonetheless, we can use Eqs.~\eqref{eq:t23effmm} and \eqref{eq:t12effmm} to find relations for the SMSW effect induced by $\dmmm$ in a low-energy regime. We find for the $2$--$3$ sector that
\begin{equation}\label{eq:t13dmmmresonance}
    2\dmmm M_+^\mu +\dmmm^2=\Dmm \cos2\theta_{23}^\mu,
\end{equation}
where we defined 
\begin{equation}
    M_+^\mu\equiv (c_{23}^\mu)^2 m_+^\mu+(s_{23}^\mu)^2 m_3 =(c_{23}^\mu)^2 (m_1 (s_{12}^\mu)^2+m_2 (c_{12}^\mu)^2)+(s_{23}^\mu)^2m_3,
\end{equation}
and the solutions read
\begin{equation} \label{eq:t13dmmmSMSW}
    \dmmm^{23,\pm}=-M_+^\mu \pm \sqrt{\left(M_+^\mu\right)^2 + \Dmm \cos2\theta_{23}^\mu}.
\end{equation}
As in the $\dmee$ case, the positive root yields a genuine SMSW resonant enhancement, whereas the negative root drives $\DmmS \to 0$ and so only describes a level crossing without mixing.
For small values of $\dmmm$, we can write the approximately linearized solution
\begin{equation} \label{eq:t13dmmmSMSWlinear}
    \dmmm^{13}=\frac{\Dmm \cos2\theta_{23}^\mu}{2M_+^\mu}.
\end{equation}
Let us again emphasize that a resonant enhancement of $\theta_{23}^\mu$ or $\theta_{12}^\mu$ would quite neatly  correspond to a resonant enhancement of the PDG parameters $\theta_{23}$ or $\theta_{12}$ respectively, as was discussed in Sec.~\ref{sec:generalizing}.

In the $1$--$2$ sector we find the SMSW resonance condition from Eq.~\eqref{eq:t12effmm} as
\begin{equation}
   \dmmm (m_3 - m_+^{\mu})(c_{23}^\mu)^2 - \AmmS (c_{23}^{\mu,\text{eff}})^2
   - \DmmS (s_{23}'^\mu)^2 = \dmsq{21} \cos2\theta_{12}^\mu
\end{equation}
where the relative sign here has flipped as compared to the $\dmee$ case in accordance with the results for $\Amm$ in Sec.~\ref{sec:Vmmtoy}. Furthermore, as was noted for $\Amm$, the fact that $\theta_{23}^\mu$ is near maximal for the Nufit6.0 values means that the $1$--$2$ and $2$--$3$ resonances are closer together. This further exacerbate the fallibility of the approximation $\theta_{23}'^\mu\approx0$, in particular for low values of $m_1$ i.e. normal mass ordering with low $m_l$. In spite of this reservation we nonetheless show the simplified result with the approximation $\theta_{23}'^\mu\approx0$ where
\begin{equation} \label{eq:t12dmmmresonance}
    -2 \dmmm m_+-\dmmm^2 = \frac{\dmsq{21} \cos2\theta_{12}^\mu}{ (c_{23}^\mu)^2}.
\end{equation}
The resonance condition in Eq.~\eqref{eq:t12dmmmresonance} yields the SMSW solutions
\begin{equation} \label{eq:t12mmSMSW}
     \dmmm^{12,{\pm}}=-m_+\pm\sqrt{m_+^2-\frac{\dmsq{21} \cos2\theta_{12}^\mu}{ (c_{23}^\mu)^2}},
\end{equation}
where one should only take the positive root, similarly to what was explained for Eq.~\eqref{eq:t12eeSMSW}, and for small $\dmmm$ we may write the linearized solution
\begin{equation}\label{eq:t12mmSMSWlinear}
    \dmmm^{12,\text{res}}=-\frac{\dmsq{21} \cos2\theta_{12}^\mu}{ 2m_+(c_{23}^\mu)^2}.
\end{equation}

\subsection{\texorpdfstring{$\dmtt$}{dmtt}} \label{sec:SNSI_tt}
We now consider the SNSI Hamiltonian in the special case $\delta m =\diag(0,0,\dmtt)$. The procedure here follows as in Sec.~\ref{sec:Vtttoy} with the PMNS matrix parameterized by $U^\tau$ as in Eq.~\eqref{eq:Utt}. To proceed, we need to assume that the standard matter potential, $\ACC$, is negligible for the same reason as in Sec.~\ref{sec:SNSI_mm}.
Let us emphasize again that the following results, although worked out in terms of the $U^\tau$ parameters, can be converted to the PDG parameters as shown in App.~\ref{app:parameterizationconversion} and ~\ref{app:approximateconversionrelations}.
Evaluating the rotated Hamiltonian, $H'=U_{a_\tau}^\dagger H_{SNSI} U_{a_\tau}$, using the auxiliary basis $U_{a_\tau}=\tilde R_{12} R_{23} U_\delta^\tau$ yields
\begin{align}
    \mathcal H_{11}' &= m_1^2 (c_{12}^\tau)^2 + m_2^2 (s_{12}^\tau)^2, \nonumber \\ 
    \mathcal H_{22}' &= m_1^2 (s_{12}^\tau)^2 + m_2^2 (c_{12}^\tau)^2 + (s_{23}^\tau)^2 \dmtt(2m_+^\tau+\dmtt),\nonumber \\ 
    \mathcal H_{33}' &= m_3^2 + (c_{23}^\tau)^2 \dmtt(2 m_3 + \dmtt) , \nonumber\\ 
    \mathcal H_{12}'=\mathcal H_{21}' &= c_{12}^\tau s_{12}^\tau \Delta m_{21}^2 + (s_{23}^\tau)^2 \dmtt m_-^\tau,\nonumber \\ 
    \mathcal H_{13}'=\mathcal H_{31}' &= - c_{23}^\tau s_{23}^\tau  \dmtt m_-^\tau, \nonumber \\ 
    \mathcal H_{23}'=\mathcal H_{32}' &=  -c_{23}^\tau s_{23}^\tau\dmtt(m_+^\tau+m_3+\dmtt),
\end{align}
where we used the shorthands $\mathcal H'=H'/2E$, $m_+^\tau=m_1 (s_{12}^\tau)^2+m_2 (c_{12}^\tau)^2$ and $m_-^\tau=s_{12}^\tau c_{12}^\tau (m_2-m_1)$. Subsequently, we perform the rotation
$H''= (R_{23}'^{\tau})^\dagger H' R_{23}'^\tau$. The angle $\theta_{23}'^\tau$ is used to set one pair of elements to zero by imposing Eq.~\eqref{eq:H23prequirementVtt}, which has the solution
\begin{equation}
    \tan 2\theta_{23}'^\tau = \frac{-\,\AttS\,\sin 2\theta_{23}^\tau}
  {\DttS + \AttS \cos 2\theta_{23}^\tau},
\end{equation}
where we defined
\begin{equation}
    \Att^S=\dmtt (m_+^\tau +m_3) + \dmtt^2
\end{equation}
and 
\begin{align}
    \DttS&=(m_3-m_1)(m_1+m_3+\dmtt)- (m_2-m_1)(m_1+m_2+\dmtt) (c_{12}^\tau)^2 \nonumber \\
    &= \Dtt + \dmtt(m_3-m_+^\tau).
\end{align}
Another pair of off-diagonal elements are set to zero by assuming
\begin{equation} \label{eq:dmttNeglect}
    H_{13}''=H_{31}'' = s_{23}'^\tau (c_{12}^\tau  s_{12}^\tau  \Delta m_{21}^2 + (s_{23}^\tau)^2 \dmtt m_-^\tau) - c_{23}'^\tau  s_{23}^\tau  c_{23}^\tau \dmtt m_-^\tau \equiv \epsilon^S_\tau,
\end{equation}
which can be motivated analogously to the assumption in Eq.~\eqref{eq:dmeeNeglect}.
The final pair of off-diagonal elements are diagonalized via the rotation $H^d= (R_{12}'^{\tau})^\dagger H'' R_{12}'^\tau$ where $\theta_{12}'^\tau$ is fixed by Eq.~\eqref{eq:t12eff}. After this, the Hamiltonian $H^d$ is diagonal and the eigenvalues can be extracted.

As in Sec.~\ref{sec:Vtttoy}, the rotations can be collected to the effective mixing matrix in Eq.~\eqref{eq:Uefftt}, which can be compared with Eq.~\eqref{eq:Utt} to conclude the relations in Eq.~\eqref{eq:primetoeff_tt} for the effective parameters. We finally arrive at the effective parameter mappings
\begin{align}
\tan 2{\theta_{23}^{\tau,\text{eff}}} &= \frac{\DttS \sin 2 \theta_{23}^\tau}{\DttS \cos 2 \theta_{23}^\tau + \Att^S}, \label{eq:t23efftt} \\
\tan2{\theta_{12}^{\tau,\text{eff}}}&= \frac{\Delta m^2_{21}\,\sin 2\theta_{12}^\tau
   c_{23}'^\tau
   + 2\,\dmtt\, m_-^\tau s_{23}^\tau s_{23}^{\tau,\text{eff}}}
  {\Delta m^2_{21}\cos 2\theta_{12}^\tau
   + \Att^S (s_{23}^{\tau,\text{eff}})^2
   + \DttS (s_{23}'^\tau)^2
   - \dmtt(m_3 - m_+^\tau) (s_{23}^\tau)^2}, \label{eq:t12efftt} \\
\Delta {m_{21}^{2,\mathrm{eff}}} &=
  \frac{\Delta m^2_{21} \sin 2\theta_{12}^\tau c_{23}'^\tau
   + 2\,\dmtt\, m_-^\tau s_{23}^\tau s_{23}^{\tau,\text{eff}}}
       {\sin 2{\theta_{12}^{\tau,\text{eff}}}} \label{eq:dm21efftt}\\
\Delta {m_{31}^{2,\mathrm{eff}}} &= \DttS\,
  \frac{\sin 2\theta_{23}^\tau}{\sin 2{\theta_{23}^{\tau,\text{eff}}}}
  + \Delta {m_{21}^{2,\mathrm{eff}}} (c_{12}^{\tau,\text{eff}})^2,\label{eq:dm31efftt}
\end{align}
where Eq.~\eqref{eq:dm31efftt} can also be interpreted as
\begin{equation}
     \Dtt\eff  =  \DttS\,
  \frac{\sin 2\theta_{23}^\tau}{\sin 2{\theta_{23}^{\tau,\text{eff}}}}.
\end{equation}
In the $2$--$3$ sector the SMSW resonance condition for $\dmtt$ follows from Eq.~\eqref{eq:t23efftt} as
\begin{equation}\label{eq:t23dmttresonance}
    2\dmtt M_+^\tau+\dmtt^2=-\Dtt\cos2\theta_{23}^\tau,
\end{equation}
where we defined
\begin{equation}
    M_+^\tau\equiv (s_{23}^\tau)^2 m_+^\tau+(c_{23}^\tau)^2 m_3=(s_{23}^\tau)^2\left(m_1(s_{12}^\tau)^2+m_2(c_{12}^\tau)^2\right)+(c_{23}^\tau)^2 m_3.
\end{equation}
The solutions of Eq.~\eqref{eq:t23dmttresonance} are
\begin{equation}\label{eq:t23dmttSMSW}
    \dmtt^{23,\pm}=-M_+^\tau\pm\sqrt{\left(M_+^\tau\right)^2-\Dtt\cos2\theta_{23}^\tau}.
\end{equation}
Is in the previous $\dmee$ and $\dmmm$ cases, the positive solution correspond to a genuine resonant enhancement of $\theta_{23}^\mu$, whereas the negative solution drives $\DttS \to 0$ only yielding a level crossing without mixing.
For small $\dmtt$ Eq.~\eqref{eq:t23dmttSMSW} can be linearized to
\begin{equation}\label{eq:t23dmttSMSWlinear}
    \dmtt^{23}\approx-\frac{\Dtt\cos2\theta_{23}^\tau}{2M_+^\tau}.
\end{equation}
In the $1$--$2$ sector, the SMSW resonance condition follows from Eq.~\eqref{eq:t12efftt} and reads as
\begin{equation}
    \dmtt(m_3-m_+^\tau)(s_{23}^\tau)^2-\Att^S(s_{23}^{\tau,\text{eff}})^2-\DttS(s_{23}'^\tau)^2
    =\dmsq{21}\cos2\theta_{12}^\tau .
\end{equation}
With the same reservation as was discussed for Eq.~\eqref{eq:t12dmmmresonance}, we here take the approximation $\theta_{23}'^\tau\approx0$ yielding
\begin{equation}\label{eq:t12dmttresonance}
    -2\dmtt m_+^\tau-\dmtt^2=\frac{\dmsq{21}\cos2\theta_{12}^\tau}{(s_{23}^\tau)^2},
\end{equation}
with solutions
\begin{equation} \label{eq:t12ttSMSW}
    \dmtt^{12,\pm}=-m_+^\tau\pm\sqrt{\left(m_+^\tau\right)^2-\frac{\dmsq{21}\cos2\theta_{12}^\tau}{(s_{23}^\tau)^2}}.
\end{equation}
As before, the positive solution is a genuine SMSW resonance whereas the negative solution, if one studies the full $\theta_{23}'^\tau\neq0$ case, can be seen to be a level crossing without mixing. For small $\dmtt$ Eq.~\eqref{eq:t12ttSMSW} linearizes to
\begin{equation}\label{eq:t12ttSMSWlinear}
    \dmtt^{12}\simeq-\frac{\dmsq{21}\cos2\theta_{12}^\tau}{2m_+^\tau(s_{23}^\tau)^2}.
\end{equation}

\section{Accuracy and Perturbative Corrections}\label{sec:approximationaccuracyandcorrections}
In this section we discuss the numerical accuracy of the resulting formulas in Sec.~\ref{sec:SNSI_ee}, \ref{sec:SNSI_mm} and \ref{sec:SNSI_tt}. We also investigate perturbative corrections and demonstrates their application for studying SNSI in realistic experiments.

\subsection{Numerical Accuracy of the SNSI Results} \label{sec:approximationaccuracy}
When we employed the Jacobi diagonalization method in Secs.~\ref{sec:standardderivation}, \ref{sec:generalizing} and \ref{sec:SNSImain}; we assumed certain elements to be negligible in order to proceed, in particular the ones $\propto \epsilon_\alpha$ and, for $\dmmm$ and $\dmtt$, also all terms $\propto \ACC$. It is important to check to what extent these approximations hold and thus determine in which experiments the resulting formulas can be used. We perform this study by comparing our analytic formulas against numerically diagonalized results exact up to machine precision. We here assume that the standard mixing parameters take the central values of the Nufit 6.0 global fit \cite{NuFit6.0}, where, unless otherwise stated, we assume normal mass ordering. We further assume a constant matter density with a value of $\rho=2.45\, \text{g}/\text{cm}^3$ for the medium-baseline JUNO experiment \cite{JUNO:2022mxj} and $\rho=2.6\, \text{g}/\text{cm}^3$ for long-baseline beam experiments.

We proceed by computing the numerical accuracy of the effective parameter mappings for SNSI in Secs.~\ref{sec:SNSI_ee}, \ref{sec:SNSI_mm} and \ref{sec:SNSI_tt}.
Throughout this section we compute the relative error 
\begin{equation}
    \frac{|\Delta P_{\alpha \beta}|}{|P_{\alpha \beta}|}= \frac{|P_{\alpha \beta}^\text{formula}-P_{\alpha \beta}^\text{numerical}|}{|P_{\alpha \beta}^\text{numerical}|}.  \label{eq:Prel}
\end{equation}
Here, $P_{\alpha \beta}^\text{numerical}$ is computed by numerically diagonalizing the effective Hamiltonian, which are subsequently used to compute the mixing matrix element values $|U_{\alpha i}|$ (as in Appendix \ref{app:parameterization}) and these are finally inserted in the general expression \cite{10.1093/acprof:oso/9780198508717.001.0001}
\begin{equation} 
    P_{\alpha \beta} = \sum_{k,j} U_{\alpha k}^* U_{\beta k} U_{\alpha j} U_{\beta j}^*  \exp(-i \frac{\Delta m^2_{kj} L}{2E}). \label{eq:PfromU}
\end{equation}
The analytical probability $P_{\alpha \beta}^\text{formula}$ is computed by extracting the effective parameters using the respective formulas in Sec.~\ref{sec:SNSImain}, after which we similarly compute the mixing matrix element and insert them into Eq.~\eqref{eq:PfromU}. 

\begin{figure}[]
    \begin{center}
    \includegraphics[width=1\linewidth]{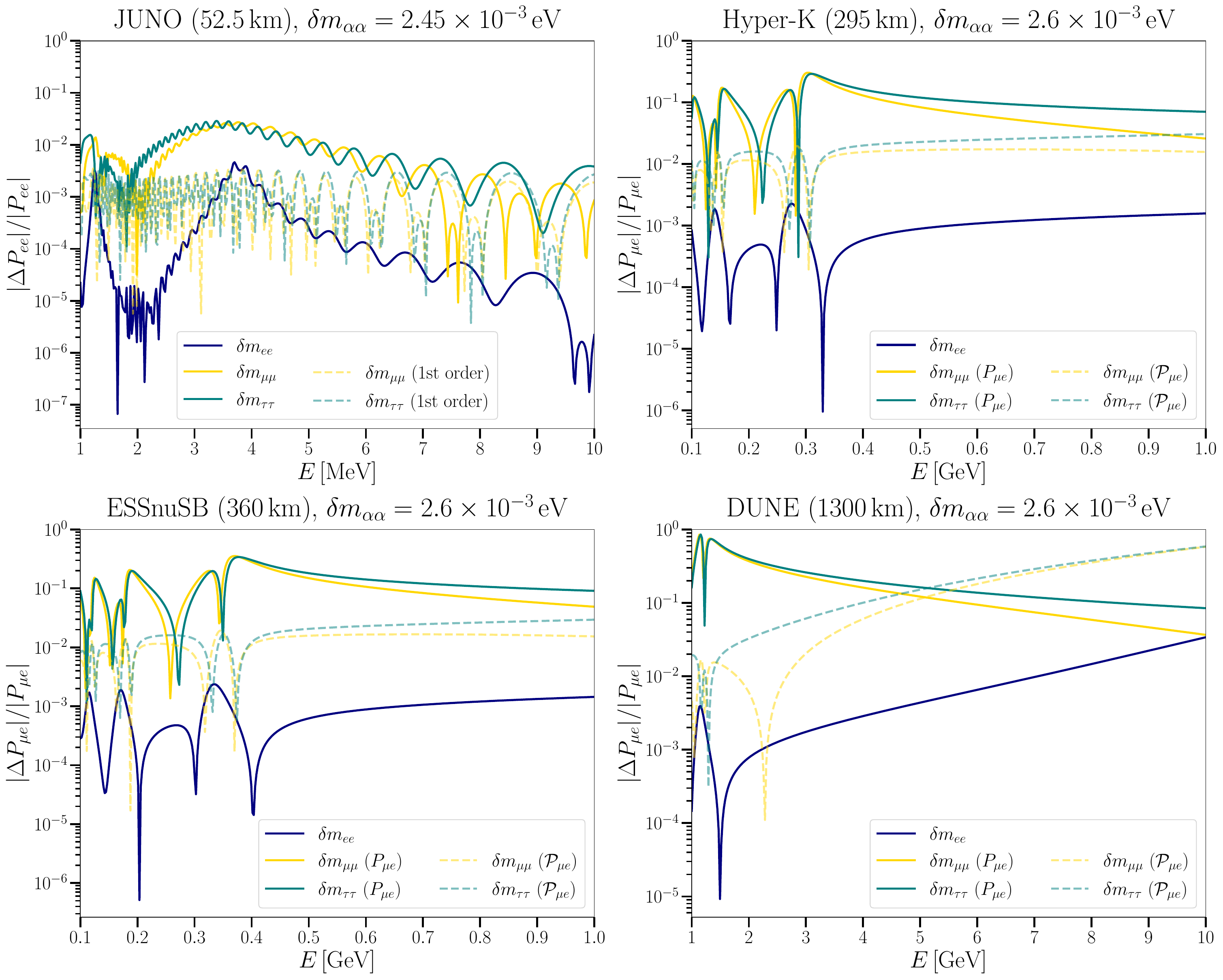}
    \caption{Relative error in the analytically obtained probability formulas using  effective parameter mappings from  Sec.~\ref{sec:SNSImain} for JUNO, Hyper-Kamiokande, ESSnuSB and DUNE experiments. In all panels we have taken the standard oscillation parameter values from Nufit 6.0 and assumed  normal mass ordering. The SNSI parameter values taken are $\dmaa/\rho=10^{-3}\,\frac{\text{eV}}{\text{g}/\text{cm}^3}$ and $m_1=0$, where $\alpha=e,\mu,\tau$. We consider a single $\dmaa$ at a time. For the long-baseline experiments, the solid lines are computed using the probability formula in Eq.~\eqref{eq:PfromU} while for the dashed lines we have used the formulas from Ref.~\cite{Martinez-Soler:2019nhb}.}
    \label{fig:experiments_spectrum}
    \end{center}
\end{figure}

In Fig.~\ref{fig:experiments_spectrum}, we show the relative error given in Eq.~\eqref{eq:Prel} as a function of the energy, $E$, for 
JUNO \cite{JUNO:2022mxj}, Hyper-Kamiokande \cite{Hyper-Kamiokande:2018ofw}, ESSnuSB \cite{ESSnuSB:2023ogw} and DUNE \cite{DUNE:2021tad} experiments.
The relative error is computed for three cases: $\dmee/\rho=10^{-3}\,\frac{\text{eV}}{\text{g}/\text{cm}^3}$, $\dmmm/\rho=10^{-3}\,\frac{\text{eV}}{\text{g/cc}}$ and $\dmtt/\rho=10^{-3}\,\frac{\text{eV}}{\text{g/cc}}$. Recall that a certain value of $\dmaa/\rho$ corresponds to a single assumption for the strength of SNSI, although it yields different magnitudes of $\dmaa$ in different experiments; for JUNO, these values correspond to $\dmaa=2.45 \times10^{-3}\,$eV, whereas in the long-baseline experiments $\dmaa=2.6 \times10^{-3}\,$eV. 
The results obtained by computing Eq.~\eqref{eq:Prel} as described above are shown by the solid lines in each panel of Fig.~\ref{fig:experiments_spectrum}. For $\dmee$, we note that the relative error is $<1\%$ for all cases except in DUNE when $E>7\,$GeV. For $\dmmm$ and $\dmtt$ we see that the accuracy is poorer in all experiments. We remind the reader that for the $\dmee$ case, standard matter potential $\ACC$ is fully incorporated in the analytical expressions, while for the $\dmmm$ and $\dmtt$ cases we assumed $\ACC=0$ when deriving the effective parameter mappings in Sec.~\ref{sec:SNSI_mm} and \ref{sec:SNSI_tt}. 

To alleviate the higher errors in the $\dmmm$ and $\dmtt$ formulas, we consider two approaches to account for $\ACC$. The first alternative, which we demonstrate for the JUNO experiment, is to compute the first order corrections to the effective parameter mappings from $H^{(1)}=\diag(\ACC,0,0)$, which will be discussed in Sec.~\ref{sec:1stordercorrections}. After this correction is applied one can use Eq.~\eqref{eq:PfromU} as before yielding higher accuracy for JUNO as is shown by the dashed lines in Fig. \ref{fig:experiments_spectrum}. 
The second alternative, which we demonstrate for the other long-baseline experiments, is that, instead of using Eq.~\eqref{eq:PfromU}, one can use formulas from the literature where the standard matter effects are accounted for via a perturbative expansion of the oscillation probability. In other words, the SNSI matter effects are accounted for using our effective parameter mappings whereas the standard matter effects are accounted for with another perturbative method that instead alters the oscillation probability expression directly.
In this work we specifically employ the probability formulas of Ref.~\cite{Martinez-Soler:2019nhb}, here referred to as $\mathcal P_{\mu e}$, which are particularly well-suited for lower energy experiments in accordance with the parameter mappings of our article. We show the resulting relative errors of this approach as the dashed lines in Fig. \ref{fig:experiments_spectrum} for the Hyper-Kamiokande, ESSnuSB and DUNE experiments. This approach reduces the relative errors down to $1\%$ for all energies in Hyper-Kamiokande and ESSnuSB, whereas for DUNE the errors decrease for lower energies but instead increase for $E>5\,$GeV. Indeed, while this approach utilizes the complementarity of having formulas from different approximation methods, we observe that the accuracy becomes worse for very high energies since one needs to take on all the approximation errors from both methods. 



\begin{figure}[]
    \begin{center}
    \includegraphics[width=1\linewidth]{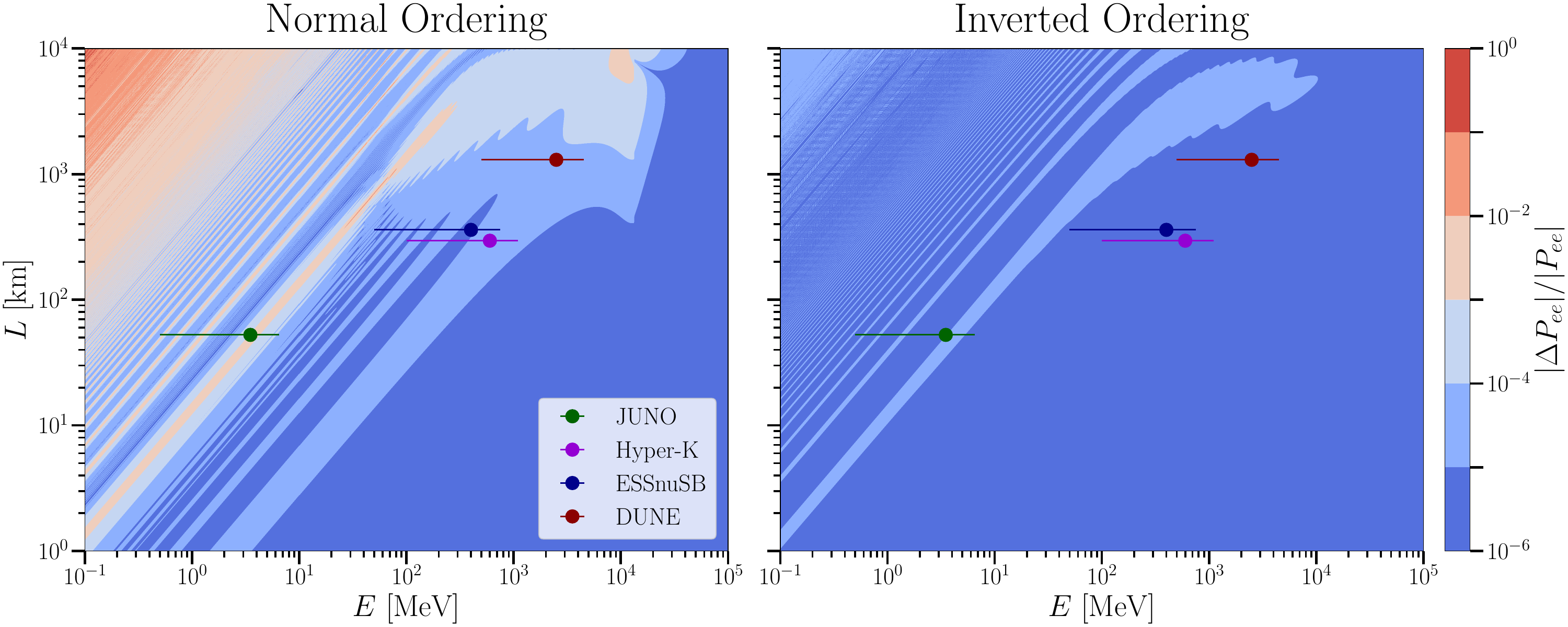}
    \caption{Relative error of the resulting formulas from Sec.~\ref{sec:SNSImain} as a function of energy and baseline. For the SNSI parameters we have taken $\delta M=\diag(\dmee,0,0)$ with $\dmee=2.45\times10^{-3}\,$eV and $m_l=0\,$eV, and for the standard oscillation parameters we take the Nufit 6.0 values. We mark out the JUNO, Hyper-Kamiokande, ESSnuSB and DUNE experiments for reference. Although the error is shown in terms of $|P_{ee}|$, which is not strictly applicable to the long-baseline experiments, the other $|P_{\alpha \beta}|$ have similar errors.}
    \label{fig:E-Lsweep0thorder}
    \end{center}
\end{figure}

In Fig. \ref{fig:E-Lsweep0thorder} we compute the relative error for the $\dmee$ parameter mappings as a function of both energy, $E$, and baseline $L$. We show this for the central values of Nufit 6.0 both for the normal ordering and inverted ordering cases. Since these parameter mappings can fully account for both $\dmee$ and $\ACC$, since they both enter in the $e$--$e$ element, the errors are remarkably small except for normal ordering with exceedingly high values of $L/E$. We observe that the errors are smaller for inverted ordering as compared to normal ordering. This can be understood by recalling Eq.~\eqref{eq:dmeeNeglect}, where we neglected terms $\propto m_-=s_{12}c_{12}(m_2-m_1) \propto \dmsq{21}/(m_1+m_2)$. For inverted ordering, $\dmsq{31}<(m_1+m_2)^2$ and so $m_-<\dmsq{21}/\sqrt{\dmsq{31}}$, whereas for normal ordering with small $m_l$ we only have $\dmsq{21}<(m_1+m_2)^2$ and so $m_-<\sqrt{\dmsq{21}}$. As a result, the neglected $\epsilon_e^S$ is largest for normal ordering with $m_l=0$.
Further note that, since these are effective parameter mappings rather than expansions of the probabilities themselves, the approximation errors are comparable for all flavor oscillations $P_{\alpha\beta}$. As such, for brevity we only scan the relative errors for $P_{ee}$ as a representative example.


\begin{figure}[]
    \begin{center}
    \includegraphics[width=1\linewidth]{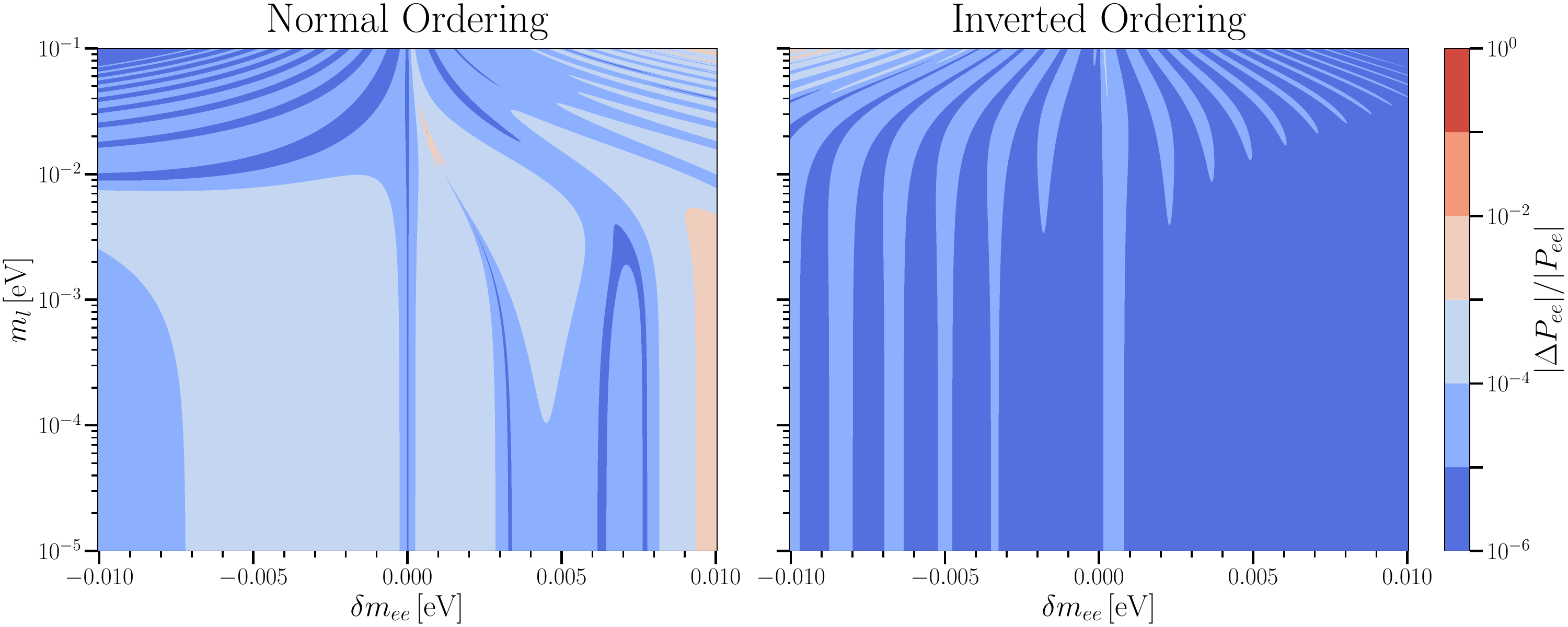}
    \caption{Relative error of the resulting formulas from Sec.~\ref{sec:SNSI_ee} applied to $|P_{ee}|$, displayed as a function of the SNSI parameters $\dmee$ and $m_l$. We here assume $E=3\,$MeV and $L=52.5\,$km, corresponding to the JUNO experiment. The corresponding plots for $\dmmm$ and $\dmtt$ are shown below in Fig. \ref{fig:eta-mlsweepmmtt0th1st}}
    \label{fig:eta-mlsweep0thorder}
    \end{center}
\end{figure}

Now, in order to understand the impact of the non-standard parameters, $\dmaa$ and $m_l$, on the error we show in Fig. \ref{fig:eta-mlsweep0thorder} the relative errors from Eq.~\eqref{eq:Prel} as a function of $\dmee$ and $m_l$ for the JUNO experiment. We also show the distinct behavior between normal ordering and inverted ordering. In both cases one might notice a distinct difference in the oscillation behavior around $m_l\approx 0.01\,$eV. This is the scale at which $ \Delta m_{21} < m_l < \Delta m_{31}$ and so manifests as a complicated transition region between the regimes $m_l \gg \Delta m_{ij}$ and $m_l \ll \Delta m_{ij}$.
One might also note that the $m_l$ dependence vanishes below the limit $m_l^2 \ll \Delta m_{21}^2$ in the case of normal mass ordering, whereas for inverted ordering this instead occurs for $m_l^2 \ll |\Delta m_{31}^2|$. Notice also the distinct behavior between positive and negative $\dmee$ values, which can be understood from Eq.~\eqref{eq:t12effee}.

\subsection{Including First Order Corrections} \label{sec:1stordercorrections}
Let us recall the two distinct assumptions taken when deriving the effective parameter mappings in Sec.~\ref{sec:SNSImain}.
First, we neglected the smallest off-diagonal element in the rotated auxiliary basis, $H''_{ij}$ as given by Eq.~\eqref{eq:dmeeNeglect}, \eqref{eq:dmmmNeglect} and Eq.~\eqref{eq:dmttNeglect} in each respective case. Second, in the context of $\dmmm$ and $\dmtt$, the matter potential, $\ACC$, was neglected. Both of these assumptions can be amended by working out higher order corrections using time independent perturbation theory \cite{Schrodinger:1926gei,Sakurai:2011zz}. It is generally relevant to account for both the matter potential and the off-diagonal element corrections simultaneously. However, it quickly becomes unwieldy to work out the corrections mainly because these corrections appear in different Hamiltonian elements in different rotated frames. Moreover, since the two corrections scale differently with energy and oscillation parameters, there are various regimes of relative smallness that one might consider depending on the experiment under consideration.

\begin{figure}[]
    \begin{center}
    \includegraphics[width=1\linewidth]{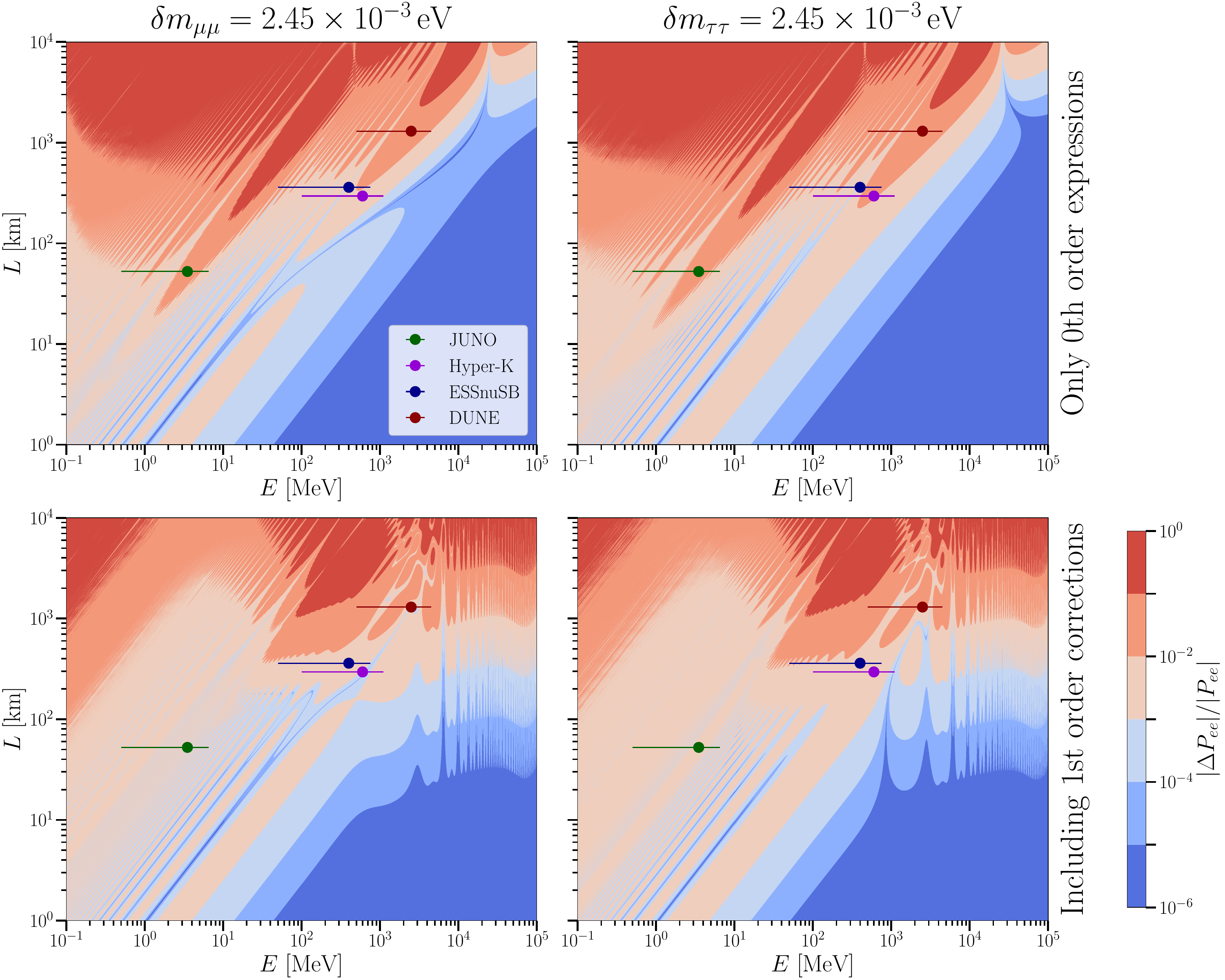}
    \caption{Relative error of the effective parameter mappings from Sec.~\ref{sec:SNSI_mm} and \ref{sec:SNSI_tt} computed on a two-dimensional grid as function of energy, $E$, and baseline, $L$. This accuracy is compared between when these 0th order formulas are applied directly as compared to when the first order corrections from equations \eqref{eq:1stordert12Ue}--\eqref{eq:1storderdm31Ue} are also accounted for. We have taken $\dmaa=2.45\times10^{-3}\,$eV, $m_l=0\,$eV, and the Nufit 6.0 values for the standard oscillation parameters}
    \label{fig:E-LsweepeeNOvsIO1st}
    \end{center}
\end{figure}

As was seen in Fig.~\ref{fig:experiments_spectrum} and Fig~\ref{fig:E-Lsweep0thorder}, the effective parameter mappings for $\delta M=\diag(0,\dmmm,0)$ and $\delta M=\diag(0,0,\dmtt)$ are inaccurate to the point of becoming unusable for a wide range of baselines and energies, which is in stark contrast to the remarkably accurate $\delta M=\diag(\dmee,0,0)$ case. Presumably, this difference stems almost entirely from the second approximation, $\ACC\approx0$. We therefore focus on applying corrections to these effective parameter mappings in the regime where $H^{(1)}=\diag(\ACC,0,0)$.
For the PDG parameters, such corrections are already well-known. Yet, since we have used three different mixing angle parameterizations throughout this work, we have worked out these corrections in all three parameterizations. We show explicitly how these corrections look in each of the three parameterizations in App.~\ref{app:perturbationtheory}. In this section, we compute our numerical results only for the $U^e$ parameter corrections, noting that the accuracy of these corrections is virtually identical with the other two parameterizations.
It is worth emphasizing that, although our discussion of accuracy centers around the SNSI results in Sec.~\ref{sec:SNSImain}, such corrections are not specific to these single-element SNSI formulas. The corrections may be expressed in terms of the general 0th order effective parameters, which could either be taken from Sec.~\ref{sec:generalizing}, from Sec.~\ref{sec:SNSImain} or from other conceivable scenario where $\ACC$ has been neglected.

\begin{figure}[]
    \begin{center}
    \includegraphics[width=1\linewidth]{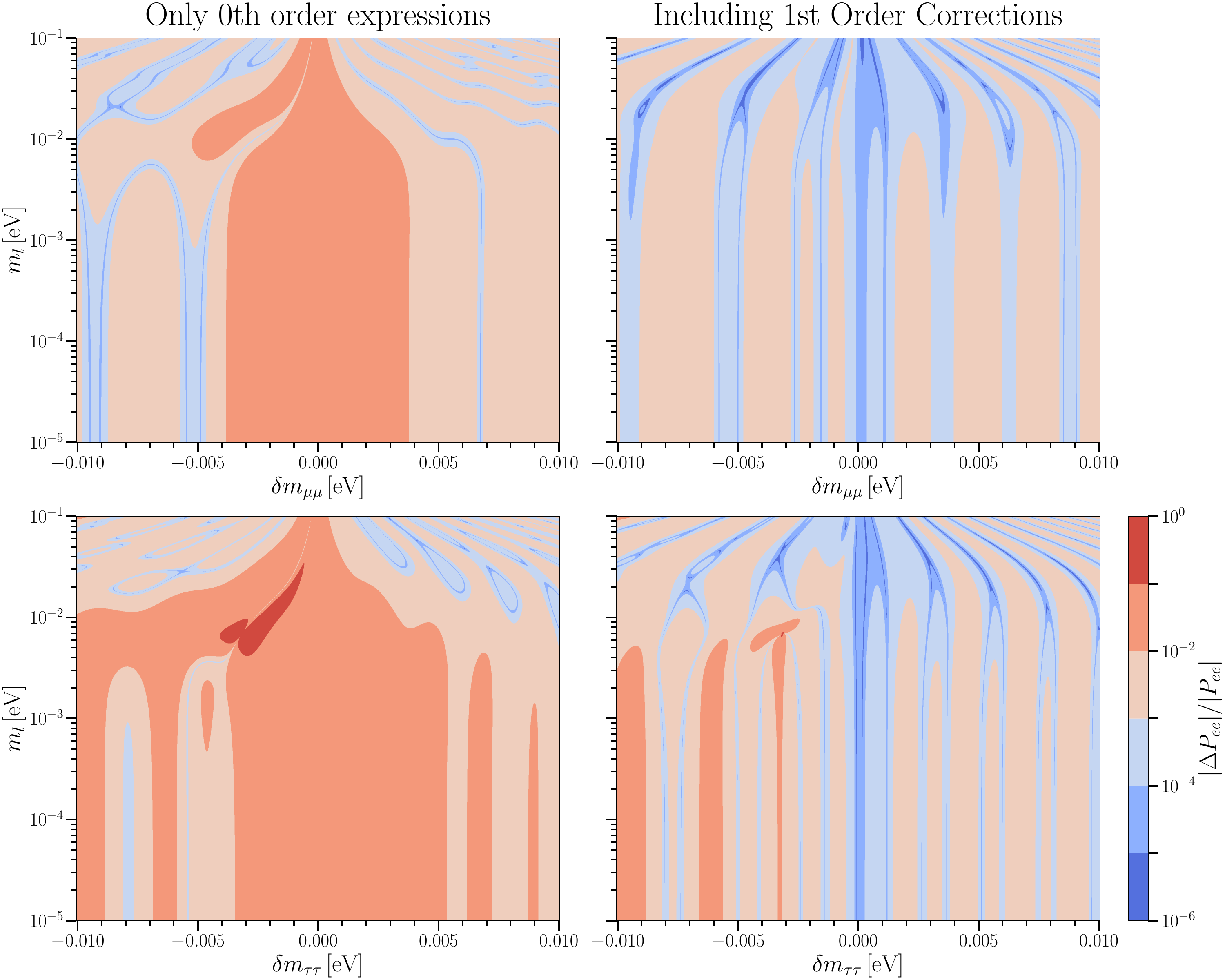}
    \caption{
    Absolute error of the resulting formulas from Sec.~\ref{sec:SNSImain} as a function of $\dmmm$ or $\dmtt$, and $m_l$. The left panels are computed using only the 0th order expressions, whereas the right panels also add the first order corrections from equations \eqref{eq:1stordert12Ue}--\eqref{eq:1storderdm31Ue}. We have assumed a baseline of $L=52.5\,$km and an energy of $E=3\,$MeV.}
    \label{fig:eta-mlsweepmmtt0th1st}
    \end{center}
\end{figure}

In Fig.~\ref{fig:E-LsweepeeNOvsIO1st} we show the relative error of the oscillation probability, as defined in Eq.~\eqref{eq:Prel}, for the cases where the first order corrections are either omitted or have been included when computing the effective parameters. The error is shown as a function of neutrino energy, $E$, and propagation baseline length, $L$. We observe that a large region of energies around $E\sim[0.1,10]\,$MeV has become significantly more accurate with the 1st order corrections, although at the same time the accuracy instead decreases for $E\gtrsim\,1$GeV . This deterioration of the accuracy corresponds to when $\ACC>\Delta m_{21}^{2,\rm{eff}}$, for which our perturbative approach here breaks down, in particular for $(\theta_{12}^{\rm{eff}})^{(1)}\propto \ACC/\Delta m_{21}^{2,\rm{eff}}$ as seen in Eq.~\eqref{eq:1stordert12Ue}. We conclude that these corrections are primarily applicable to the case of JUNO where the accuracy improves by an order of magnitude, thus achieving a sub $1\%$ analytical accuracy that is particularly important due to the level of precision required in JUNO \cite{JUNO:2022mxj}.

In Fig. \ref{fig:eta-mlsweepmmtt0th1st} examine the analytical accuracy of JUNO with and without the 1st order corrections. We explicitly show how the relative error of the effective parameter mappings in Sec.~\ref{sec:SNSI_mm} and Sec.~\ref{sec:SNSI_tt} vary as a function of $m_l$ and $\dmmm$ or $\dmtt$. In all cases for a baseline of $L=52.5\,$km and an energy of $E=3\,$MeV. Importantly, we observe that the accuracy of the formula is rather poor when one only accounts for the 0th order expression, whereas the 1st order corrected formulas mostly remain well below $1\%$ error for the considered ranges of SNSI parameter values with the exception of negative values $\dmtt$ with low values $m_l$. The displayed results here assume normal ordering, which yields worse accuracy in all cases as compared to inverted ordering, which was previously discussed for Fig. \ref{fig:E-LsweepeeNOvsIO1st}. In this way, the results shown are conservative estimates.



\section{Discussion and Conclusion} \label{sec:discussionandconclusion}

Analytic expressions for neutrino oscillation probabilities help in getting a deeper understanding of the physics from the data of neutrino experiments. These calculations can become detrimentally complicated in a three-generation framework with substantial matter effects. In this work, we used the Jacobi diagonalization method to obtain the analytic formulas for the effective parameter mappings of neutrino mixing angles and mass squared differences the presence of matter effects. These effective oscillation parameters can be directly plugged into the standard formulas for neutrino oscillation probabilities. Neutrino propagation in matter can produce effective potentials both via standard and non-standard interactions. While standard interactions bring matter effects that show up as a potential term in the $e$--$e$ flavor element, the NSI coupling can result in additional contributions in all elements of the neutrino Hamiltonian matrix. We worked with matter effects appearing along each of the diagonal $e$--$e$, $\mu$--$\mu$ and $\tau$--$\tau$ elements of the neutrino Hamiltonian matrix, considering a single element at a time, and obtained the effective neutrino mass and mixing parameters. We performed these calculations for both VNSI-like and SNSI-like contributions. 

A major highlight in this work was the choice of the parametrization for the neutrino mixing matrix. We argued that, although the mixing matrix is parameterized by three rotation angles and a Dirac phase, the specific parametrization itself is merely a matter of choice. We observed that the standard PDG parameterization, $U^e(\theta_{12},\theta_{13}, \theta_{23}, \dCP) = R_{23} U_\delta R_{13} U_\delta^\dagger R_{12}$, works perfectly well in the Jacobi diagonalization method for the standard matter effects because the matter potential term comes in the $e$--$e$ element. However, the PDG choice for the mixing matrix would not work out elegantly in the Jacobi diagonalization method if the matter potential term came in the $\mu$--$\mu$ or $\tau$--$\tau$ elements of the neutrino Hamiltonian matrix. We discussed the choice of the parametrization of the mixing matrix for such matter potentials, and concluded that it should be governed by two major considerations. The first consideration came from the symmetry of the matter potential matrix. We noted that the matter potential matrix with term in the $e$--$e$ element is symmetric with respect to rotations with respect to the x-axis ($R_{23}$), while those with terms in the $\mu$--$\mu$ or $\tau$--$\tau$ elements are symmetric under rotations with respect to the y-axis ($R_{13}$) and z-axis ($\tilde R_{12}$), respectively. The rotation matrix about which we have a symmetry is always put left-most in the parametrization of the neutrino mixing matrix. 
The second consideration comes from the fact that, in the Jacobi method, one small Hamiltonian element needs to be approximated to zero. For the method to work, this term should depend only on the smaller mass squared difference $\Delta m_{21}^2$ so that it can be neglected in comparison to the other terms that depend on $\Delta m_{31}^2$. To ensure this, we must always have the rotation matrix $R_{12}$ right-most in the parametrization of the mixing matrix. Therefore, we concluded that when matter term appears in the $\mu$--$\mu$ element, the appropriate choice for the neutrino mixing matrix is $U^{\mu}(\theta_{13}^\mu, \theta_{23}^\mu,\theta_{12}^\mu, \delta_{\text{CP}}^\mu)=R_{13} U_\delta R_{23} {U_\delta}^\dagger R_{12}$ while when the matter term comes in the $\tau$--$\tau$ element it is $U^{\tau}(\tilde \theta_{12}^\tau,\theta_{23}^\tau, \theta_{12}^\tau, \dCP^\tau) = \tilde R_{12} R_{23} U_\delta^\tau R_{12}$. Since the magnitude of the mixing angles depend on the parameterization chosen, we put superscripts $\mu$ and $\tau$ for the proposed parameterizations of $U^{\mu}$ and $U^{\tau}$ to distinguish it from the parametrization of the PDG parameterization. One can naturally relate the parameters of the mixing matrix in the $U^\mu$ and $U^\tau$ parameterizations to the conventional $U^e$ parameterization, as has been outlined in App.~\ref{app:parameterization}. 

What we achieve by invoking these tailor-made parameterizations of the neutrino mixing matrix is that we can take matter effects fully into account in our analytical calculations without any assumptions for them to be small. Therefore, the results we obtain for the effective mixing angles in matter are valid for large enhancement of neutrino oscillations due to matter effects. In particular, we can study the resonances of neutrino oscillations in the full three-generation formalism without having to assume two generations. We showed that, not only can the standard MSW resonances be studied in the full three-generation picture in our formalism, but also NSI-induced MSW resonances. For the first time, we show the resonance conditions of a scalar induced resonant enhancement of the neutrino mixing angles, i.e. the SMSW effect, in the presence of every diagonal scalar NSI coupling. The relations we obtain between the effective mixing parameters in presence of matter with the intrinsic mixing parameters and the matter potential were presented for six cases in total. 
\begin{enumerate}
    \item $\ACC$ in the $e$--$e$ element: This is the standard matter effects coming from $W$-mediated scattering of the neutrinos. We obtained the resonance condition for both $\theta_{12}$ and $\theta_{13}$ in the three-generation picture. These results already exists in the literature \cite{Agarwalla:2013tza,Ioannisian:2018qwl}. 
    \item $A_{\mu\mu}$ in the $\mu$--$\mu$ element: This is a toy case when a vector-like NSI is taken with contribution in the $\mu$--$\mu$ element, but no contribution of the standard matter effects is taken in the $e$--$e$ element. We obtained MSW-like resonance conditions for the mixing angle $\theta_{12}^\mu$ and $\theta_{23}^\mu$ in the three-generation picture.
    \item $A_{\tau\tau}$ in the $\tau$--$\tau$ element: This is a toy case when a vector-like NSI is taken with contribution in the $\tau$--$\tau$ element, but no contribution of the standard matter effects is taken in the $e$--$e$ element. We obtained MSW-like resonance conditions for the mixing angle $\theta_{12}^\tau$ and $\theta_{23}^\tau$ in the three-generation picture.
    \item $\dmee$ in the $e$--$e$ element: This is the case for a scalar-like NSI coupling appearing in the $e$--$e$ element. We obtained the resonance condition for both $\theta_{12}$ and $\theta_{13}$ in the three-generation picture. We also included the standard $W$-mediated matter term in the calculations, revealing the interplay between the MSW and SMSW effect.
    \item $\delta m_{\mu\mu}$ in the $\mu$--$\mu$ element: This is the case for scalar-like NSI coupling appearing in the $\mu$--$\mu$ element. We obtained SMSW-like resonance conditions for both $\theta_{12}^\mu$ and $\theta_{23}^\mu$ in the three-generation picture.
    \item $\delta m_{\tau\tau}$ in the $\tau$--$\tau$ element: This is the case for a scalar-like NSI coupling appearing in the $\tau$--$\tau$ element. We obtained SMSW-like resonance conditions for both $\theta_{12}^\tau$ and $\theta_{23}^\tau$ in the three-generation picture.
\end{enumerate}
The neutrino oscillation probabilities can be simply obtained in our method by replacing the oscillation parameters by the effective parameters in the exact three-generation oscillation probabilities. 

We checked the robustness of our analytical results by comparing the results for the case of SNSI with the exact numerical probabilities. We showed these comparisons for the on-going JUNO experiment, and the planned Hyper-Kamiokande, ESSnuSB and DUNE experiments. We found that the relative error was at the 0.5\% (5\%) level for JUNO, the 0.1\% (10\%) level for Hyper-Kamiokande and ESSnuSB, and the 1\% (100\%) level for DUNE when we consider $\dmee$ ($\dmmm$ or $\dmtt$). We noted that the matching for $\dmee$ was very good for both JUNO (low $\ACC$) and DUNE (high $\ACC$). Indeed, in this case the analytical formulas are nearly exact barring the throwing away of extremely small terms of order $O((\epsilon_e^S)^2)$, as we discussed in section Sec.~\ref{sec:SNSI_ee}. However, for the cases of $\delta m_{\mu\mu}$ and $\delta m_{\tau\tau}$, we could not include the matter effect from the standard matter term $\ACC$ since our method of Jacobi diagonalization can only be applied when there is exactly one dominant element in the matter potential matrix. Therefore, we found that for these cases, the discrepancy between our analytical formulas and the exact numerical probability was significantly higher. We could still account for $\ACC$ in our formulas for $\delta m_{\mu\mu}$ and $\delta m_{\tau\tau}$ via perturbation theory. This was done firstly by applying first order corrections applied directly to the effective parameter mappings, as shown in App~\ref{app:perturbationtheory}, and secondly by combining our effective parameter mappings with perturbatively corrected probability formulas, here taken from Ref.~\cite{Martinez-Soler:2019nhb}. 
The first approach was shown to improve the accuracy of the formulas for $\delta m_{\mu\mu}$ and $\delta m_{\tau\tau}$ in the JUNO experiment such that the difference with the numerical results were reduced to the level of only 0.1\%. The second approach was shown to improve the accuracy of these formulas for ESSnuSB and Hyper-Kamiokande down to the level of $1\%$, whereas for DUNE they remained highly inaccurate in these cases.

Regarding phenomenological application, in this work we have primarily focused on scalar non-standard interactions, so let us also briefly comment on the application of this method for vector non-standard interactions, which has previously been studied in Ref.~\cite{Ge:2016dlx}.
For $\varepsilon_{ee}$ one can trivially extend the formula derived in Sec.~\ref{sec:standardderivation}. However, the difficulty with treating any other $\varepsilon_{\alpha \beta}$ with our generalized method, is that they have the same energy dependence as $\ACC$ and so must always be taken together. This is unlike the case of SNSI where one can move to the energy regime where $\ACC$ is subdominant (e.g. in JUNO). Thus, one will have two non-zero elements from the effective potential matrix, and the method developed in this article cannot be applied elegantly. If one considers a $\varepsilon_{\mu \mu}$ or $\varepsilon_{\tau \tau}$ dominant vector NSI contribution, the $U^\mu$ and $U^\tau$ parameterizations shown here could be useful. The complication is that vector NSI is clearly constrained to be subdominant \cite{Ohlsson:2012kf, Coloma:2023ixt}, and so these parameters should be considered only as perturbative corrections and can clearly not be taken as larger than $\ACC$. 

In conclusion, we have reviewed and extended the Jacobi diagonalization method for diagonalizing the neutrino propagation Hamiltonian. We have particularly shown a way to generalize the method by choosing specific parameterizations of the PMNS mixing matrix. The parameterization is deliberately chosen based on the fact that $\Delta m_{21}^2 \ll \Delta m_{31}^2$ and that single element potentials carry certain symmetries. This enables exploiting the Jacobi method, previously only used to derive formulas for the standard matter density $V\propto \diag(1,0,0)$, to elegantly diagonalize the neutrino Hamiltonian with matter potentials of the form $V\propto \diag(0,1,0)$ and $V\propto \diag(0,0,1)$.
Interestingly, these three single element matter potential cases all generate parametric resonant enhancement for different parameters and different signs of the matter potential.
Furthermore, we have shown that the phenomenology of diagonal couplings arising from scalar non-standard interactions can be understood with this same method and generalization. We used our generalized method to derive analogous results for SNSI and show that they also reveal resonant enhancement of the mixing angles, in other words a scalar MSW effect. This effect distinguishes itself from the standard MSW effect in that it may induce resonances that are independent of energy and can thus only be distinguished by studying neutrino oscillations in environments of different matter densities. We proceeded by studying the numerical accuracy of the SNSI formulas. From this we concluded that the formulas for $\delta M=\diag(\dmee,0,0)$ is remarkably accurate in all cases, whereas the $\delta M=\diag(0,\dmmm,0)$ and $\delta M=\diag(0,0,\dmtt)$ formulas are lacking in accuracy for relevant baselines and energies. This difference in accuracy is due to the fact that the latter formulas assume a subdominant $\ACC$. We demonstrated how one may compensate for this lack in accuracy with perturbative corrections, which were shown to significantly improve the accuracy in certain regimes.

\acknowledgments

We would like to thank Tommy Ohlsson for a useful discussion during the early stage of this work. This work is supported by the Swedish Research Council (Vetenskapsrådet) through grant 2023-05141. 



\newpage 
\appendix
\section{Mixing Matrix Parameterization} \label{app:parameterization}
\begin{table}[h]
    \centering
    \caption{The twelve possible decompositions of three-dimensional rotations into Euler rotation matrices. We indicate the decompositions used to parameterize the PMNS matrix in this work.}
    \begin{tabular}{|c|c|} 
        \hline
         Proper Euler angles & Tait-Bryan angles  \\
         \hline
         $\tilde R_{23} R_{12}  R_{23}$ & $R_{23} R_{12} R_{13}$ \\
         \hline
         $\tilde  R_{23} R_{13} R_{23}$ & $R_{23} R_{13} R_{12} \sim U^e$ \\
         \hline
         $\tilde  R_{13} R_{23} R_{13}$ & $R_{13} R_{23} R_{12} \sim U^\mu$ \\
         \hline
         $ \tilde R_{13} R_{12} R_{13}$ & $R_{13} R_{12} R_{23}$ \\
         \hline
         $ \tilde R_{12} R_{13} R_{12} (\sim U^\tau)$ & $R_{12} R_{13} R_{23} (\sim {U^e}^\dagger)$ \\
         \hline
         $\tilde R_{12} R_{23}  R_{12} \sim U^\tau$ & $R_{12} R_{23} R_{13} (\sim {U^\mu}^\dagger)$ \\
         \hline
    \end{tabular}
    \label{tab:parameterizations}
\end{table}
\noindent In Tab.~\ref{tab:parameterizations}, we list the 12 parameterizations of three-dimensional rotations in terms of Euler rotation matrices. The CP violating phase is omitted from this table since its assignment is not unique (for further details see Ref.~\cite{Fritzsch:1999ee}). We remark that not all of the 12 listed options are structurally different due to two distinct symmetries. First, there is an equivalence between $R_{ij} R_{kl} R_{ik}$ and $R_{ij} R_{mn} R_{kl}$ (with $ij\neq kl \neq mn$) by redefining certain angles by $\pi/2$ rotations \cite{Fritzsch:1997st}. Second, noting that the only physical observable in neutrino oscillations is the quartic products, $U_{\alpha i} U_{\beta _i}^* U_{\alpha j}^* U_{\beta j}$, it is possible to redefine the mixing such that $U \mapsto U^\dagger, \, \dCP \mapsto - \dCP$. The method used in this article is still applicable under this transformation since $U U^\dagger = U^\dagger U$. As such, $R_{ij} R_{kl} R_{mn}$ and $R_{mn} R_{kl} R_{ij}$ (with $ij\neq kl \neq mn$) are structurally equivalent as far as this method is concerned. There remains 6 structurally distinct parameterizations. Three of these are utilized in this work and are shown explicitly below:
\begin{align}
    U^e = &R_{23} U_\delta R_{13} {U_\delta}^\dagger R_{12} =
\begin{pmatrix}
c_{12}c_{13} & s_{12} c_{13} & s_{13} e^{-i \delta }\\
-s_{12} c_{23}-c_{12}s_{13}s_{23} e^{i \delta} & c_{12}c_{23}-s_{12}s_{13}s_{23} e^{i\delta}  & c_{13}s_{23}\\
s_{12}s_{23}-c_{12}s_{13}c_{23}e^{i \delta} & -c_{12}s_{23}-s_{12}s_{13}c_{23}e^{i \delta} & c_{13}c_{23}
\end{pmatrix} \\
U^\mu = &R_{13} U_{\delta} R_{23} {U_\delta}^\dagger R_{12} =
\begin{pmatrix}
c_{12}c_{13}+s_{12}s_{23}s_{13}e^{i \delta} & s_{12} c_{13}-c_{12}s_{23} s_{13}e^{i \delta} & c_{23} s_{13}\\
-s_{12} c_{23}  & c_{12} c_{23}  & s_{23} e^{-i \delta}\\
-c_{12} s_{13}+s_{12} s_{23} c_{13}e^{i \delta} & -s_{12}s_{13}-c_{12}s_{23} c_{13}e^{i \delta} & c_{23} c_{13}
\end{pmatrix} \\
U^\tau =& \tilde R_{12} R_{23} U_\delta^\tau R_{12} =
\begin{pmatrix} 
c_{12}\tilde c_{12}e^{i \delta}-s_{12}c_{23}\tilde s_{12} & s_{12}\tilde c_{12} e^{i \delta}+c_{12}c_{23}\tilde s_{12} & s_{23}\tilde s_{12}\\
-c_{12}\tilde s_{12}e^{i \delta}-s_{12}c_{23}\tilde c_{12} & -s_{12}\tilde s_{12}e^{i \delta}+c_{12}c_{23}\tilde c_{12} & s_{23}\tilde c_{12}\\
s_{12}s_{23} & -c_{12}s_{23} & c_{23}
\end{pmatrix}
\end{align}
for which we defined
\begin{align}
    &R_{12} = \begin{pmatrix}
        c_{12} & s_{12} & 0 \\
        -s_{12} & c_{12} & 0 \\
        0 & 0 & 1 \\
    \end{pmatrix}, \quad 
    R_{13} = \begin{pmatrix}
        c_{13} & 0 & s_{13} \\
        0 & 1 & 0 \\
        -s_{13} & 0 & c_{13} \\
    \end{pmatrix}, \quad 
    R_{23} = \begin{pmatrix}
        1 & 0 & 0 \\
        0 & c_{23} & s_{23} \\
        0 & -s_{23} & c_{23} \\
    \end{pmatrix}, \nonumber \\
    & 
    \tilde R_{12} = \begin{pmatrix}
        \tilde c_{12} & \tilde s_{12} & 0 \\
        - \tilde s_{12} & \tilde c_{12} & 0 \\
        0 & 0 & 1 \\
    \end{pmatrix},\quad U_\delta^\tau = \begin{pmatrix}
        e^{i \dCP} & 0 & 0 \\
        0 & 1 & 0 \\
        0 & 0 & 1 \\
    \end{pmatrix}
    , \quad U_\delta = \begin{pmatrix}
        1 & 0 & 0 \\
        0 & 1 & 0 \\
        0 & 0 & e^{i \dCP} \\
    \end{pmatrix}
\end{align}
where $c_{ij}\equiv\cos\theta_{ij}$,$s_{ij}\equiv \sin\theta_{ij}$ and when two distinct angles rotate in the same plane we use the notation $\tilde c_{ij}\equiv\cos\tilde \theta_{ij}$,$\tilde s_{ij}\equiv \sin\tilde \theta_{ij}$.

\subsection{Conversion formulas} \label{app:parameterizationconversion}

This work relies on different parameterizations to derive different results, and so it is worth reviewing how one can convert these formulas from one parameterization to another. Such conversion formula follow from the invariance of the absolute values of the PMNS matrix elements, $|U_{\alpha i}|$, and the invariance of quartics of the form $U_{\alpha i}U_{\alpha j}^*U_{\beta i}^*U_{\beta j}$.
To arrive at the simplest relations, one may utilize the matrix element combinations that have simple forms in both of the two parameterizations being converted between.
Specifically, with the $U^e$ parameterization we have $\sin\theta_{13}= |U_{e3}|$ and $\tan \theta_{23}=\frac{|U_{\mu3}|}{|U_{\tau 3}|}$. The $\theta_{12}$ sector conversions are more complicated since one cannot utilize the simplest rows/columns in both parameterizations simultaneously. Some options for extracting this angle are $\tan^2(\theta_{12}) = \frac{|U_{e2}|^2}{|U_{e1}|^2}$, $ \sin^2  \theta_{12} = \frac{|U_{e2}|^2}{|U_{e1}|^2+|U_{e2}|^2}$ or $\cos 2 \theta_{12} = \frac{|U_{e1}|^2-|U_{e2}|^2}{|U_{e1}|^2+|U_{e2}|^2}$. We adopt the $\cos 2 \theta_{12}$ type relations since they yield relatively simple expressions.
Finally, to extract CP violating phases we use the invariance of the imaginary part of the quartic product $J=\Im{(U_{e2}U_{e3}^*U_{\mu2}^*U_{\mu3})}\propto \sin\delta$. This does not fix the quadrant of $\dCP$ however. As such, one would need to find $\cos \dCP$ to uniquely determine $\dCP$. The cosine can either be extracted as the real part of the same quadrant, $K=\Re{(U_{e2}U_{e3}^*U_{\mu2}^*U_{\mu3})}$, or by using the invariance of $|U_{\mu 2}|^2$, the latter of which is simpler and will be utilized here.

The extraction of mixing angles from the $U^\mu$ and $U^\tau$ parameterizations follows similarly, except that different matrix elements are chosen based on which yield the simplest relations. The resulting conversion formulas in each case is shown below.

\paragraph{ \underline{$U^{e} \rightarrow U^\mu$}}
\begin{align}
    &\tan\theta_{13}^{\mu} = { \tan \theta_{13} \over c_{23}} \\ 
    &\sin\theta_{23}^{\mu} = {c_{13} \sin \theta_{23} } \\ 
      & \cos 2\theta_{12}^\mu
      = \frac{\cos 2\theta_{12}\,\big(c_{23}^2 - s_{13}^2 s_{23}^2\big)
       - \sin 2\theta_{12}\, s_{13} \sin 2\theta_{23} \cos\delta}{c_{23}^2 + s_{13}^2 s_{23}^2} \\
    &\left\{
    \begin{aligned}
        &\sin \delta^{\mu}_{CP} = -\sin \dCP { s_{12} c_{12} s_{13} c_{13}^2 s_{23} c_{23} \over s_{12}^{\mu} c_{12}^{\mu}  s_{23}^{\mu} (c_{23}^{\mu})^2 s_{13}^{\mu} c_{13}^{\mu}} \\  &\cos \dCP^{\mu}
    = \frac{ c_{12}^2 c_{13}^2 - (c_{12}^\mu)^2 (c_{13}^\mu)^2
              - (s_{12}^\mu)^2 (s_{23}^\mu)^2 (s_{13}^\mu)^2 }
            { 2 s_{12}^\mu c_{12}^\mu s_{23}^\mu s_{13}^\mu c_{13}^\mu }
    \end{aligned} \right.
\end{align}
\raggedbottom

\paragraph{ \underline{$U^\mu \rightarrow U^e$}}

\begin{align}
    &\sin\theta_{13} = c_{23}^\mu  \sin \theta_{13}^\mu \\ 
&\tan\theta_{23} = { \tan \theta_{23}^\mu \over c_{13}^\mu} \\
&\cos 2\theta_{12}
      = \frac{\cos 2\theta_{12}^\mu
         \left(\left(c_{13}^\mu\right)^2 - (s_{23}^\mu)^2 (s_{13}^\mu)^2\right)
       + \sin 2\theta_{12}^\mu\, s_{23}^\mu
         \sin 2\theta_{13}^\mu \cos\dCP^\mu}{(c_{13}^\mu)^2 + (s_{23}^\mu)^2  (s_{13}^\mu)^2} \\
&\left\{
    \begin{aligned}
&\sin \dCP = -\sin \dCP^\mu { s_{12}^\mu c_{12}^\mu s_{23}^\mu (c_{23}^\mu)^2 s_{13}^\mu c_{13}^\mu  \over s_{12} c_{12} s_{13} c_{13}^2 s_{23} c_{23} } \\
&\cos \dCP
    = \frac{ (s_{12}^\mu)^2 (c_{23}^\mu)^2 - s_{12}^2 c_{23}^2
              - c_{12}^2 s_{13}^2 s_{23}^2 }
            { 2 s_{12} c_{12} s_{13} s_{23} c_{23} }
\end{aligned}\right.
\end{align}
\newline

\paragraph{ \underline{$U^e \rightarrow U^\tau$}}
\begin{align}
    &\cos\theta_{23}^{\tau} = c_{13} \cos \theta_{23} \\
    &\tan\tilde{\theta}_{12}^{\tau} = { \tan \theta_{13} \over s_{23} } \\
    &\cos 2\theta_{12}^\tau
      = \frac{\cos 2\theta_{12}\,\big(s_{23}^2 - s_{13}^2 c_{23}^2\big)
       + \sin 2\theta_{12}\, s_{13} \sin 2\theta_{23} \cos\dCP}{s_{23}^2 + s_{13}^2 c_{23}^2} \\
    &\left\{
    \begin{aligned}
    &\sin\dCP^{\tau} = \sin\dCP { c_{12} s_{12} c_{13}^2 s_{13} c_{23} s_{23} \over s_{12}^{\tau} c_{12}^{\tau}  (s_{23}^{\tau})^2 c_{23}^{\tau} \tilde{s}_{12}^{\tau} \tilde{c}_{12}^{\tau}} \\
    &\cos \dCP^{\tau}
    = \frac{ (c_{12}^\tau)^2 (\tilde c_{12}^\tau)^2
              + (s_{12}^\tau)^2 (c_{23}^\tau)^2 (\tilde s_{12}^\tau)^2 - c_{12}^2 c_{13}^2 }
            { 2 s_{12}^\tau c_{12}^\tau c_{23}^\tau
              \tilde s_{12}^\tau \tilde c_{12}^\tau }
    \end{aligned}\right.
\end{align}

\paragraph{ \underline{$U^\tau \rightarrow U^e$}}
\begin{align}
    &\sin\theta_{13} = \sin \theta_{23}^\tau \tilde{s}_{12}^\tau \\
    &\tan \theta_{23} = \tan \theta_{23}^\tau \tilde{c}_{12}^\tau \\
    &\cos 2\theta_{12}
      = \frac{\cos 2\theta_{12}^\tau
         \left( (\tilde c_{12}^\tau)^2 - \left(c_{23}^\tau\right)^2 (\tilde s_{12}^\tau)^2 \right)
       - \sin 2\theta_{12}^\tau\, c_{23}^\tau
         \sin 2\tilde\theta_{12}^\tau \cos\dCP^\tau}{(\tilde c_{12}^\tau)^2 + (c_{23}^\tau)^2 (\tilde s_{12}^\tau)^2} \\
    &\left\{
    \begin{aligned}
    &\sin\dCP = \sin\dCP^\tau { c_{12}^\tau s_{12}^\tau (s_{23}^\tau)^2 c_{23}^\tau \tilde{c}_{12}^\tau \tilde{s}_{12}^\tau  \over s_{12} c_{12} c_{13}^2 s_{13} c_{23} s_{23} } \\
    &\cos \dCP = \frac{ s_{12}^2 s_{23}^2 + c_{12}^2 s_{13}^2 c_{23}^2
              - (s_{12}^\tau)^2 (s_{23}^\tau)^2 }
            { 2 s_{12} c_{12} s_{13} s_{23} c_{23} }
    \end{aligned}\right.
\end{align}
\raggedbottom

\paragraph{ \underline{$U^\mu \rightarrow U^\tau$}}
\begin{align}
    &\cos\theta_{23}^{\tau} = c_{13}^\mu c_{23}^\mu \\ 
    &\tan\tilde\theta_{12}^{\tau} = s_{13}^\mu \cot\theta_{23}^\mu \\ 
    &\cos 2\theta_{12}^\tau
      = \frac{\cos 2\theta_{12}^\mu
         \left( (c_{13}^\mu)^2 (s_{23}^\mu)^2 - (s_{13}^\mu)^2 \right)
       + \sin 2\theta_{12}^\mu\, s_{23}^\mu
         \sin 2\theta_{13}^\mu \cos\dCP^\mu}{(c_{13}^\mu)^2 (s_{23}^\mu)^2 + (s_{13}^\mu)^2} \\
    &\left\{
    \begin{aligned}
    &\sin\dCP^{\tau} = -\sin\dCP^\mu
      { s_{12}^\mu c_{12}^\mu s_{13}^\mu c_{13}^\mu s_{23}^\mu (c_{23}^\mu)^2
        \over s_{12}^{\tau} c_{12}^{\tau} (s_{23}^{\tau})^2 c_{23}^{\tau}
              \tilde s_{12}^{\tau} \tilde c_{12}^{\tau}} \\
    &\cos \dCP^{\tau}
      = \frac{ (s_{12}^\mu)^2 (c_{23}^\mu)^2
               - (c_{12}^\tau)^2 (\tilde s_{12}^\tau)^2
               - (s_{12}^\tau)^2 (c_{23}^\tau)^2 (\tilde c_{12}^\tau)^2 }
             { 2 s_{12}^\tau c_{12}^\tau c_{23}^\tau
               \tilde s_{12}^\tau \tilde c_{12}^\tau }
    \end{aligned}\right.
\end{align}

\paragraph{ \underline{$U^\tau \rightarrow U^\mu$}}
\begin{align}
    &\sin\theta_{23}^{\mu} = \tilde c_{12}^\tau \sin\theta_{23}^\tau \\ 
    &\tan\theta_{13}^{\mu} = \tilde s_{12}^\tau \tan\theta_{23}^\tau \\ 
    &\cos 2\theta_{12}^\mu
      = \frac{\cos 2\theta_{12}^\tau
         \left( (\tilde c_{12}^\tau)^2 (c_{23}^\tau)^2 - (\tilde s_{12}^\tau)^2 \right)
       - \sin 2\theta_{12}^\tau\, c_{23}^\tau
         \sin 2\tilde\theta_{12}^\tau \cos\dCP^\tau}
        {(\tilde c_{12}^\tau)^2 (c_{23}^\tau)^2 + (\tilde s_{12}^\tau)^2} \\
    &\left\{
    \begin{aligned}
    &\sin\dCP^{\mu} = -\sin\dCP^\tau
      { s_{12}^{\tau} c_{12}^{\tau} (s_{23}^{\tau})^2 c_{23}^{\tau}
        \tilde s_{12}^{\tau} \tilde c_{12}^{\tau}
        \over s_{12}^\mu c_{12}^\mu s_{13}^\mu c_{13}^\mu s_{23}^\mu (c_{23}^\mu)^2 } \\
    &\cos \dCP^{\mu}
      = \frac{ (c_{12}^\mu)^2 (s_{13}^\mu)^2
               + (s_{12}^\mu)^2 (s_{23}^\mu)^2 (c_{13}^\mu)^2
               - (s_{12}^\tau)^2 (s_{23}^\tau)^2 }
             { 2 s_{12}^\mu c_{12}^\mu s_{23}^\mu s_{13}^\mu c_{13}^\mu }
    \end{aligned}\right.
\end{align}

\subsection{Approximate Conversion Formulas} \label{app:approximateconversionrelations}
Below are conversion relations series expanded in terms of $\sin \theta_{13}$ or the corresponding small mixing angle in the $U^\mu$ and $U^\tau$ parameterizations. With the central Nufit 6.0 values, these expressions are quite accurate at first order. The reported error values assume normal ordering. These relations may provide an easier short-hand for relating the different parameters used in this work.

\paragraph{\underline{$U^{e}\rightarrow U^{\mu}$}}
\begin{align}
    \theta_{13}^{\mu} &= \frac{\theta_{13}}{\cos\theta_{23}}+O(\theta_{13}^{3})
      && [0.665\%] \\
    \theta_{23}^{\mu} &= \theta_{23}+O(\theta_{13}^{2})
      && [1.408\%] \\
    \theta_{12}^{\mu} &= \theta_{12}+\tan\theta_{23}\,\theta_{13}\cos\dCP
       +O(\theta_{13}^{2})
      && [0.915\%] \\
    \dCP^{\mu} &= -\dCP+2\cot 2\theta_{12}\tan\theta_{23}\,
       \theta_{13}\sin\dCP+O(\theta_{13}^{2})
      && [1.089\%]
\end{align}

\paragraph{\underline{$U^{\mu}\rightarrow U^{e}$}}
\begin{align}
    \theta_{13} &= \theta_{13}^{\mu}\cos\theta_{23}^{\mu}
       +O\big((\theta_{13}^{\mu})^{3}\big)
      && [0.328\%] \\
    \theta_{23} &= \theta_{23}^{\mu}
       +O\big((\theta_{13}^{\mu})^{2}\big)
      && [1.389\%] \\
    \theta_{12} &= \theta_{12}^{\mu}-\sin\theta_{23}^{\mu}\,
       \theta_{13}^{\mu}\cos\dCP^{\mu}
       +O\big((\theta_{13}^{\mu})^{2}\big)
      && [0.823\%] \\
    \dCP &= -\dCP^{\mu}-2\cot 2\theta_{12}^{\mu}\sin\theta_{23}^{\mu}\,
       \theta_{13}^{\mu}\sin\dCP^{\mu}
       +O\big((\theta_{13}^{\mu})^{2}\big)
      && [0.876\%]
\end{align}
\raggedbottom

\paragraph{\underline{$U^{e}\rightarrow U^{\tau}$}}
\begin{align}
    \theta_{23}^{\tau} &= \theta_{23}+O(\theta_{13}^{2})
      && [1.517\%] \\
    \tilde\theta_{12}^{\tau} &= \frac{\theta_{13}}{\sin\theta_{23}}
       +O(\theta_{13}^{3})
      && [0.819\%] \\
    \theta_{12}^{\tau} &= \theta_{12}-\cot\theta_{23}\,\theta_{13}\cos\dCP
       +O(\theta_{13}^{2})
      && [0.055\%] \\
    \dCP^{\tau} &= \dCP+2\cot 2\theta_{12}\cot\theta_{23}\,
       \theta_{13}\sin\dCP+O(\theta_{13}^{2})
      && [0.728\%]
\end{align}

\paragraph{\underline{$U^{\tau}\rightarrow U^{e}$}}
\begin{align}
    \theta_{13} &= \tilde\theta_{12}^{\tau}\sin\theta_{23}^{\tau}
       +O\big((\tilde\theta_{12}^{\tau})^{3}\big)
      && [0.403\%] \\
    \theta_{23} &= \theta_{23}^{\tau}
       +O\big((\tilde\theta_{12}^{\tau})^{2}\big)
      && [1.540\%] \\
    \theta_{12} &= \theta_{12}^{\tau}+\cos\theta_{23}^{\tau}\,
       \tilde\theta_{12}^{\tau}\cos\dCP^{\tau}
       +O\big((\tilde\theta_{12}^{\tau})^{2}\big)
      && [0.051\%] \\
    \dCP &= \dCP^{\tau}-2\cot 2\theta_{12}^{\tau}\cos\theta_{23}^{\tau}\,
       \tilde\theta_{12}^{\tau}\sin\dCP^{\tau}
       +O\big((\tilde\theta_{12}^{\tau})^{2}\big)
      && [0.628\%]
\end{align}

\paragraph{\underline{$U^{\mu}\rightarrow U^{\tau}$}}
\begin{align}
    \theta_{23}^{\tau} &= \theta_{23}^{\mu}
       +O\big((\theta_{13}^{\mu})^{2}\big)
      && [2.885\%] \\
    \tilde\theta_{12}^{\tau} &= \cot\theta_{23}^{\mu}\,\theta_{13}^{\mu}
       +O\big((\theta_{13}^{\mu})^{3}\big)
      && [2.292\%] \\
    \theta_{12}^{\tau} &= \theta_{12}^{\mu}
       -\frac{\theta_{13}^{\mu}\cos\dCP^{\mu}}{\sin\theta_{23}^{\mu}}
       +O\big((\theta_{13}^{\mu})^{2}\big)
      && [1.373\%] \\
    \dCP^{\tau} &= -\dCP^{\mu}
       -\frac{2\cot 2\theta_{12}^{\mu}\,\theta_{13}^{\mu}\sin\dCP^{\mu}}
             {\sin\theta_{23}^{\mu}}
       +O\big((\theta_{13}^{\mu})^{2}\big)
      && [3.593\%]
\end{align}

\paragraph{\underline{$U^{\tau}\rightarrow U^{\mu}$}}
\begin{align}
    \theta_{23}^{\mu} &= \theta_{23}^{\tau}
       +O\big((\tilde\theta_{12}^{\tau})^{2}\big)
      && [2.970\%] \\
    \theta_{13}^{\mu} &= \tan\theta_{23}^{\tau}\,\tilde\theta_{12}^{\tau}
       +O\big((\tilde\theta_{12}^{\tau})^{3}\big)
      && [2.212\%] \\
    \theta_{12}^{\mu} &= \theta_{12}^{\tau}
       +\frac{\tilde\theta_{12}^{\tau}\cos\dCP^{\tau}}{\cos\theta_{23}^{\tau}}
       +O\big((\tilde\theta_{12}^{\tau})^{2}\big)
      && [2.371\%] \\
    \dCP^{\mu} &= -\dCP^{\tau}
       +\frac{2\cot 2\theta_{12}^{\tau}\,\tilde\theta_{12}^{\tau}\sin\dCP^{\tau}}
             {\cos\theta_{23}^{\tau}}
       +O\big((\tilde\theta_{12}^{\tau})^{2}\big)
      && [1.738\%]
\end{align}

\section{Results Converted to the PDG Parameters} \label{app:SNSIinUe}
The PDG parameterization, referred to in this work as $U^{e}$, has become the convention for discussing neutrino oscillation phenomenology. Although this work demonstrates the concrete benefits of not being restricted to a single convention when different non-standard scenarios are under consideration, it has not escaped us that many readers still would want to see the resulting formulas of this work expressed in the PDG parameterization. As such, we proceed in showing the main results of Sec.~\ref{sec:SNSI_mm} and \ref{sec:SNSI_tt} expressed instead in the $U^{e}$ parameters. The below derivation holds for both the $\dmmm$ and the $\dmtt$ case, which reduce to a single set of relations where we employ the superscript $\alpha=\mu,\tau$ for the corresponding parameterizations, $U^{\alpha}$. Note that these converted formula are more obfuscated than their original form, yet the overarching phenomenology of the SMSW effect can still be discerned.
We start by introducing
\begin{align}
  &\mathcal{S}_{\mu}=1-c_{13}^{2}s_{23}^{2},\qquad
   \mathcal{F}_{\mu}=c_{23}^{2}-s_{13}^{2}s_{23}^{2},\qquad
   \mathcal{C}_{\mu}=c_{13}c_{23},\qquad \eta_{\mu}=+1, \label{eq:geommu}\\
  &\mathcal{S}_{\tau}=1-c_{13}^{2}c_{23}^{2},\qquad
   \mathcal{F}_{\tau}=s_{23}^{2}-s_{13}^{2}c_{23}^{2},\qquad
   \mathcal{C}_{\tau}=c_{13}s_{23},\qquad \eta_{\tau}=-1, \label{eq:geomtau}
\end{align}
together with the quantities common to both cases, $\hat Y\equiv s_{13}\sin 2\theta_{23}$, and
$\kappa_{\alpha}\equiv 2\mathcal{S}_{\alpha}-s_{13}^{2}$. These satisfy
\begin{equation}
  \mathcal{S}_{\alpha}^{2}=\mathcal{F}_{\alpha}^{2}+\hat Y^{2},
  \qquad
  \mathcal{C}_{\alpha}^{2}=\mathcal{S}_{\alpha}-s_{13}^{2}.
  \label{eq:geomident}
\end{equation}
Geometrically $\mathcal{S}_{\mu}=(c_{23}^{\mu})^{2}$ and $\mathcal{S}_{\tau}=(s_{23}^{\tau})^{2}$. Furthermore, if we define
\begin{align}
  \mathcal{P}_{\alpha} &\equiv \mathcal{F}_{\alpha}\cos 2\theta_{12}
      -\eta_{\alpha}\hat Y\cos\dCP\,\sin 2\theta_{12}, \\
  \mathcal{Q}_{\alpha} &\equiv \mathcal{F}_{\alpha}\sin 2\theta_{12}
      +\eta_{\alpha}\hat Y\cos\dCP\,\cos 2\theta_{12}, \\
  \mathcal{N}_{\alpha} &\equiv \sqrt{\mathcal{Q}_{\alpha}^{2}+\hat Y^{2}\sin^{2}\dCP},
\end{align}
then the solar angles in each respective $U^{\alpha}$ is
\begin{equation}
  \cos 2\theta_{12}^{\alpha}=\frac{\mathcal{P}_{\alpha}}{\mathcal{S}_{\alpha}},
  \qquad
  \sin 2\theta_{12}^{\alpha}=\frac{\mathcal{N}_{\alpha}}{\mathcal{S}_{\alpha}},
\end{equation}
and the CP-violating phase reads
\begin{equation}
  \cos\dCP^{\alpha}
    =\frac{\eta_{\alpha}\hat Y\cos 2\theta_{12}
           +\mathcal{F}_{\alpha}\cos\dCP\,\sin 2\theta_{12}}{\mathcal{N}_{\alpha}},
  \qquad
  \sin\dCP^{\alpha}
    =-\,\eta_{\alpha}\,
      \frac{\mathcal{S}_{\alpha}\sin 2\theta_{12}\sin\dCP}{\mathcal{N}_{\alpha}} .
\end{equation}
With $\Delta m_{ij}=m_i-m_j$ and the shorthands
\begin{equation}
  m_{1+3+\alpha}\equiv m_{1}+m_{3}+\dmaa,
  \qquad
  m_{1+2+\alpha}\equiv m_{1}+m_{2}+\dmaa,
\end{equation}
the $U^{\alpha}$ solar mass combination of Sec.~\ref{sec:SNSI_mm} and Sec.~\ref{sec:SNSI_tt} reads
\begin{equation}
  m^{+,\alpha}=m_{1}(s_{12}^{\alpha})^{2}+m_{2}(c_{12}^{\alpha})^{2}
             =m_{1}+\Delta m_{21}\,
              \frac{\mathcal{S}_{\alpha}+\mathcal{P}_{\alpha}}{2\mathcal{S}_{\alpha}},
\end{equation}
where the second equality follows from
$(c_{12}^{\alpha})^{2}=(\mathcal{S}_{\alpha}+\mathcal{P}_{\alpha})/2\mathcal{S}_{\alpha}$.
The two matter parameters then become
\begin{align}
  \DaaS &= \Delta m_{31}\,m_{1+3+\alpha}
          -\Delta m_{21}\,m_{1+2+\alpha}\,(c_{12}^{\alpha})^{2} \nonumber \\
         &= \Delta m_{31}\,m_{1+3+\alpha}
          -\Delta m_{21}\,m_{1+2+\alpha}\,
           \frac{\mathcal{S}_{\alpha}+\mathcal{P}_{\alpha}}{2\mathcal{S}_{\alpha}},
  \label{eq:DSpdg}\\
  \AaaS &= \dmaa\big(m^{+,\alpha}+m_{3}+\dmaa\big) \nonumber \\
         &= \dmaa\!\left(m_{1+3+\alpha}
           +\Delta m_{21}\frac{\mathcal{S}_{\alpha}+\mathcal{P}_{\alpha}}
                              {2\mathcal{S}_{\alpha}}\right).
  \label{eq:ASpdg}
\end{align}
The atmospheric sector is then governed by
\begin{align}
  h_{\alpha}&\equiv\DaaS(1-2\mathcal{S}_{\alpha})+\AaaS,
  \qquad
  g_{\alpha}\equiv\DaaS+\AaaS(1-2\mathcal{S}_{\alpha}),
  \qquad \\
  \lambda_{\alpha}&\equiv\sqrt{\big(\DaaS+\AaaS\big)^{2}
        -4\,\mathcal{S}_{\alpha}\,\DaaS\AaaS}\;>0,
\end{align}
which obey $\DaaS g_{\alpha}+\AaaS h_{\alpha}=\lambda_{\alpha}^{2}$ and
$\lambda_{\alpha}^{2}-h_{\alpha}^{2}=4\mathcal{S}_{\alpha}(1-\mathcal{S}_{\alpha})(\DaaS)^{2}$.
For the solar sector put
\begin{align}
  c_{23}'^{\alpha}&=\sqrt{\frac{\lambda_{\alpha}+g_{\alpha}}{2\lambda_{\alpha}}},
  &
  \chi_{\alpha}&\equiv\frac{\mathcal{S}_{\alpha}
      \big(\lambda_{\alpha}+\DaaS-\AaaS\big)}{2\lambda_{\alpha}\,c_{23}'^{\alpha}}, \\
  \mathcal{A}_{\alpha}&\equiv\mathcal{N}_{\alpha}
      \Big[\dmsq{21}\,c_{23}'^{\alpha}+\dmaa\,\Delta m_{21}\,\chi_{\alpha}\Big],
  &
  \mathcal{B}_{\alpha}&\equiv\dmsq{21}\,\mathcal{P}_{\alpha}
      +\mathcal{S}_{\alpha}\,\mathcal{K}_{\alpha},
\end{align}
with
\begin{equation}
  \mathcal{K}_{\alpha}\equiv\frac{\DaaS+\AaaS-\lambda_{\alpha}}{2}
      -\mathcal{S}_{\alpha}\,\dmaa\big(m_{3}-m^{+,\alpha}\big),
  \qquad
  \Lambda_{\alpha}\equiv\sqrt{\mathcal{A}_{\alpha}^{2}+\mathcal{B}_{\alpha}^{2}} .
\end{equation}
The $\dmaa\Delta m_{21}\chi_{\alpha}$ term in $\mathcal{A}_{\alpha}$ and the
$\mathcal{S}_{\alpha}\dmaa(m_{3}-m^{+,\alpha})$ term in $\mathcal{K}_{\alpha}$ are the two SNSI-specific contributions that arise in Sec.~\ref{sec:SNSImain}. Both vanish in the $\delta M =0$ limit.

Finally, writing
\begin{equation}
  \varrho_{\alpha}\equiv\frac{\lambda_{\alpha}+h_{\alpha}}{2\lambda_{\alpha}},
  \qquad\text{so that}\qquad
  \varrho_{\mu}=\sin^{2}\theta_{23}^{\mu,\rm eff},
  \qquad
  \varrho_{\tau}=\cos^{2}\theta_{23}^{\tau,\rm eff},
\end{equation}
the effective PDG parameters are the reactor angle
\begin{equation}
  \sin^{2}\theta_{13}\eff
    =\frac{s_{13}^{2}}{\mathcal{S}_{\alpha}}\big(1-\varrho_{\alpha}\big)
    =\frac{s_{13}^{2}}{2\mathcal{S}_{\alpha}}\,
      \frac{\lambda_{\alpha}-h_{\alpha}}{\lambda_{\alpha}},
  \label{eq:s2t13effpdg}
\end{equation}
the atmospheric angle
\begin{equation}
  \cos 2\theta_{23}\eff
    =-\eta_{\alpha}\,\frac{s_{13}^{2}\lambda_{\alpha}+\kappa_{\alpha}h_{\alpha}}
                          {\kappa_{\alpha}\lambda_{\alpha}+s_{13}^{2}h_{\alpha}},
  \qquad
  \sin 2\theta_{23}\eff
    =\frac{2\mathcal{S}_{\alpha}c_{13}^{2}\sin 2\theta_{23}\,\DaaS}
          {\kappa_{\alpha}\lambda_{\alpha}+s_{13}^{2}h_{\alpha}},
  \label{eq:t23effpdg}
\end{equation}
and the solar angle
\begin{equation}
  \cos 2\theta_{12}\eff
    =\frac{\big(\mathcal{C}_{\alpha}^{2}-s_{13}^{2}\varrho_{\alpha}\big)\mathcal{B}_{\alpha}
          +2\eta_{\alpha}s_{13}\mathcal{C}_{\alpha}\sqrt{\varrho_{\alpha}}
           \cos\dCP^{\alpha}\,\mathcal{A}_{\alpha}}
          {\big(\mathcal{C}_{\alpha}^{2}+s_{13}^{2}\varrho_{\alpha}\big)\Lambda_{\alpha}} .
  \label{eq:c2t12effpdg}
\end{equation}
The mass-squared splittings are
\begin{equation}
  \dmsqeff{21}=\frac{\Lambda_{\alpha}}{\mathcal{S}_{\alpha}},
  \qquad
  \dmsqeff{31}=\lambda_{\alpha}
      +\frac{\Lambda_{\alpha}+\mathcal{B}_{\alpha}}{2\mathcal{S}_{\alpha}},
  \label{eq:dmsqeffpdg}
\end{equation}
and finally $\dCP\eff$ can be extracted from the effective Jarlskog invariant given by
\begin{align}
  J\eff&= s_{12}\eff c_{12}\eff s_{23}\eff c_{23}\eff s_{13}\eff (c_{13}\eff)^{2}\sin\dCP\eff \nonumber\\
  &= -\frac{\eta_{\alpha}}{2}\,\frac{\mathcal{A}_{\alpha}}{\Lambda_{\alpha}}\,
        \frac{s_{13}\,\mathcal{C}_{\alpha}}{\mathcal{S}_{\alpha}}\,
        \big(1-\varrho_{\alpha}\big)\sqrt{\varrho_{\alpha}}\;\sin\dCP^{\alpha} .
  \label{eq:Jeffpdg}
\end{align}
We now have the converted relations for all the effective parameter mappings in Eqs.~\eqref{eq:s2t13effpdg}, \eqref{eq:t23effpdg}, \eqref{eq:c2t12effpdg}, \eqref{eq:dmsqeffpdg} and \eqref{eq:Jeffpdg}; where one should take $\alpha=\mu, \tau$ for converting the final relations in Sec.~\ref{sec:SNSI_mm}, \ref{sec:SNSI_tt} respectively. We emphasize that, even though $\dmaa$ only affects 4 parameters leaving 2 angles unchanged in the respective $U^\alpha$ parameterization, in the $U^e$ parameterization all 6 parameters are significantly affected by $\dmaa$.

We finally remark that the conversion of the $A_{\mu \mu}$ and $A_{\tau \tau}$ toy model results have been omitted here because they do not correspond to any contemporary phenomenological scenario. We leave it to the interested reader to follow the above procedure to straightforwardly derive the analogous formulas in these simpler scenarios.



\section{First order perturbative corrections} \label{app:perturbationtheory}
We here review the method for deriving first order corrections to the effective mixing parameters. We subsequently derive the first order corrections for the specific regime where $H^{(1)}=\diag(\ACC,0,0)$. These were applied to improve the accuracy of the effective parameter mappings in Sec.~\ref{sec:SNSI_mm} and Sec.~\ref{sec:SNSI_tt} when considering the JUNO experiment.

The expressions for first-order perturbative corrections to eigenvalues, $\lambda_i^{(1)}$, and eigenvectors, $\ket{\psi_i^{(1)}}$, are written as
\begin{align} 
    \lambda_i^{(1)}&={\bra{\psi^{(0)}_i} H^{(1)}\ket{\psi^{(0)}_i}} \label{eq:eigvaluepert} \\
    \ket{\psi_i^{(1)}} &= \sum_{j\neq i} { {\bra{\psi^{(0)}_j} H^{(1)}\ket{\psi^{(0)}_i}} \over \lambda^{(0)}_i - \lambda^{(0)}_j} \ket{\psi_j^{0}} \label{eq:eigvectorpert}
\end{align}
Using $H^{(1)}=\diag(\ACC,0,0)$ we first compute the eigenvalue corrections using Eq.~\eqref{eq:eigvaluepert} as
\begin{equation}
     \lambda_i^{(1)} = \left(U^{\rm{eff}} \, \rm{diag}(\ACC,0,0)  U^{\rm{eff}} \right)_{ii} = \ACC |U_{ei}^{\rm{eff}}]^2.
\end{equation}
and then get the mass squared difference corrections as 
\begin{equation} \label{eq:masssquareddiffcorrection}
    {\Delta m_{ij}^{2,\rm{eff}}}^{(1)}=\lambda_i^{(1)}-\lambda_j^{(1)}.
\end{equation}
Next, the eigenvector corrections in Eq.~\eqref{eq:eigvectorpert} imply corrections to the squared mixing matrix elements
\begin{align}
    &{U_{\alpha i}^{\rm{eff}}}^{(1)} = \sum_{j\neq i} { {\bra{\psi^{(0)}_j} H^{(1)}\ket{\psi^{(0)}_i}} \over \lambda^{(0)}_i - \lambda^{(0)}_j} U_{\alpha j}^{(0)}  =   \ACC \sum_{j\neq i} {{U_{ej}^{(0)}}^* U_{ei}^{(0)}  \over \lambda^{(0)}_i - \lambda^{(0)}_j } U_{\alpha j}^{(0)}, \label{eq:Ueffcorrections}  \\
    &{|U_{\alpha i}^{\rm{eff}}|^2}^{(1)} = 2 \ACC \sum_{j\neq i} {\Re( U_{ei}^{(0)} {U_{ej}^{(0)}}^*  {U_{\alpha i}^{(0)}}^* U_{ \alpha j}^{(0)}  )\over \lambda^{(0)}_i - \lambda^{(0)}_j }.  \label{eq:Ueff2corrections}
\end{align}
Subsequently, corrections of the form ${\sin^2 \theta_{ij}}^{(1)}$ can be derived from Eq.~\eqref{eq:Ueff2corrections} via the respective relations to the squared mixing matrix elements shown in Sec.~\ref{app:parameterizationconversion}. The mixing angle corrections then follow from the approximation
\begin{equation} \label{eq:mixinganglecorrection}
    \theta_{ij}^{(1)} \approx \frac{1}{\sin(2 \theta_{ij}^{(0)})} {\sin^2(\theta_{ij})}^{(1)} \approx \frac{\cos^4(\theta_{ij}^{(0)})}{\sin(2\theta_{ij}^{(0)})} {\tan^2(\theta_{ij})}^{(1)},
\end{equation}
which is accurate up to the second order in the series expansion.

Finally, the correction to $\dCP$ can be extracted from its relation to the quartic $\Pi \equiv U_{\alpha i} U_{\beta j} U_{\alpha j}^* U_{\beta i}^*$ and $\kappa\equiv \frac{|U_{ \alpha i}|^2 |U_{\beta j}|^2 |U_{\alpha j}|^2}{1-|U_{\beta j}|^2}$ as 
\begin{equation} \label{eq:dCPcorrection}
    \dCP^{(1)}= \frac{\Im\left((\Pi^{(1)} + \kappa^{(1)})e^{-i \dCP})\right)}{J/\sin \dCP}
\end{equation}
where the variations $\Pi^{(1)},\, \kappa^{(1)}$ are evaluated using Eqs.~\eqref{eq:Ueffcorrections} and \eqref{eq:Ueff2corrections}. The quartet is chosen here based on the parameterization of $\PMNS$ so that $\delta_{CP}$ only enters through one factor. For $U^{e}$ we use $(\alpha,\beta,i,j)=(e,\mu,2,3)$, for $U^\mu$ we use $(\alpha,\beta,i,j)=(\mu,\tau,2,3)$, and for $U^\tau$ we use $(\alpha,\beta,i,j)=(e,\tau,1,3)$

\subsection{Corrections in the $U^e$ basis} \label{app:Uecorrections}
Let us now consider the first order corrections from $H^{(1)}=\diag(\ACC,0,0)$ on the mixing parameters as expressed in the PDG parameterization, $U^e$. We reiterate that this is not a new result, but is worth reviewing in order to compare corrections between the three parameterizations used in this work.
Working out the expressions in App.~\ref{app:perturbationtheory} in terms of the $U^e$ parameters yields
\begin{align}
(\theta_{12}^{\rm{eff}})^{(0)+(1)} &= 
    \theta_{12}^{\rm{eff}} + \frac{\ACC}{2 \Delta m_{21}^{2,\rm{eff}}} \sin2\theta_{12}^{\rm{eff}} c_{13}^{2,\rm{eff}} \label{eq:1stordert12Ue}\\
(\theta_{13}^{\rm{eff}})^{(0)+(1)} &=
    \theta_{13}^{\rm{eff}} + \frac{\ACC}{2} \sin2\theta_{13}^{\rm{eff}}(\frac{c_{12}^{2,\rm{eff}}}{\Delta m_{31}^{2,\rm{eff}}} + \frac{s_{12}^{2,\rm{eff}}}{\Delta m_{32}^{2,\rm{eff}}}) \\
(\theta_{23}^{\rm{eff}})^{(0)+(1)} &= 
    \theta_{23} + \frac{\ACC}{2} \sin2\theta_{12}^{\rm{eff}} s_{13}^{\rm{eff}} \cos\dCP  (\frac{1}{\Delta m_{32}^{2,\rm{eff}}} -\frac{1}{\Delta m_{31}^{2,\rm{eff}}} ) \\
(\dCP\eff)^{(0)+(1)}& = 
    \dCP\eff-\frac{\ACC \Delta m_{21}^{2,\rm{eff}}}{\Delta m_{31}^{2,\rm{eff}} \Delta m_{32}^{2,\rm{eff}}} \sin 2\theta_{12}\eff s_{13}\eff \cot 2\theta_{23}\eff \sin\dCP\eff \\
(\Delta m_{21}^{2,\rm{eff}})^{(0)+(1)}
    &={\Delta m_{21}^{2,\rm{eff}}} - \ACC \cos2 \theta_{13}^{\rm{eff}} c_{13}^{2,\rm{eff}}  \\
(\Delta m_{31}^{2,\rm{eff}})^{(0)+(1)} 
    &= {\Delta m_{31}^{2,\rm{eff}}} + \ACC (s_{13}^{2,\rm{eff}} - c_{12}^{2,\rm{eff}} c_{13}^{2,\rm{eff}}) \label{eq:1storderdm31Ue}
\end{align}
We note that these results have been slightly simplified. In particular, the $\theta_{12}\eff$ and $\delta_{\text{CP}}\eff$ corrections yield expressions of the form $A/\Delta m_{21}^{2,\rm{eff}}+B/\Delta m_{3i}^{2,\rm{eff}}$ where we have omitted the terms $\propto 1/\Delta m_{3i}^{2,\rm{eff}}$ treating them as second order corrections. This simplification is implemented consistently in the numerical results below.

These first order corrections are expressed most simply in the $U^e$ parameterization as stated here. This is so because $H^{(1)}=\diag(\ACC,0,0)$ perturbs only the $e-e$ element, for which the $U^e$ parameterization is particularly well suited as was discussed in Sec.~\ref{sec:generalizing}. The disadvantage of applying corrections in the $U^e$ parameterization is that the 0th order results, which were worked out in the $U^\mu$ and $U^\tau$ parameterization respectively, must first be converted to the $U^e$ parameters in order to be added with these corrections. The alternative is to instead compute the corrections directly in the $U^\mu$ or $U^\tau$ parameterization so that no conversion is needed for the 0th order results. Conversely, the minor disadvantage with this approach is that the first order corrections will become a bit more complicated when expressed in the $U^\mu$ or $U^\tau$ parameters. With this trade-off in mind, in the following sections we work out the same first order corrections in terms of the $U^{\mu}$ and $U^{\tau}$ parameters.

\raggedbottom

\subsection{Corrections in the $U^\mu$ basis} \label{app:Umcorrections}
We now consider the first order corrections from $H^{(1)}=\diag(\ACC,0,0)$ to the $U^{\mu}$ mixing parameters. Using the formulas from App.~\ref{app:perturbationtheory} then yields
\begin{align}
    (\theta_{12}^{\mu,\text{eff}})^{(0)+(1)} = \theta_{12}^{\mu,\text{eff}}
        &+ \frac{\ACC}{\Delta m_{21}^{2,\rm{eff}}} \Big[
          s_{12}^{\mu,\rm{eff}} c_{12}^{\mu,\rm{eff}} \left( (c_{13}^{\mu,\rm{eff}})^2 - (s_{23}^{\mu,\rm{eff}})^2 (s_{13}^{\mu,\rm{eff}})^2 \right) \nonumber \\
        & - \cos 2\theta_{12}^{\mu,\text{eff}}  s_{23}^{\mu,\rm{eff}} s_{13}^{\mu,\rm{eff}} c_{13}^{\mu,\rm{eff}}  \cos\dCP^{\mu,\text{eff}} \Big] \\
    (\theta_{23}^{\mu,\text{eff}})^{(0)+(1)} = \theta_{23}^{\mu,\text{eff}}
        &+ \ACC \Big[ - s_{23}^{\mu,\rm{eff}} c_{23}^{\mu,\rm{eff}} (s_{13}^{\mu,\rm{eff}})^2
          \left( \frac{(s_{12}^{\mu,\rm{eff}})^2}{\Delta m_{31}^{2,\rm{eff}}}
               + \frac{(c_{12}^{\mu,\rm{eff}})^2}{\Delta m_{32}^{2,\rm{eff}}} \right) \nonumber \\
        &+ s_{12}^{\mu,\rm{eff}} c_{12}^{\mu,\rm{eff}} c_{23}^{\mu,\rm{eff}} s_{13}^{\mu,\rm{eff}} c_{13}^{\mu,\rm{eff}}   \cos \dCP^{\mu,\text{eff}}
          \left( \frac{1}{\Delta m_{31}^{2,\rm{eff}}} - \frac{1}{\Delta m_{32}^{2,\rm{eff}}} \right) \Big] \\
    (\theta_{13}^{\mu,\text{eff}})^{(0)+(1)} = \theta_{13}^{\mu,\text{eff}}
        &+ \ACC \Big[ s_{13}^{\mu,\rm{eff}} c_{13}^{\mu,\rm{eff}}
          \left( \frac{(c_{12}^{\mu,\rm{eff}})^2}{\Delta m_{31}^{2,\rm{eff}}}
               + \frac{(s_{12}^{\mu,\rm{eff}})^2}{\Delta m_{32}^{2,\rm{eff}}} \right) \nonumber \\
        & + s_{12}^{\mu,\rm{eff}} c_{12}^{\mu,\rm{eff}} s_{23}^{\mu,\rm{eff}}  (s_{13}^{\mu,\rm{eff}})^2 \cos\dCP^{\mu,\text{eff}}
          \left( \frac{1}{\Delta m_{31}^{2,\rm{eff}}} - \frac{1}{\Delta m_{32}^{2,\rm{eff}}} \right) \Big] \\
    (\dCP^{\mu,\text{eff}})^{(0)+(1)} = \dCP^{\mu,\text{eff}} 
        &+ \frac{\ACC}{\Delta m^{2,\text{eff}}_{21}} \frac{s_{23}^{\mu,\text{eff}} \sin 2 \theta_{13}^{\mu,\text{eff}}  \sin\dCP^{\mu,\text{eff}}}{\sin 2\theta_{12}^{\mu,\text{eff}}}   \\
    (\Delta m_{21}^{2,\rm{eff}})^{(0)+(1)} = \Delta m_{21}^{2,\rm{eff}}
        &+ \ACC \Big[ \cos 2\theta_{12}^{\mu,\text{eff}}
          \left( (s_{23}^{\mu,\rm{eff}})^2 (s_{13}^{\mu,\rm{eff}})^2 - (c_{13}^{\mu,\rm{eff}})^2 \right) \nonumber \\
        & - \sin 2\theta_{12}^{\mu,\text{eff}} s_{23}^{\mu,\rm{eff}}  \sin 2\theta_{13}^{\mu,\text{eff}} \cos\dCP^{\mu,\text{eff}} \Big] \\
    (\Delta m_{31}^{2,\rm{eff}})^{(0)+(1)} = \Delta m_{31}^{2,\rm{eff}}
        &+ \ACC \Big[ (c_{23}^{\mu,\rm{eff}})^2 (s_{13}^{\mu,\rm{eff}})^2 
        - (s_{12}^{\mu,\rm{eff}})^2  (s_{23}^{\mu,\rm{eff}})^2 (s_{13}^{\mu,\rm{eff}})^2 \nonumber \\
        & - (c_{12}^{\mu,\rm{eff}})^2 (c_{13}^{\mu,\rm{eff}})^2
        - \frac{1}{2}  \sin 2\theta_{12}^{\mu,\text{eff}} s_{23}^{\mu,\rm{eff}} \sin 2 \theta_{13}^{\mu,\text{eff}} \cos \dCP^{\mu,\text{eff}} \Big]
\end{align}
where $s_{ij}^{\mu,\rm{eff}}=\sin \theta_{ij}^{\mu,\rm{eff}}$, $c_{ij}^{\mu,\rm{eff}}=\cos \theta_{ij}^{\mu,\rm{eff}}$ and $\Delta m_{ij}^{2,\rm{eff}}$ are the 0th order results given by equations \eqref{eq:t23effmm}--\eqref{eq:dm31effmm}. 

Notice that the correction to the CP violating phase in the $U^\mu$ parameterization scheme contains terms $\propto 1/\Delta m_{21}^{2,\rm{eff}}$, which is different from the $U^e$ CP violating phase correction results that is $\propto \Delta m_{21}^{2,\rm{eff}}/(\Delta m_{31}^{2,\rm{eff}} \Delta m_{32}^{2,\rm{eff}})$. As a result, the corrections to $\dCP$ remain relatively small, whereas the corrections to $\dCP^\mu$ can fluctuate more substantially. The order of magnitude of the correction to $\theta_{12}\eff$ is similar to that of $\theta_{12}^{\mu, \text{eff}}$, which are both $\propto \ACC/\Delta m_{21}^{2,\rm{eff}}$. These are considerably larger than the corrections to $\theta_{13}\eff,\theta_{23}\eff$ and $\theta_{23}^{\mu,\text{eff}},\theta_{13}^{\mu,\text{eff}}$, which are all $\propto \frac{\ACC}{\Delta m_{3i}^{2,\rm{eff}}}$.
\raggedbottom

\subsection{Corrections in the $U^\tau$ basis} \label{app:Utcorrections}
Moving on to the $U^\tau$ parameters with the same first order correction, $H^{(1)}=\diag(\ACC,0,0)$, we obtain
\begin{align}
    (\theta_{12}^{\tau,\text{eff}})^{(0)+(1)} = \theta_{12}^{\tau,\text{eff}}
        &+ \frac{\ACC}{\Delta m_{21}^{2,\rm{eff}}} \Big[
          s_{12}^{\tau,\rm{eff}} c_{12}^{\tau,\rm{eff}}
          \left( (\tilde c_{12}^{\tau,\rm{eff}})^2 - (c_{23}^{\tau,\rm{eff}})^2 (\tilde s_{12}^{\tau,\rm{eff}})^2 \right) \nonumber \\
        & +  \cos 2\theta_{12}^{\tau,\text{eff}} c_{23}^{\tau,\rm{eff}} \tilde c_{12}^{\tau,\rm{eff}} \tilde s_{12}^{\tau,\rm{eff}}
           \cos\dCP^{\tau,\text{eff}} \Big] \\
    (\theta_{23}^{\tau,\text{eff}})^{(0)+(1)} = \theta_{23}^{\tau,\text{eff}}
        &+ \ACC \Big[s_{23}^{\tau,\rm{eff}} c_{23}^{\tau,\rm{eff}} (\tilde s_{12}^{\tau,\rm{eff}})^2 
          \left( \frac{(s_{12}^{\tau,\rm{eff}})^2}{\Delta m_{31}^{2,\rm{eff}}}
               + \frac{(c_{12}^{\tau,\rm{eff}})^2}{\Delta m_{32}^{2,\rm{eff}}} \right) \nonumber \\
        & + s_{12}^{\tau,\rm{eff}} c_{12}^{\tau,\rm{eff}} s_{23}^{\tau,\rm{eff}}
          \tilde s_{12}^{\tau,\rm{eff}} \tilde c_{12}^{\tau,\rm{eff}} \cos\dCP^{\tau,\text{eff}}
          \left( \frac{1}{\Delta m_{32}^{2,\rm{eff}}} - \frac{1}{\Delta m_{31}^{2,\rm{eff}}} \right) \Big] \\
    (\tilde{\theta}_{12}^{\tau,\text{eff}})^{(0)+(1)} = 
        \tilde{\theta}_{12}^{\tau,\text{eff}}
        &+ \ACC \Big[ \tilde c_{12}^{\tau,\rm{eff}} \tilde s_{12}^{\tau,\rm{eff}}
          \left( \frac{(c_{12}^{\tau,\rm{eff}})^2}{\Delta m_{31}^{2,\rm{eff}}}
               + \frac{(s_{12}^{\tau,\rm{eff}})^2}{\Delta m_{32}^{2,\rm{eff}}} \right) \nonumber \\
        & + s_{12}^{\tau,\rm{eff}} c_{12}^{\tau,\rm{eff}} c_{23}^{\tau,\rm{eff}} (\tilde s_{12}^{\tau,\rm{eff}})^2 \cos\dCP^{\tau,\text{eff}}
          \left( \frac{1}{\Delta m_{32}^{2,\rm{eff}}} - \frac{1}{\Delta m_{31}^{2,\rm{eff}}} \right) \Big] \\
    (\dCP^{\tau,\text{eff}})^{(0)+(1)} = \dCP^{\tau,\text{eff}} 
    &- \frac{\ACC}{\Delta m^{2,\text{eff}}_{21}} \frac{\cos\theta_{23}^{\tau,\text{eff}} \sin 2\tilde\theta_{12}^{\tau,\text{eff}} \sin\dCP^{\tau,\text{eff}}}{\sin 2\theta_{12}^{\tau,\text{eff}}} \\
    (\Delta m_{21}^{2,\rm{eff}})^{(0)+(1)} = \Delta m_{21}^{2,\rm{eff}}
        &+ \ACC \Big[ \left( (c_{12}^{\tau,\rm{eff}})^2 - (s_{12}^{\tau,\rm{eff}})^2 \right)
          \left( (c_{23}^{\tau,\rm{eff}})^2 (\tilde s_{12}^{\tau,\rm{eff}})^2 - (\tilde c_{12}^{\tau,\rm{eff}})^2 \right) \nonumber \\
        &+ \sin 2 \theta_{12}^{\tau,\text{eff}} c_{23}^{\tau,\rm{eff}}
           \sin 2 \tilde \theta_{12}^{\tau,\text{eff}} \cos\dCP^{\tau,\text{eff}} \Big] \\
    (\Delta m_{31}^{2,\rm{eff}})^{(0)+(1)} = \Delta m_{31}^{2,\rm{eff}}
        &+ \ACC \Big[ (s_{23}^{\tau,\rm{eff}})^2 (\tilde s_{12}^{\tau,\rm{eff}})^2 
        - (s_{12}^{\tau,\rm{eff}})^2 (c_{23}^{\tau,\rm{eff}})^2 (\tilde s_{12}^{\tau,\rm{eff}})^2 \nonumber \\ 
        &- (c_{12}^{\tau,\rm{eff}})^2 (\tilde c_{12}^{\tau,\rm{eff}})^2
        + 2 s_{12}^{\tau,\rm{eff}} c_{12}^{\tau,\rm{eff}}  c_{23}^{\tau,\rm{eff}} 
          \tilde c_{12}^{\tau,\rm{eff}} \tilde s_{12}^{\tau,\rm{eff}} \cos\dCP^{\tau,\text{eff}} \Big]
\end{align}
where $s_{12}^{\tau,\rm{eff}}=\sin \theta_{12}^{\tau,\rm{eff}}$, $s_{23}^{\tau,\rm{eff}}=\sin \theta_{23}^{\tau,\rm{eff}}$, $\tilde s_{12}^{\tau,\rm{eff}}=\sin \tilde \theta_{12}^{\tau,\rm{eff}}$ and analogous definitions for the cosines, with $\Delta m_{32}^{2,\rm{eff}} = \Delta m_{31}^{2,\rm{eff}} - \Delta m_{21}^{2,\rm{eff}}$. These trigonometric expressions, as well as $\Delta m_{ij}^{2,\rm{eff}}$, refer to the 0th order results from equations \eqref{eq:t23efftt}--\eqref{eq:dm31efftt}. As in the previous cases, the terms suppressed by $1/\Delta m_{3i}^{2,\rm{eff}}$ have been dropped in the corrections to $\theta_{12}^{\tau,\text{eff}}$ and $\dCP^{\tau,\text{eff}}$.

Overall, the above expressions for the corrections to the $U^\mu$ and $U^\tau$ parameters are both slightly more convoluted than the ones for the $U^e$ parameters. Regardless, the corrections exhibit very similar dependence on the respective parameters. We compute the all relative errors in this section using the $U^e$ parameter corrections, noting that the numerical difference between when the corrections are applied to the $U^e$ or $U^\mu$ or $U^\tau$ parameters is practically negligible. As noted before, the advantage of taking these slightly more convoluted relations is that they may directly be applied to the effective parameters in Sec.~\ref{sec:SNSI_mm} and \ref{sec:SNSI_tt} and subsequently used to compute the probability in Eq.~\eqref{eq:PfromU} without first arbitrarily converting the 0th order parameters to the $U^e$ basis. 
\raggedbottom




\bibliographystyle{JHEP}
\bibliography{biblio.bib}
\end{document}